\newcommand*{\ATLASLATEXPATH}{latex/}

\documentclass[cernpreprint,texlive=2014,txfonts,UKenglish]{latex/atlasdoc}
\pdfoutput=1

\usepackage{\ATLASLATEXPATH atlaspackage}
\usepackage{\ATLASLATEXPATH atlasbiblatex}
\usepackage{\ATLASLATEXPATH atlasphysics}
\usepackage{\ATLASLATEXPATH atlasheavyion}
\usepackage{\ATLASLATEXPATH atlasbiblatex}

\usepackage{\ATLASLATEXPATH atlascontribute}
\usepackage{float}
\usepackage{\ATLASLATEXPATH atlasphysics}
\renewcommand{\eqref}[1]{\textup{{\normalfont(\ref{#1}}\normalfont)}}

\graphicspath{{logos/}{figures/}}

\AtlasTitle{Measurement of multi-particle azimuthal correlations in $pp$, $p$+Pb and low-multiplicity Pb+Pb collisions with the ATLAS detector}

\author{The ATLAS Collaboration}

 \PreprintIdNumber{CERN-EP-2017-048}

\HepDataRecord{https://www.hepdata.net/record/77996}

\AtlasJournalRef{Eur. Phys. J. C 77 (2017) 428}
\AtlasDOI{10.1140/epjc/s10052-017-4988-1}

\AtlasAbstract{%
Multi-particle cumulants and corresponding Fourier harmonics are measured for azimuthal angle distributions of charged particles in \pp collisions at \sqs = 5.02 and 13~TeV and in \pPb collisions at \sqn = 5.02~TeV, and compared to the results obtained for low-multiplicity \PbPb collisions at \sqn = 2.76~TeV. These measurements aim to assess the collective nature of particle production. The measurements of multi-particle cumulants confirm the evidence for collective phenomena in \pPb and low-multiplicity \PbPb collisions.  On the other hand, the \pp results for four-particle cumulants do not demonstrate collective behaviour, indicating that they may be biased by contributions from non-flow correlations.  A comparison of multi-particle cumulants and derived Fourier harmonics across different collision systems is presented as a function of the charged-particle multiplicity. For a given multiplicity, the measured Fourier harmonics are largest in \PbPb, smaller in \pPb and smallest in \pp collisions. The \pp results show no dependence on the collision energy, nor on the multiplicity.  
}

\hypersetup{pdftitle={ATLAS document},pdfauthor={The ATLAS Collaboration}}

\begin{document}

\maketitle

\section{Introduction}
\label{sec:intro}
One of the signatures of the collective behaviour of the hot, dense medium produced in heavy-ion collisions is the azimuthal anisotropy of produced particles. This anisotropy results from spatial asymmetry in the initial interaction region in off-centre ion--ion collisions. The initial asymmetry activates strong pressure gradients along the shorter axis of the overlap region, leading to increased production of particles within the reaction plane, defined by the impact parameter vector (the vector separation of the barycentres of the two nuclei) and the beam axis.  The azimuthal anisotropy is commonly characterized by Fourier harmonics $\mathrm{v}_n$, referred to as single-particle harmonic flow coefficients: $\mathrm{v}_n= \langle \cos [n(\phi-\Phi_R)] \rangle$, where $\phi$ is the azimuthal angle of a produced particle and $\Phi_R$ is the azimuthal angle of the reaction plane~\cite{Poskanzer:1998yz}.  This anisotropic, collective  enhancement of particle production is a global long-range phenomenon extending over a wide pseudorapidity range. 

The anisotropy of charged-particle azimuthal angle distributions in \NucNuc collisions has been a subject of extensive experimental studies at RHIC  \cite{flowReview,flowReview1,brahms1,phobos1,star1,phenix1} and at the LHC  \cite{Alice1,Alice2,Atlas1,Alice4,Atlas2,Cms1,Cms2,Atlas3,Cms4,Cms5,
Alice9,Atlas5,Atlas6,Alice10,Alice11}.  In non-central heavy-ion collisions, the large and dominating $\mathrm{v}_2$ coefficient is mainly associated with the elliptic shape of the nuclear overlap. The $\mathrm{v}_2$ coefficient in ultra-central collisions and other $\mathrm{v}_n$ coefficients in all collisions are related to various geometric configurations arising from fluctuations of the nucleon positions in the overlap region \cite{roland,fluctCGC}.  The reported results are consistent with model calculations based on a hydrodynamic description of the system evolution and  provide conclusive evidence that the hot and dense matter produced in \NucNuc collisions behaves collectively in accordance with a hydrodynamic flow and has properties resembling those of a nearly perfect fluid~\cite{hydro1,hydro2,hydro3,Kovtun:2004de}. 

The study of \pNuc collisions was thought to provide information on cold nuclear matter effects, relevant for understanding the hot and dense system produced in \NucNuc collisions. In $p$+A collisions, the size of the produced system is small compared to the mean free path of its constituents. Therefore, it might be expected that the collective flow, if any, generated in $p$+A collisions is much weaker than in heavy-ion interactions. Contrary to these expectations, significant $\mathrm{v}_n$ coefficients, only about 40\% smaller in magnitude than those obtained in \PbPb collisions, have been measured in \pPb collisions at the LHC energy of \sqn = 5.02~TeV \cite{pPbcms1,pPbalice1,pPbatlas1,pPbatlas2,pPbcms2,pPbalice2,pPbatlas3,pPbcms3,pPbLHCb1,pPbcms4}. Observations of azimuthal anisotropies were also reported recently for $d$+Au \cite{rhicdAu1,rhicdAu2} and $^{3}$He+Au \cite{rhicHeAu} collisions at the RHIC energy of \sqn = 200~GeV. 

Interestingly, long-range two-particle azimuthal correlations have also been observed in high-multiplicity \pp collisions at the LHC energies \cite{ppcms1,ppatlas1,ppcms2,ppcms3,ppatlas2}. It was found that the measured azimuthal correlations, which extend over a wide range in pseudorapidity, can be explained by the $\cos (n\phi)$ modulation of the single-particle azimuthal angle distribution. The extracted Fourier harmonics $\mathrm{v}_n$ for $n=$ 2--4 \cite{ppatlas2} are generally much smaller than those measured in \pPb and \PbPb collisions, and show no dependency on the charged-particle multiplicity. On the other hand, they display a similar dependence on particle transverse momenta, suggesting that the same underlying mechanism may be responsible for the long-range azimuthal correlations. These observations in \pp collisions, together with the results from the \pNuc  system described above, are among the most challenging and pressing problems in the domain of soft quantum chromodynamics.  Various models have been proposed to explain the source of the observed long-range correlations in small collision systems \cite{werner,ryskin,wong,avsar,deng,bozek,strikman,bozek1,dusling1,kovchegov,dusling,BzdakSchenke,v468Ollitrault,MaBzdak,schenke1,chesler1,chesler2}, but the origin of the effect is still under intense debate.  It is not yet known whether the mechanism responsible for the observed collective behaviour in A+A collisions is also relevant for the smaller systems. The main purpose of this paper is to contribute to this debate by providing new experimental results.

Several differing analysis methods are applied to measure Fourier harmonics in high-energy collisions. They differ principally in their sensitivity to correlations not related to the initial collision geometry (referred to as non-flow correlations), which can result from resonance decays, jet production, Bose--Einstein correlations or energy--momentum conservation. For small collision systems and low-multiplicity final states, the most common method uses two-particle correlation functions \cite{phenix2p,pPbcms1,pPbalice1,pPbatlas1,pPbcms2,pPbatlas3,pPbcms3,pPbLHCb1,pPbcms4,
ppcms1,ppatlas1,ppcms2,ppcms3,ppatlas2}.  In this method,  the non-flow correlations are suppressed by requiring a large pseudorapidity separation,  $|\Delta\eta|$, between particles forming a pair.  This requirement eliminates most of the short-range correlations including intra-jet correlations. The jet--jet correlations are subtracted from the two-particle correlation function using the correlations measured in low-multiplicity events (see  e.g. \cite{ppatlas1,ppatlas2}). 

The multi-particle cumulant method \cite{Borghini:2000,Borghini:2001,Snellings} was proposed to suppress the non-flow correlations. The method aims to measure correlations between a large number of particles, from which the correlations between a small number of particles are subtracted. Since non-flow correlations typically involve a low number of particles, they are suppressed in many-particle cumulants. The drawback of the method is the statistical limitation in calculating the cumulants of more than two particles. Furthermore, the multi-particle cumulants in small collision systems, derived from correlations between low number of particles, can be biased by non-flow jet and dijet correlations, which dominate the azimuthal correlation signal.  The cumulant method has been applied to measure global correlations and Fourier harmonics in \PbPb and \pPb collisions \cite{Alice9,Atlas6,pPbatlas2,pPbcms2,pPbcms3}. Recently, the four- and six-particle cumulants were also  measured by the CMS Collaboration in \pp collisions at 5, 7 and 13 TeV \cite{ppcms3}. 

In this paper, the ATLAS measurements of multi-particle cumulants are presented for \pp collisions at 5.02 and 13 TeV and for \pPb collisions at \sqn= 5.02~TeV. For comparison, the results for low-multiplicity (peripheral)  \PbPb collisions at \sqn= 2.76 TeV are also shown. The results are averaged over large ranges in \pT and pseudorapidity. Results obtained from different collision systems are compared as a function of the charged-particle multiplicity.

The paper is organized as follows. The analysis method is described in the next section, followed by the description of the detector (Section~\ref{sec:detector}) and presentation of the analysed data samples and event and track selections in Sections~\ref{sec:sets} and \ref{sec:selections}. The analysis details are given in Section~\ref{sec:analysis} while Section~\ref{sec:syst} contains a discussion of systematic uncertainties and cross-checks. The results for cumulants and the corresponding Fourier harmonics are shown in Section~\ref{sec:result}. A summary and concluding remarks are given in Section~\ref{sec:conclusion}. 

\section{Multi-particle cumulants}
\label{sec:method}
The multi-particle cumulant method is useful in studying the global nature of correlations observed in azimuthal angles of particles produced in high-energy collisions. 
The cumulant method involves the calculation of $2k$-particle azimuthal correlations, $\mathrm{corr}_n\{2k\}$,  and cumulants, $c_n\{2k\}$, for $n^{\mathrm{th}}$ Fourier harmonics, where $n= 2, 3, 4$  and $k= 1, 2, 3, 4$ for the analysis presented in this paper. The $\mathrm{corr}_n\{2k\}$ are defined as \cite{Borghini:2000,Snellings}: 
\begin{eqnarray}
  &&   \langle \left\langle \mathrm{corr}_n\{2\}\right\rangle\rangle \equiv \langle \langle \mathrm{e}^{\mathrm{i}n(\phi_1-\phi_2)} \rangle\rangle,  \nonumber \\
  &&   \langle\left\langle \mathrm{corr}_n\{4\}\right\rangle\rangle \equiv \langle\langle \mathrm{e}^{\mathrm{i}n(\phi_1 +\phi_2-\phi_3-\phi_4)} \rangle\rangle, \nonumber \\
  &&   \langle\left\langle \mathrm{corr}_n\{6\}\right\rangle\rangle \equiv \langle\langle \mathrm{e}^{\mathrm{i}n(\phi_1 +\phi_2+\phi_3-\phi_4 -\phi_5-\phi_6)} \rangle\rangle, \nonumber \\
  &&   \langle\left\langle \mathrm{corr}_n\{8\}\right\rangle\rangle \equiv \langle\langle \mathrm{e}^{\mathrm{i}n(\phi_1 +\phi_2+\phi_3+\phi_4 -\phi_5-\phi_6-\phi_7-\phi_8)} \rangle\rangle, \nonumber 
\end{eqnarray}
\noindent
where the brackets ``$\left\langle \langle \right\rangle \rangle$'' denote double averaging, performed first over particles in an event and then over all events within a given event class.  For every event, the average is taken over all possible of the combinations of the azimuthal angles $\phi_{i} ( i= 1,\ldots,8)$ of the  $2k$ particles. 

With the calculated multi-particle azimuthal correlations,  the cumulants $c_n\{2k\}$  are obtained after subtracting the correlations between $2(k-1)$ particles according to the following formulae~\cite{Borghini:2000,Snellings}:
\begin{eqnarray}
  &&   c_n\{2\}=\langle\left\langle \mathrm{corr}_n\{2\}\right\rangle\rangle,  \nonumber \\
  &&   c_n\{4\}=\langle\left\langle \mathrm{corr}_n\{4\}\right\rangle\rangle -2\langle\left\langle \mathrm{corr}_n\{2\}\right\rangle\rangle ^2, \nonumber \\
  &&   c_n\{6\}=\langle\left\langle \mathrm{corr}_n\{6\}\right\rangle\rangle -9\langle\left\langle \mathrm{corr}_n\{2\}\right\rangle\rangle \langle\left\langle \mathrm{corr}_n\{4\}\right\rangle\rangle +12 \langle\left\langle \mathrm{corr}_n\{2\}\right\rangle\rangle ^3, \nonumber \\
  &&   c_n\{8\}=\langle\left\langle \mathrm{corr}_n\{8\}\right\rangle\rangle -16\langle\left\langle \mathrm{corr}_n\{2\}\right\rangle\rangle \langle\left\langle \mathrm{corr}_n\{6\}\right\rangle\rangle -18 \langle\left\langle \mathrm{corr}_n\{4\}\right\rangle\rangle ^2 \nonumber\\&& \hspace{1.4cm} +144 \langle\left\langle \mathrm{corr}_n\{2\}\right\rangle\rangle ^2\langle\left\langle \mathrm{corr}_n\{4\}\right\rangle\rangle -144\langle\left\langle \mathrm{corr}_n\{2\}\right\rangle\rangle^4. \nonumber 
\end{eqnarray} 

The Q-cumulant  method ~\cite{Snellings}, used in this analysis, relies on the idea of expressing the multi-particle correlations in terms of powers of the flow vector $Q_{n}$. This approach allows multi-particle correlations and cumulants to be calculated in a single pass over data events. The flow vector is defined for each collision event with multiplicity $M$ as: 
\begin{equation}
Q_{n,j}\equiv \sum_{i=1}^{M} w_{i}^j  \mathrm{e}^{\mathrm{i}n\phi_i},
\label{eq:qnweighted}
\end{equation}
\noindent
where the subscript $n$ denotes the order of the flow harmonic, $j$ is the power of the flow vector, and the sum runs over all particles in an event with $w_{i}$ being the weight of the $i^{\mathrm{th}}$ particle.  The weight accounts for detector effects including the tracking efficiency and is defined in Section~\ref{sec:analysis}. 

If the measured $c_n\{2k\}$ cumulants are free of non-flow correlations, they can be used to estimate Fourier harmonics $\mathrm{v}_n$. Furthermore, assuming that the event-by-event fluctuations of $\mathrm{v}_n$ are negligibly small, the Fourier harmonics denoted by $\mathrm{v}_n\{2k\}$ can be determined ~\cite{Borghini:2000}:
\begin{eqnarray}
  &&  \label{eq:vn2} \mathrm{v}_n\{2\}=\sqrt{c_n\{2\}},\\
  &&  \label{eq:vn4} \mathrm{v}_n\{4\}=\sqrt[4]{-c_n\{4\}},\\
  &&  \label{eq:vn6} \mathrm{v}_n\{6\}=\sqrt[6]{c_n\{6\}/4},\\
  &&  \label{eq:vn8} \mathrm{v}_n\{8\}=\sqrt[8]{-c_n\{8\}/33}.
\end{eqnarray}
From the above definitions it is evident that determination of real values of Fourier harmonics  requires negative (positive) $c_n\{4\}$ and $c_n\{8\}$ ($c_n\{2\}$ and $c_n\{6\}$) values. 
\section{ATLAS detector}
\label{sec:detector}
The data were collected with the ATLAS detector \cite{atlas-detector}.\footnote{ATLAS uses a right-handed coordinate system with its origin at the nominal interaction point (IP)
in the centre of the detector and the $z$-axis along the beam pipe.
The $x$-axis points from the IP to the centre of the LHC ring,
and the $y$-axis points upwards.
Cylindrical coordinates $(r,\phi)$ are used in the transverse plane, 
$\phi$ being the azimuthal angle around the $z$-axis.
The pseudorapidity is defined in terms of the polar angle $\theta$ as $\eta = -\ln \tan(\theta/2)$.}
The detector consists of three main systems: an inner tracking detector (ID) surrounded by a thin superconducting solenoid, electromagnetic and hadronic calorimeters,
and a muon spectrometer.  The ID is immersed in a \SI{2}{\tesla} axial magnetic field and provides charged-particle tracking in the range $|\eta| < 2.5$. It consists of silicon pixel, silicon microstrip (SCT), and straw-tube transition radiation tracking detectors. Since 2015 the pixel detector includes an additional layer at smaller radius, the “insertable B-layer” (IBL) \cite{ATLAS-TDR-19}. The calorimeter system covers the pseudorapidity range up to $|\eta| = 4.9$. The muon spectrometer surrounds the calorimeters and is based on
three large air-core toroid superconducting magnets with eight coils each.
The field integral of the toroids ranges between \SI{2}  to \SI{6}{\tesla\metre} across most of the detector.
Measurements presented in this document  use signals from the ID while other components are used for triggering. 

Events are selected with a trigger system \cite{atlas-trigger-2010}.
The first-level (L1) trigger is implemented in hardware and uses a subset of the detector information. For this analysis the information from calorimeters, minimum bias trigger scintillator (MBTS) counters (covering the range $2.1 < |\eta| < 3.8$) and zero degree calorimeters (ZDCs) with the range $|\eta| > 8.3$ is used at L1.
The L1 trigger is followed by two software-based trigger levels: level-2 (L2) and Event Filter (EF). In \pp data-taking in 2015, the L2 and EF trigger levels are combined in a common high-level trigger (HLT) framework. 

\section{Data sets}
\label{sec:sets}

The \sqs= 5.02~TeV \pp data were recorded in November 2015 and correspond to an integrated luminosity of about 28~pb$^{-1}$. The average number of additional interactions in the same bunch crossing, $\mu$, ranges from 0.4 to 1.3.  For the low-multiplicity event selections, three minimum-bias triggers were used:  the first required a hit in at least one MBTS counter, the second required a hit in at least one MBTS counter on each side, and the third required at least one reconstructed track at the HLT seeded by a random trigger at L1.  In order to enhance the number of high-multiplicity events, dedicated high-multiplicity triggers (HMTs) were implemented. Three HMTs required at L1 more than 5~GeV, 10~GeV and 20~GeV in the total transverse energy ($\sum E_{\mathrm{T}}$) recorded in the calorimeters, and at the HLT more than 60, 90 and 90 reconstructed charged-particle tracks with $\pT > 0.4$~GeV and $|\eta|<2.5$, respectively. 

The \sqs= 13~TeV \pp data were taken over two running periods in June and August of 2015. For the first running period, $\mu$ varied between 0.002 and 0.03, while for the second $\mu$ ranged from 0.05 to 0.6. The total integrated luminosity collected over these two periods is approximately 0.075~pb $^{-1}$. In addition to the minimum-bias event trigger, HMTs were implemented seeded by a L1 requirement of $\sum E_{\mathrm{T}} > 10$~GeV. For the low-$\mu$ running period, the requirement of  more than 60 reconstructed charged-particle tracks at the HLT was imposed. For the moderate-$\mu$ data (the second data-taking period), two requirements on the number of online reconstructed charged-particle tracks at the HLT, of more than 60 and 90, were employed. 

The \pPb data were collected during the LHC run at the beginning of 2013. The  LHC operated in two configurations during this running period, by reversing the directions of the proton and lead beams. The proton beam with the  energy of 4 TeV collided with a Pb beam of energy  $1.57$~TeV per nucleon. This leads to \sqn = 5.02~TeV in the nucleon--nucleon centre-of-mass frame, which is shifted by 0.465 in rapidity in the proton direction.  The total integrated luminosity corresponds to approximately 0.028~\ipb. The data were recorded with the minimum-bias trigger and several HMTs, seeded by L1 thresholds on the total transverse energy recorded in the forward calorimeters ($\sum E_{\mathrm{T}}^{\mathrm{FCal}}, 3.1 < |\eta| < 4.9$) and HLT thresholds on the number of online reconstructed charged-particle tracks, $N_{\mathrm{ch}}^{\mathrm{online}}$ \cite{ATLAS-CONF-2013-104}. Six different combinations of the L1 and HLT thresholds were implemented: ($\sum E_{\mathrm{T}}^{\mathrm{FCal}} [\mathrm{GeV}] >, N_{\mathrm{ch}}^{\mathrm{online}} >$) = (10,100), (10,130), (50,150), (50,180), (65,200) and (65,225). More details can be found in Ref.~\cite{pPbatlas3}. For the \pPb data, $\mu \approx 0.03$. 

The \sqn= 2.76~TeV \PbPb data set used in this analysis consists of the data collected in 2010 and then reprocessed in 2014 with the same reconstruction software as used for \pPb data. The number of additional interactions per bunch crossing is negligibly small, of the order of $10^{-4}$. 

Monte Carlo (MC) simulated event samples are used to determine the track reconstruction efficiency (Section~\ref{sec:selections}) and to perform  closure tests, as described in Section~\ref{sec:syst}. For the  13~TeV and 5.02 TeV  \pp data the baseline MC event generator used is \PYTHIA 8  \cite{pythia8} with parameter values set according to the ATLAS A2 tune \cite{ATL-PHYS-PUB-2011-009} and with MSTW2008LO parton distribution functions \cite{MSTW2008}. The \textsc{Hijing}
 event generator \cite{hijing} is used to produce \pPb and \PbPb collisions with the same energy  as in the data. 
The detector response is simulated  \cite{AtlasInfra} with \GEANT 4 \cite{geant} and with detector conditions matching those during the data-taking. The simulated events are reconstructed with the same algorithms as data events, including track reconstruction.  
\section{Event and track selections}
\label{sec:selections}
Additional event selections are implemented in the offline analysis.
Events are required to have a reconstructed vertex.   For the \pPb and \PbPb data, only events with a reconstructed vertex for which $|z_{\mathrm{vtx}}| < 150$~mm are selected while for \pp data sets this requirement is not applied. 

In order to suppress additional interactions per bunch crossing (referred to as pile-up) in \pp data sets, only tracks associated with the vertex for which the $\sum \pT^{2}$ is the largest are used. In addition, all events with a second vertex reconstructed from at least four  tracks are disregarded. 
For the \pPb data, even though the average number of interactions per bunch crossing is small ($\sim 0.03)$, it can be significantly larger in events with a high multiplicity. Therefore, events containing more than one interaction per bunch crossing are rejected if they contain more than one good reconstructed vertex, where a good vertex is defined as that with the scalar sum of the tracks transverse momenta $\sum{\pT} > 5$~GeV. The remaining pile-up events are further suppressed using the ZDC signal on the Pb-fragmentation side, calibrated to the number of recorded neutrons \cite{pPbatlas3}. In order to suppress beam backgrounds in \pPb and \PbPb data, a requirement on the time difference between signals from MBTS counters on opposite sides of the interaction region is imposed, $|\Delta t| < 10$~ns and < 3~ns, respectively.  

For the \pp data, charged-particle tracks are reconstructed in the ID with the tracking algorithm optimized for Run-2 data \cite{ATL-PHYS-PUB-2015-006}. The tracks are required to have
$|\eta| < 2.5 $ and $\pT > 0.1$~GeV. At least one pixel hit is required and a hit in the IBL is also required if the track passes through the active region of the IBL. If a track passes through an inactive area of the IBL, then a hit is required in the next pixel layer if one is expected. The requirement on the minimum number of SCT hits depends on \pT:
$\geq 2$ for $0.1 < \pT < 0.3$~GeV, $\geq 4$ for $0.3 < \pT < 0.4$~GeV and $\geq 6$ for $\pT > 0.4$~GeV. Additional selection requirements are imposed on the transverse, $|d_0|$, and longitudinal, $|z_0 \sin  \theta| $, impact parameters. The transverse impact parameter is measured with respect to the beam line, and $z_0$ is the difference between the longitudinal position (along the beam line) of the track at the point where $d_0$ is measured and the primary vertex. Both must be smaller than 1.5~mm. In order to reject tracks with incorrectly measured \pT due to interactions with the detector material, the track-fit probability must be larger than 0.01 for tracks with $\pT >10$~GeV.  

For the reconstruction of \pPb and \PbPb data, the same tracking algorithms are used. The track selection requirements are modified slightly from those applied in the \pp reconstruction. Specifically, the same requirements are imposed on the impact parameters, although $|d_0|$ is determined with respect to the primary vertex. To suppress falsely reconstructed charged-particle tracks, additional requirements are imposed on the significance of the transverse and longitudinal impact parameters: $|d_0|/\sigma_{d_0}<3$ and $|z_0\sin \theta| /\sigma_{z_0}<3$, where $\sigma_{d_0}$ and $\sigma_{z_0}$ are the uncertainties in the transverse and longitudinal impact parameter values, respectively, as obtained from the covariance matrix of the track fit. 

The tracking efficiencies are estimated using the MC samples reconstructed with the same tracking algorithms and the same track selection requirements. Efficiencies, $\epsilon(\eta, \pT)$, are evaluated as a function of track $\eta$, \pT  and the number of reconstructed charged-particle tracks, but averaged over the full range in azimuth. For all collision systems, the efficiency increases by about 4\% with \pT increasing from 0.3~GeV to 0.6~GeV. Above 0.6~GeV, the efficiency is independent of \pT and reaches 86\% (72\%) at $\eta\approx 0$ ($|\eta|>2$),
83\% (70\%) and 83\% (70\%) for \pp, \pPb and peripheral \PbPb collisions, respectively. The efficiency is independent of the event multiplicity for $N_{\mathrm{ch}} > 40$.  For lower-multiplicity events the efficiency is smaller by a few percent. The rate of falsely reconstructed charged-particle tracks, $f(\pT,\eta)$, is also estimated and found to be small; even at the lowest transverse momenta it stays below 1\% (3\%) at  $\eta\approx 0$ ($|\eta|>2$).

Residual detector defects (not accounted for by tracking efficiencies), which may arise on a run-by-run basis and could lead to a non-uniformity of the azimuthal angle distribution, are corrected for by a data-driven approach, the so-called flattening procedure  described in Section~\ref{sec:analysis}. 

The analysis is performed as a function of the charged-particle multiplicity. Three measures of the event multiplicity are defined based on counting the number of particles observed in different transverse momentum ranges: $0.3 < \pT < 3$~GeV,  $0.5 < \pT < 5$~GeV and $\pT > 0.4$~GeV  (see next section for details). For each multiplicity definition, only events with multiplicity $\geq 10$ are used to allow a robust calculation of the multi-particle cumulants. Furthermore, in order to avoid potential biases due to HMT inefficiencies, events selected by the HMTs are accepted only if the trigger efficiency for each multiplicity definition exceeds 90\%. The only exception is the \pp 13 TeV data collected in August 2015 with the HMT requiring more than 90 particles reconstructed at the HLT, for which the 90\% efficiency is not reached. It was carefully checked that inclusion of this data set does not generate any bias in the calculation of multi-particle cumulants.

\section{Overview of the analysis}
\label{sec:analysis}
\begin{figure}[ht!]
\begin{center}
\includegraphics[width=60mm]{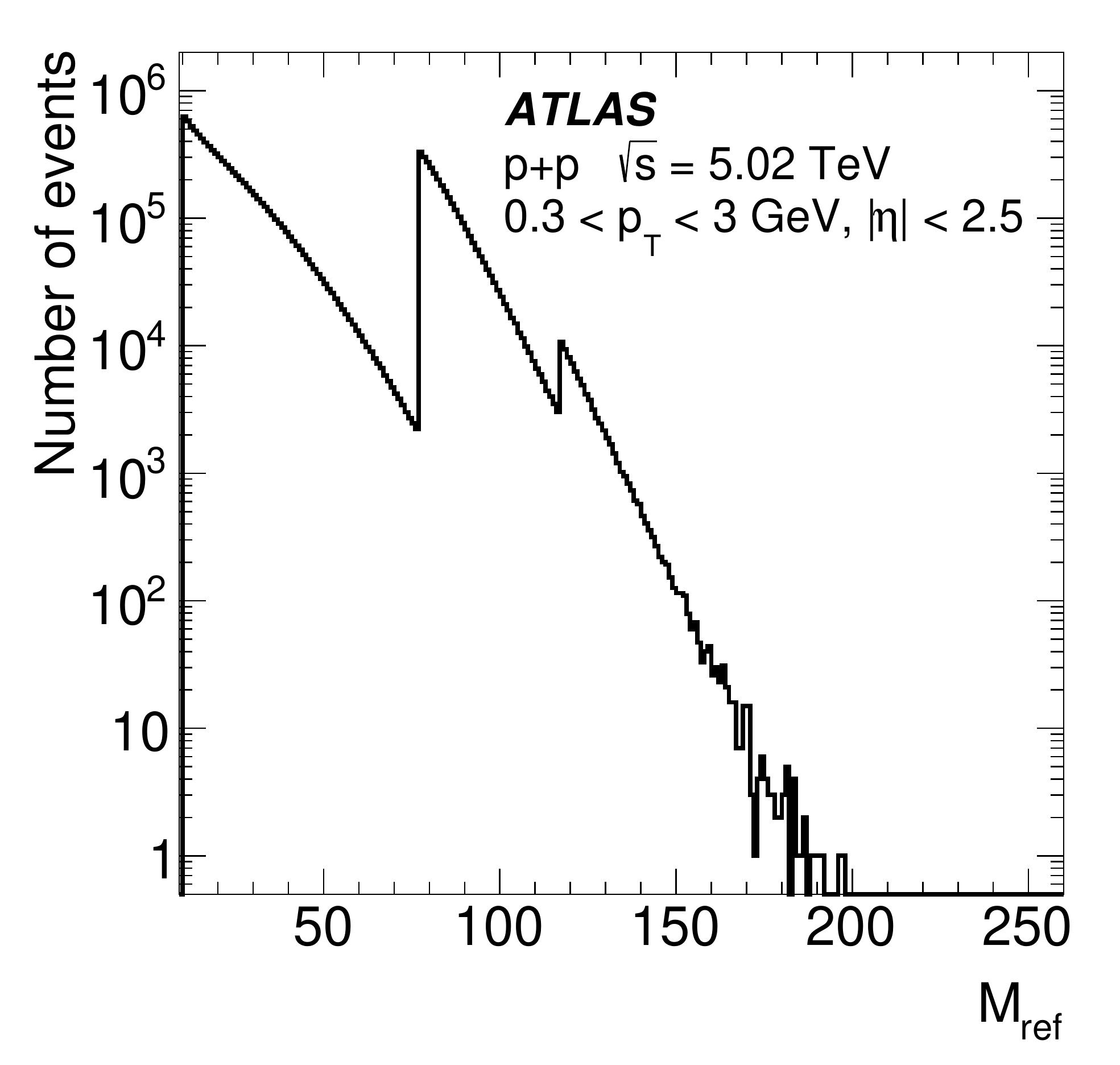}
\includegraphics[width=60mm]{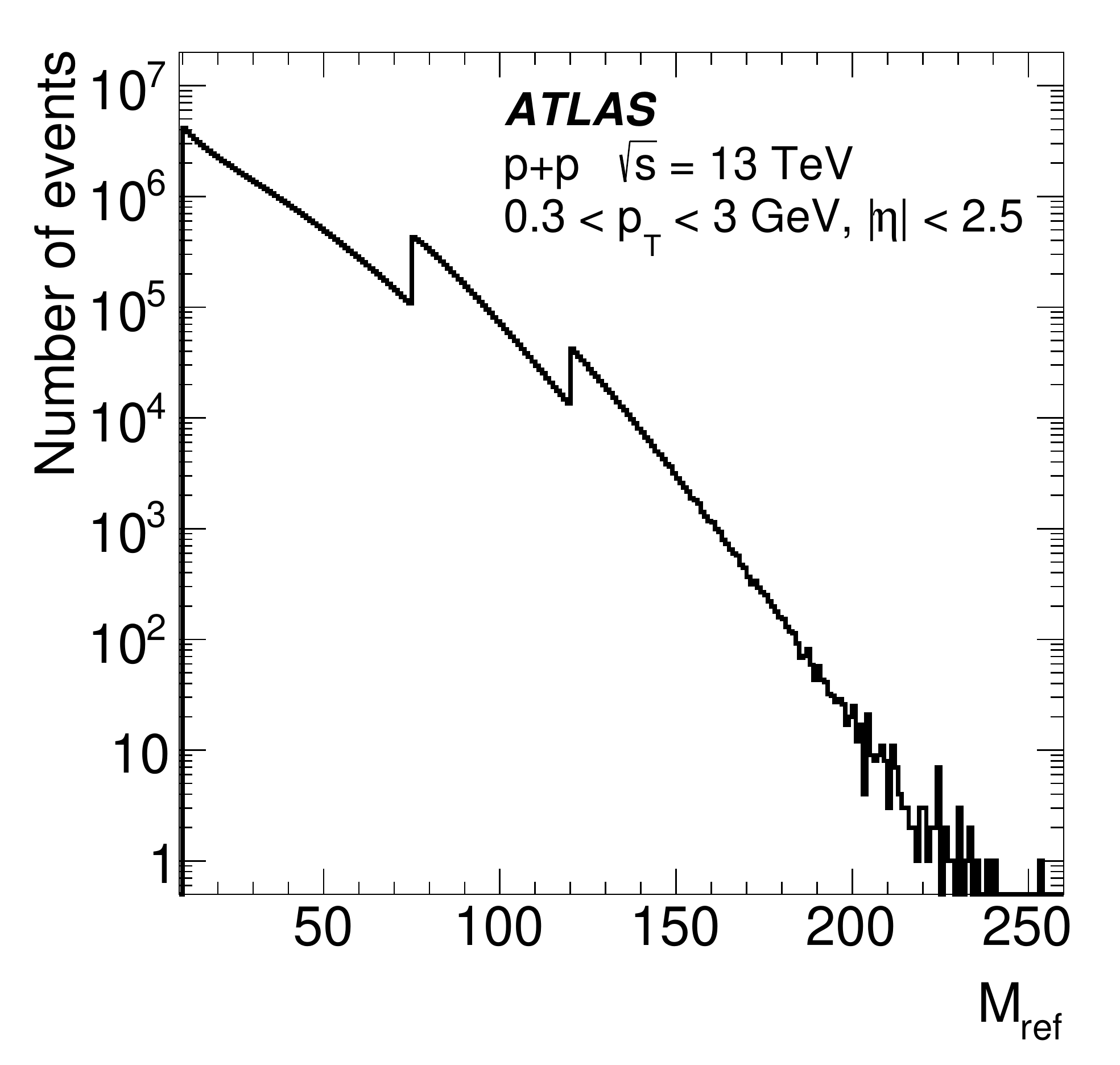}
\includegraphics[width=60mm]{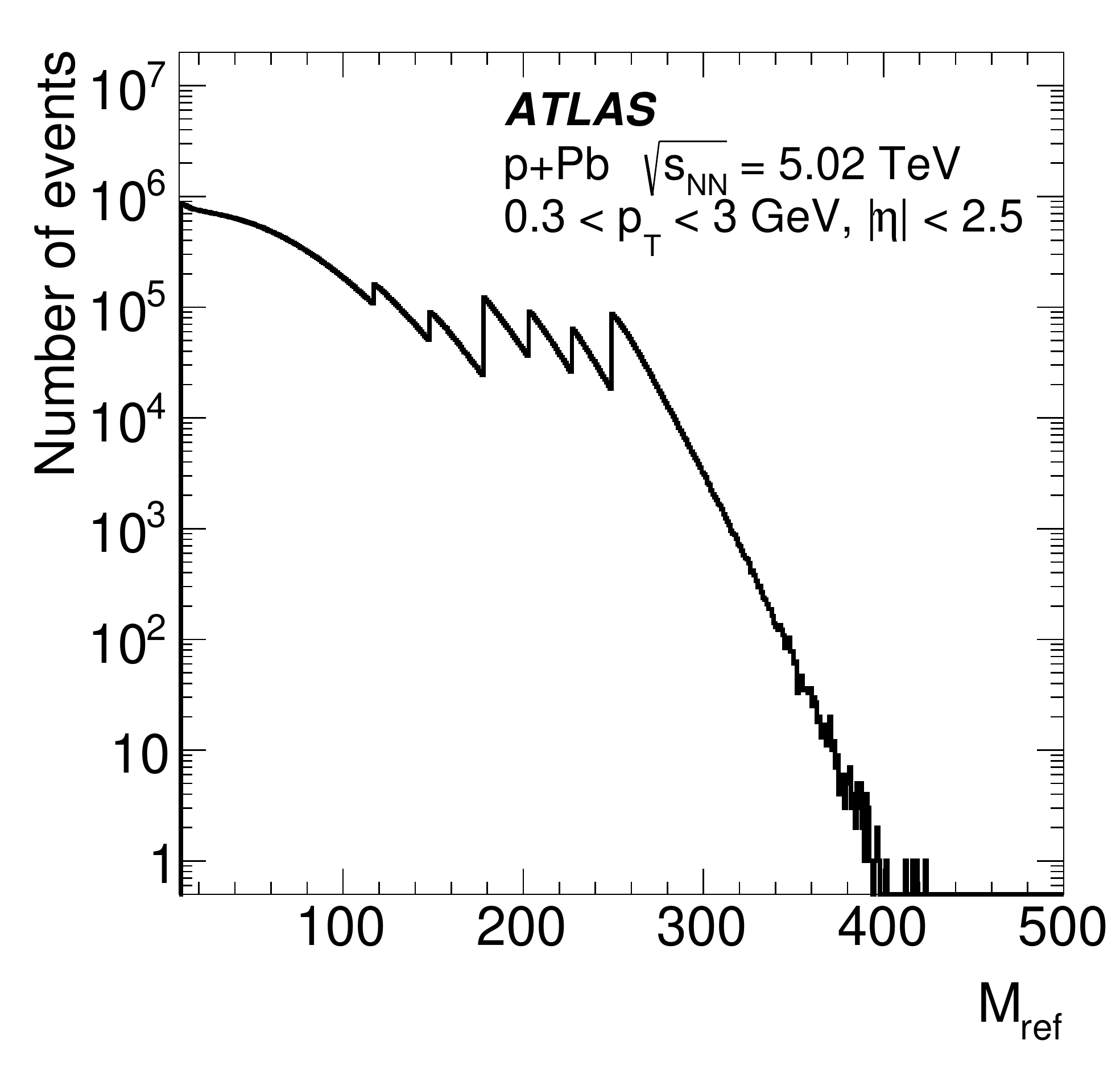}
\includegraphics[width=60mm]{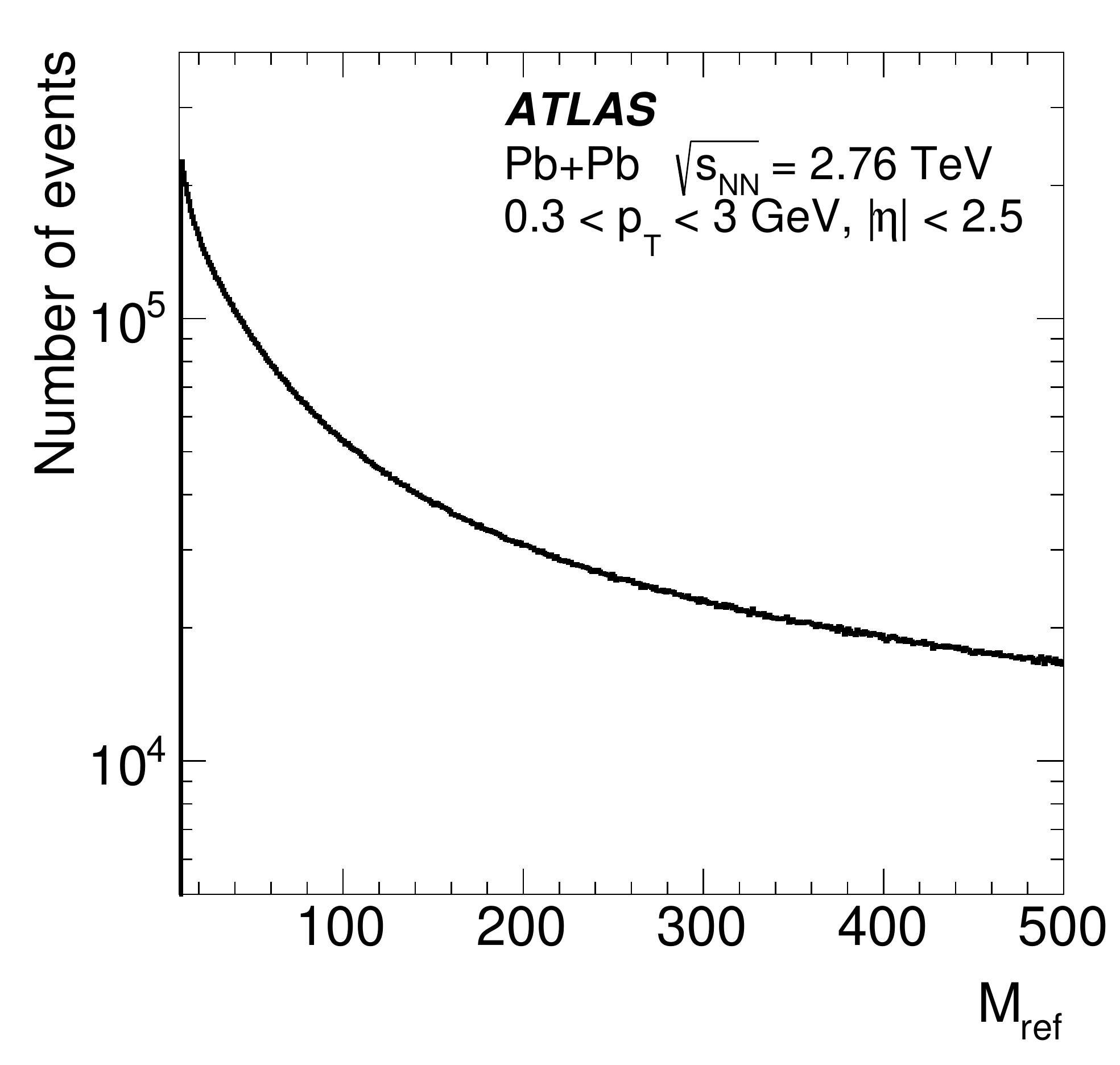}
\caption{Distributions of the reference particle multiplicity, $M_{\mathrm{ref}}$, for the selected reference particles with $ 0.3 < \pT <3$~GeV for \pp collisions at \sqs= 5.02 and 13 TeV, \pPb collisions at \sqn= 5.02 TeV and low-multiplicity \PbPb collisions at \sqn= 2.76 TeV. The discontinuities in the upper and lower-left distributions correspond to different high-multiplicity trigger thresholds.}
\label{fig:ntrkAllsystems033} 
\end{center}
\end{figure} 

For each collision system, the multi-particle cumulants are calculated using the so-called reference particles.  Two selections of reference particles are considered, for which the multiplicity $M_{\mathrm{ref}}$ in a given event is the number of reconstructed charged particles with $|\eta|<2.5$ and with corresponding \pT ranges: $0.3 < \pT < 3$~GeV or $0.5 < \pT < 5$~GeV. Figure~\ref{fig:ntrkAllsystems033} shows the uncorrected $M_{\mathrm{ref}}$ multiplicity distributions for the reconstructed charged-particle tracks with $ 0.3 < \pT <3$~GeV for all collision systems. The observed discontinuities  reflect the offline selection requirement of at least 90\% efficiency for the HMT thresholds. Event weights are introduced to account for the trigger efficiency and the trigger prescale factors \cite{pPbatlas3}. 

Particle weights (see Eq.~\eqref{eq:qnweighted}) are applied to account for detector effects via $w_{\phi}(\eta,\phi)$, the tracking efficiency $\epsilon(\eta,\pT)$ and the rate of fake tracks $f(\eta,\pT)$, and are defined as:
\begin{equation}
w_{i}(\eta,\phi,\pT) = \frac{w_{\phi,i}(\eta,\phi)(1-f_i(\eta,\pT))}{\epsilon_i(\eta,\pT)}. \nonumber 
\end{equation}
The tracking efficiencies and fake rates are determined as described in Section~\ref{sec:selections}. The weights $w_{\phi}(\eta,\phi)$ are determined from the data by the procedure of azimuthal-angle flattening in order  to correct for non-uniformity of the azimuthal acceptance of the detector.  The flattening procedure uses the $\eta$--$\phi$ map of all reconstructed charged-particle tracks. For each small interval $(\delta\eta,\delta\phi)$, a ``flattening'' weight is calculated as $w_{\phi}(\eta,\phi)=\langle N(\delta\eta) \rangle/N(\delta\eta,\delta\phi)$ where $\langle N(\delta\eta) \rangle$ is the event-averaged number of tracks in the $\delta\eta$ slice, averaged over the full range in $\phi$, while $N(\delta\eta,\delta\phi)$ is the number of tracks within this interval. 

\begin{figure}[ht!]
\begin{center}
\includegraphics[width=60mm]{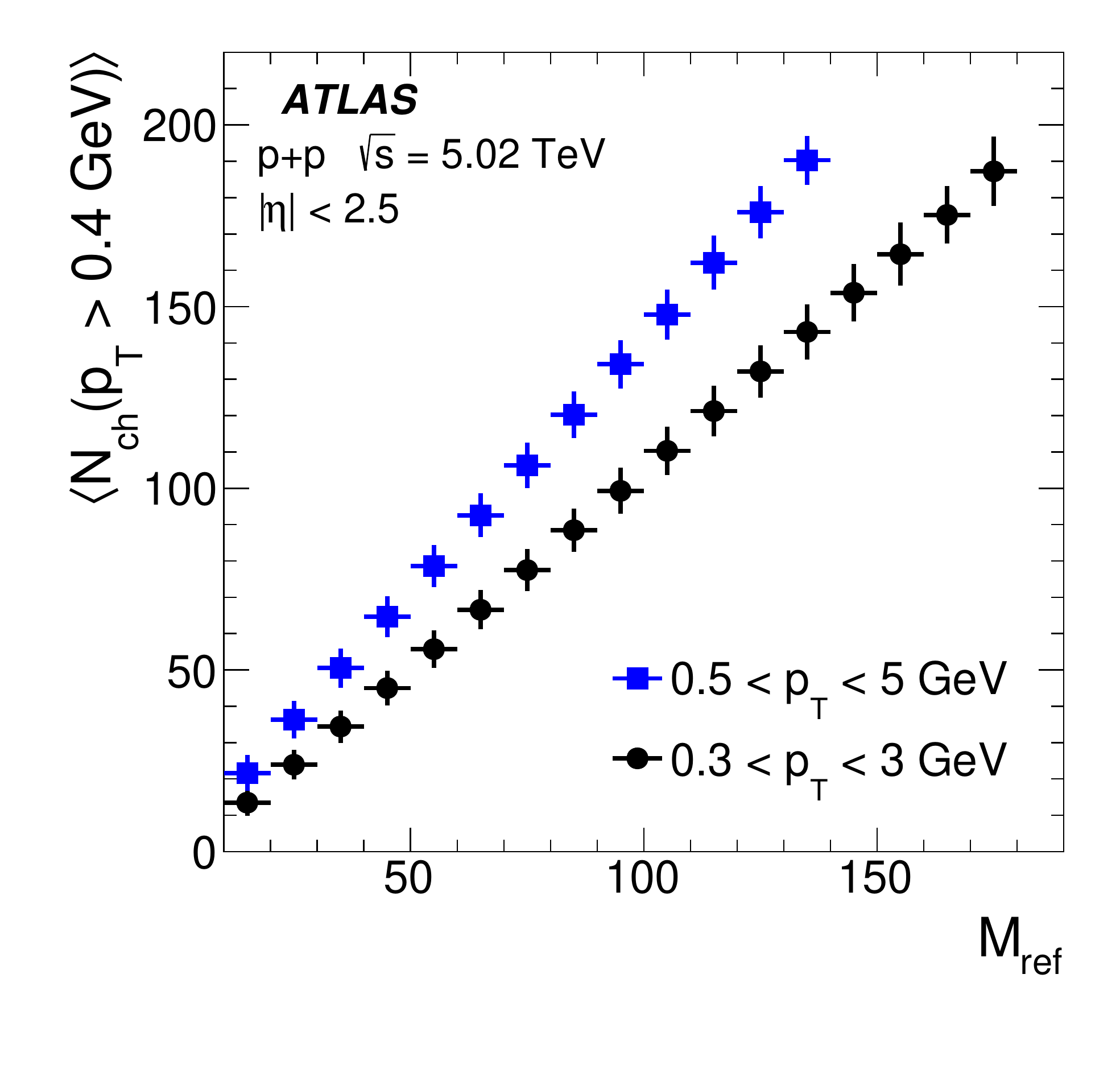}
\includegraphics[width=60mm]{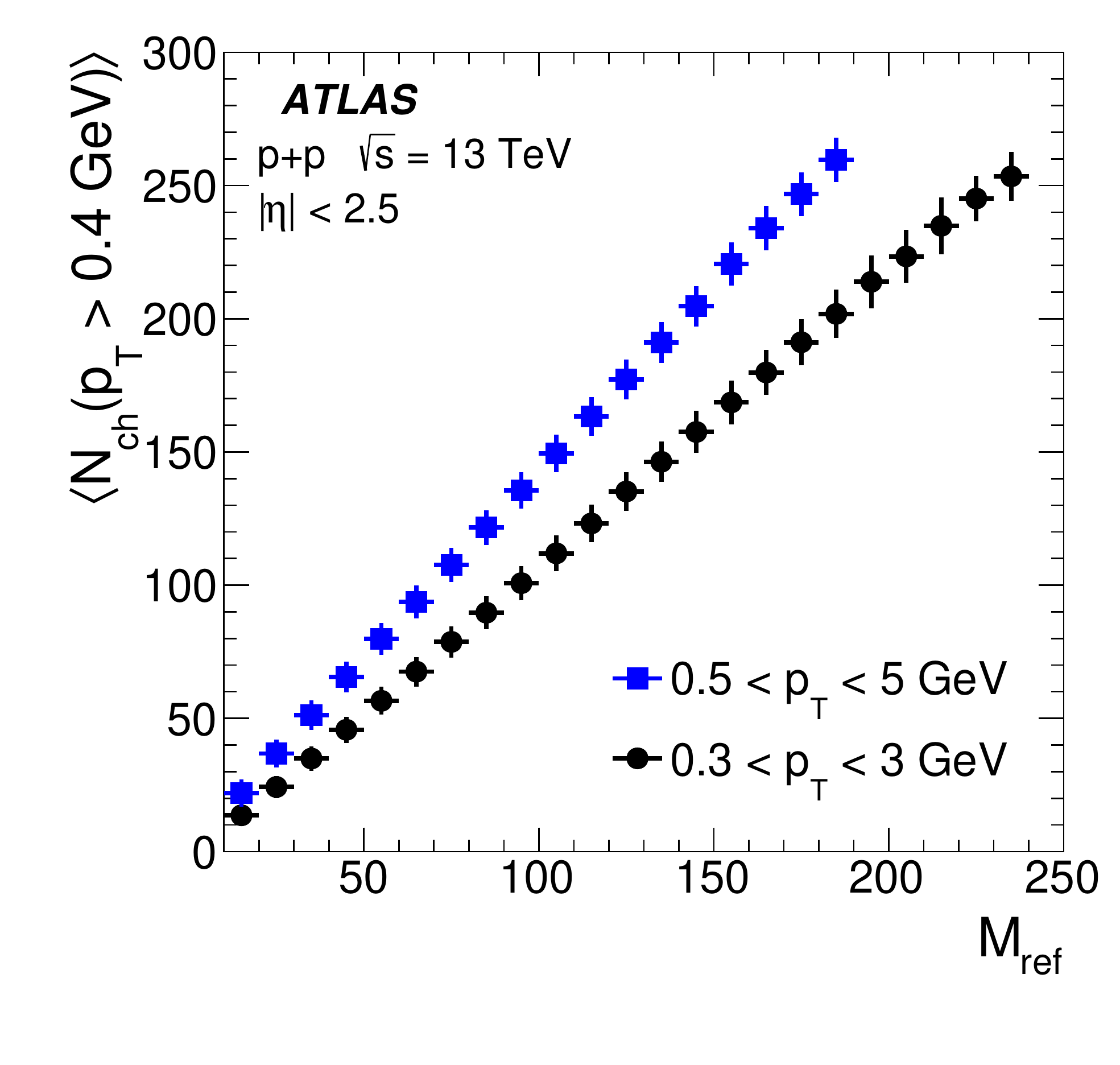}
\includegraphics[width=60mm]{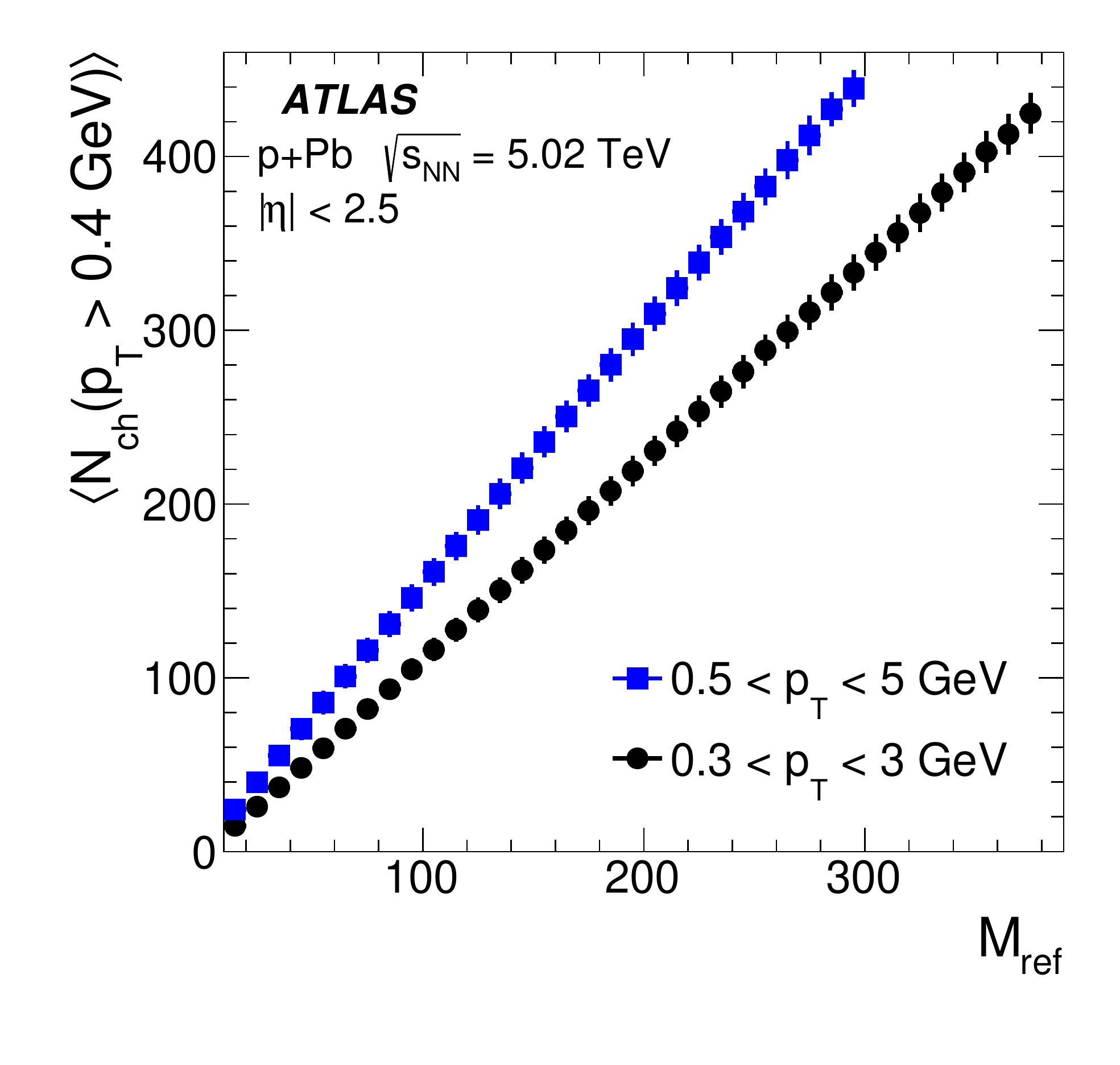}
\includegraphics[width=60mm]{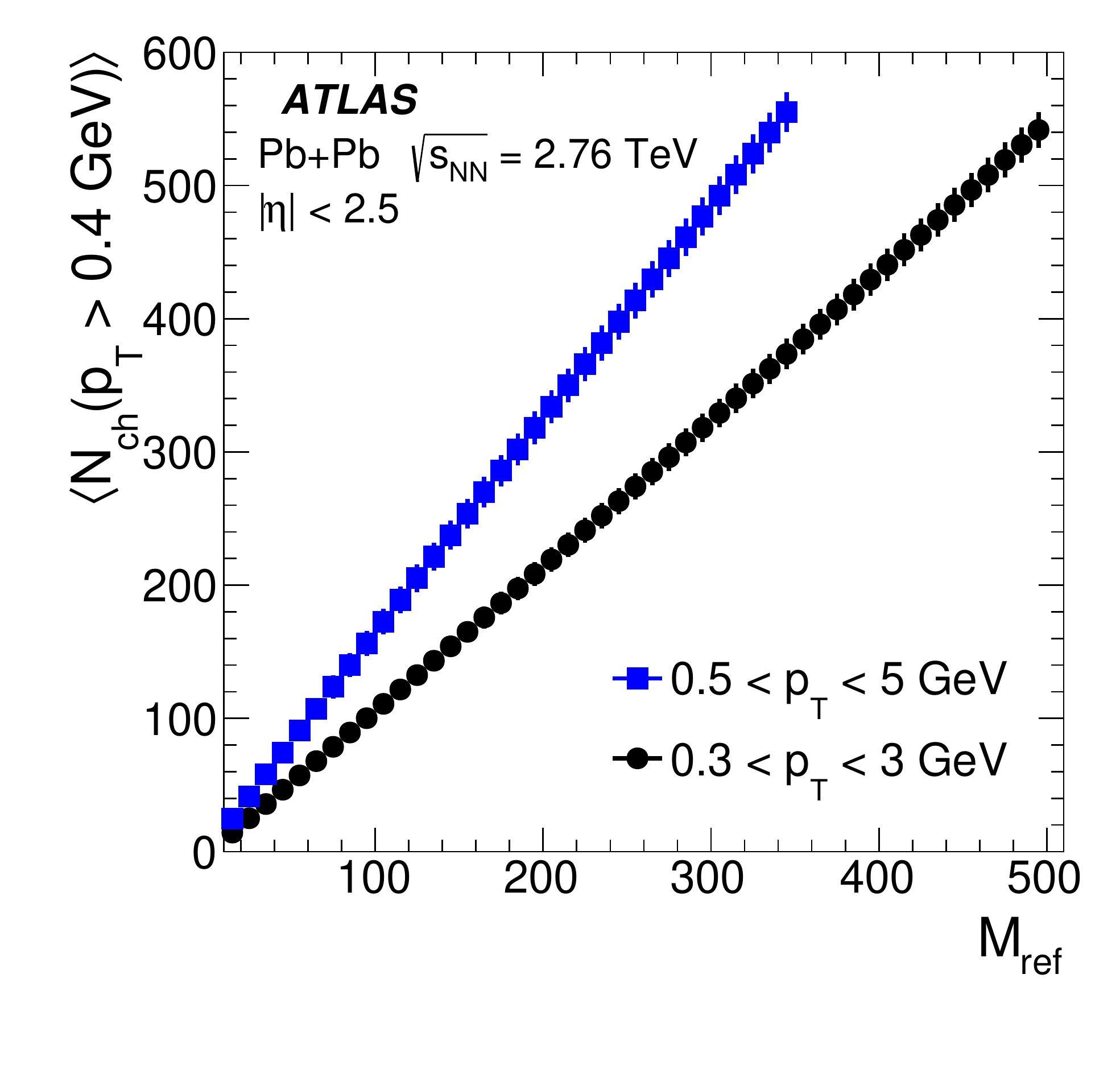}
\caption{ The average number of charged particles per event with $\pT >0.4$~GeV  as a function of reference particle multiplicity for reference particles with $ 0.5 < \pT <5$~GeV  and $ 0.3 < \pT <3$~GeV  for \pp collisions at \sqs= 5.02 and 13 TeV, \pPb collisions at \sqn= 5.02 TeV and low-multiplicity \PbPb collisions at \sqn= 2.76 TeV. The error bars show one standard deviations on $\langle N_{\mathrm{ch}}(\pT > 0.4$~GeV)$\rangle$. }
\label{fig:CorrNtrk} 
\end{center}
\end{figure}  

The cumulants and corresponding Fourier harmonics are studied as a function of the charged-particle multiplicity. Two ways of selecting events according to the event multiplicity are considered. The first one is to select events with a given $M_{\mathrm{ref}}$, which is referred to as EvSel\_$M_{\mathrm{ref}}$. An alternative way (EvSel\_$N_{\mathrm{ch}}$) is to apply the event-selection on the basis of the number of reconstructed charged particles with $\pT > 0.4$~GeV, $N_{\mathrm{ch}}^{\mathrm{rec}}$, and then for such selected events calculate the cumulants using reference particles. For both event selections,  the cumulants are calculated in unit-size bins in either $M_{\mathrm{ref}}$ or $N_{\mathrm{ch}}^{\mathrm{rec}}$, which are then combined into broader, statistically significant multiplicity intervals by averaging the cumulants, $c_n\{2k\}$. 

For the purpose of a direct comparison of results obtained with different event selections, the standard multiplicity variable measuring the event activity is used. The $N_{\mathrm{ch}} (\pT >$ 0.4~GeV)  multiplicity, corrected for tracking efficiency and the rate of falsely reconstructed charged-particle tracks as well as for trigger efficiencies, is used to present the results. When selecting events according to $M_{\mathrm{ref}}$ multiplicity, the correlation  between $M_{\mathrm{ref}}$ and the $N_{\mathrm{ch}} (\pT >$ 0.4~GeV) is employed. Figure~\ref{fig:CorrNtrk} shows mean $N_{\mathrm{ch}} (\pT > 0.4~\mathrm{GeV})$ multiplicities calculated in $M_{\mathrm{ref}}$ intervals, which are used in the analysis. The correlation is shown for each collision system and for two \pT ranges of reference particles. In the case of EvSel\_$N_{\mathrm{ch}}$, a similar mapping of  $N_{\mathrm{ch}}^{\mathrm{rec}}$ intervals into $\langle N_{\mathrm{ch}} (\pT >$ 0.4~GeV)$\rangle$ is made. 
\begin{figure}[ht!]
\begin{center}
\includegraphics[width=75mm]{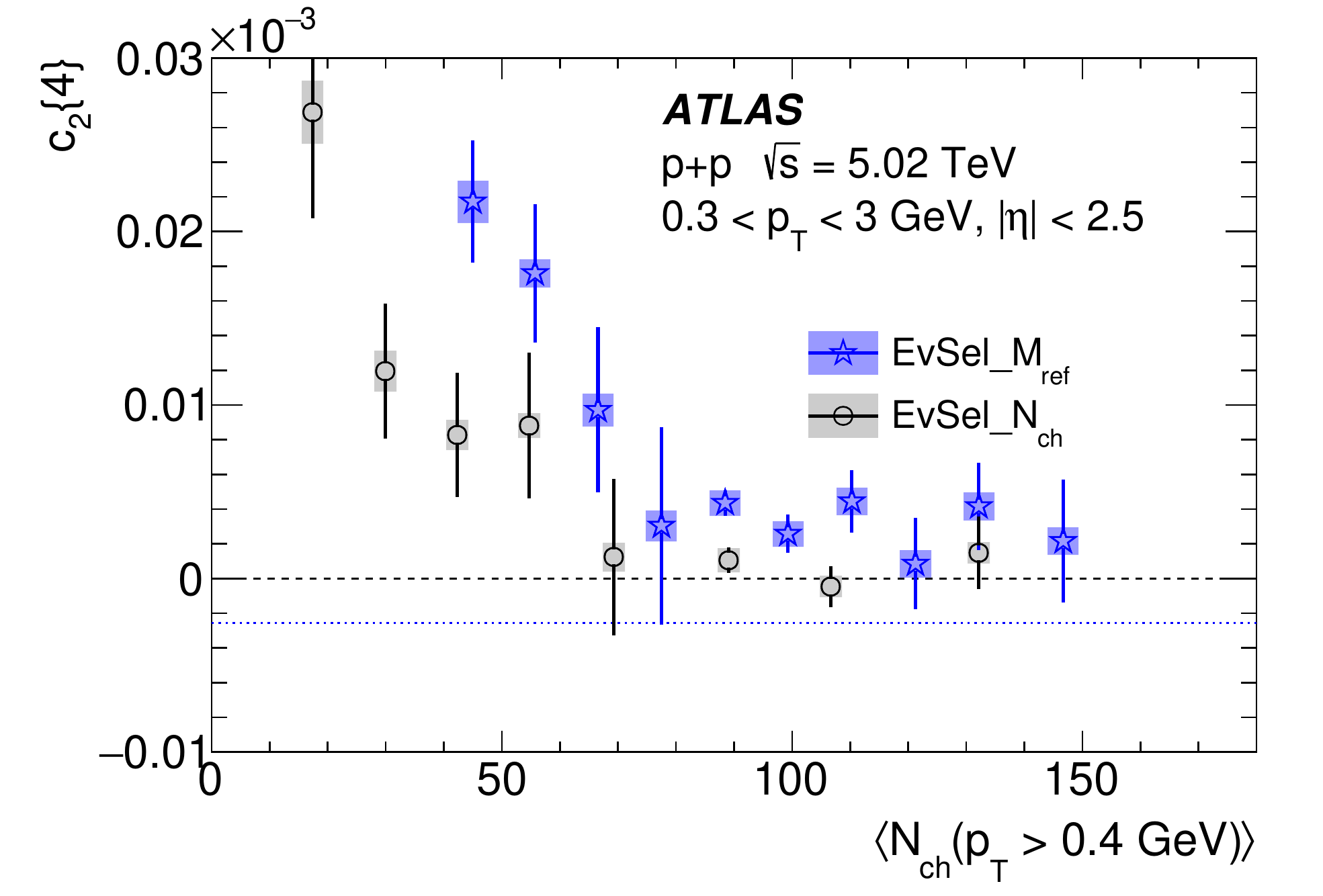}
\includegraphics[width=75mm]{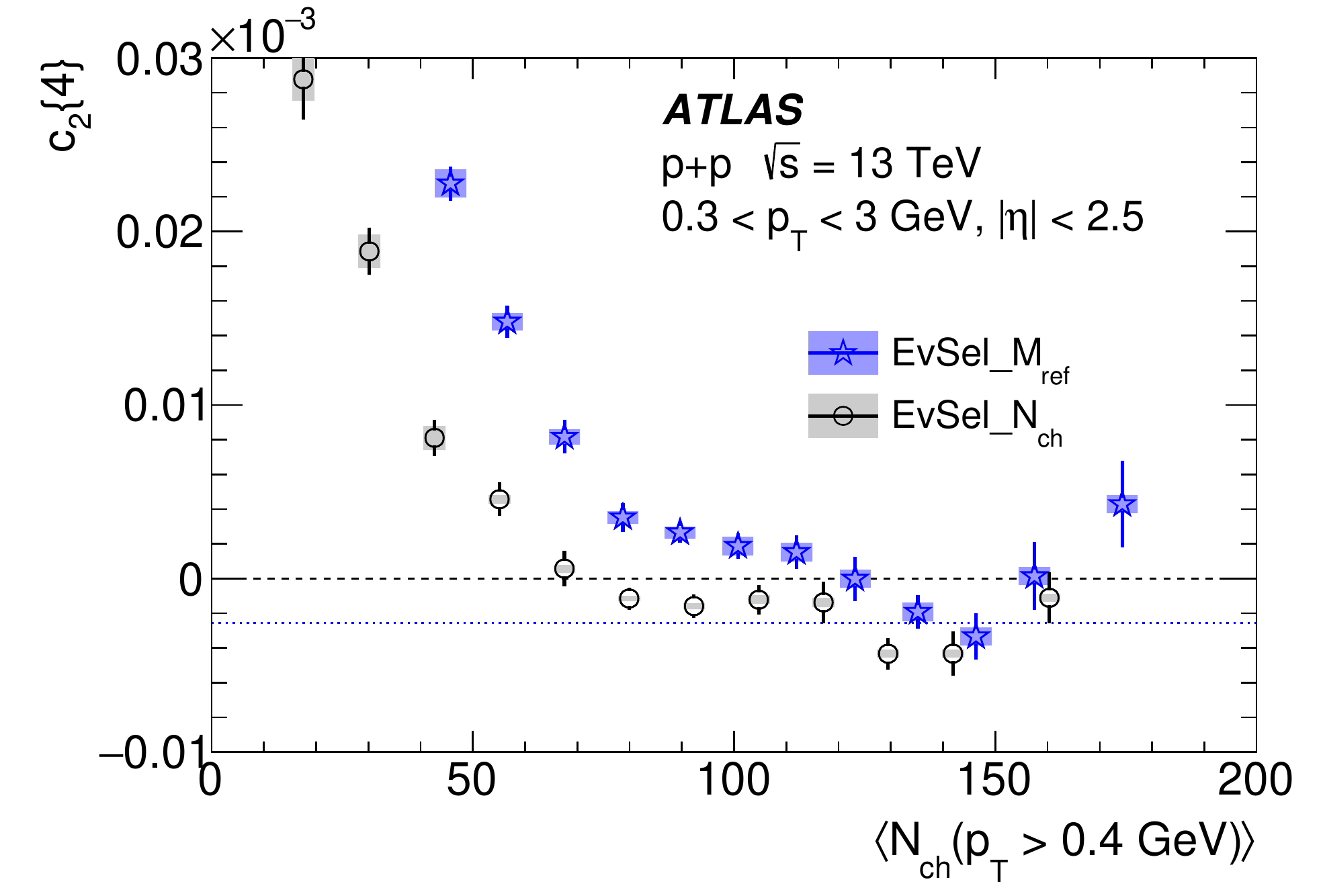}
\includegraphics[width=75mm]{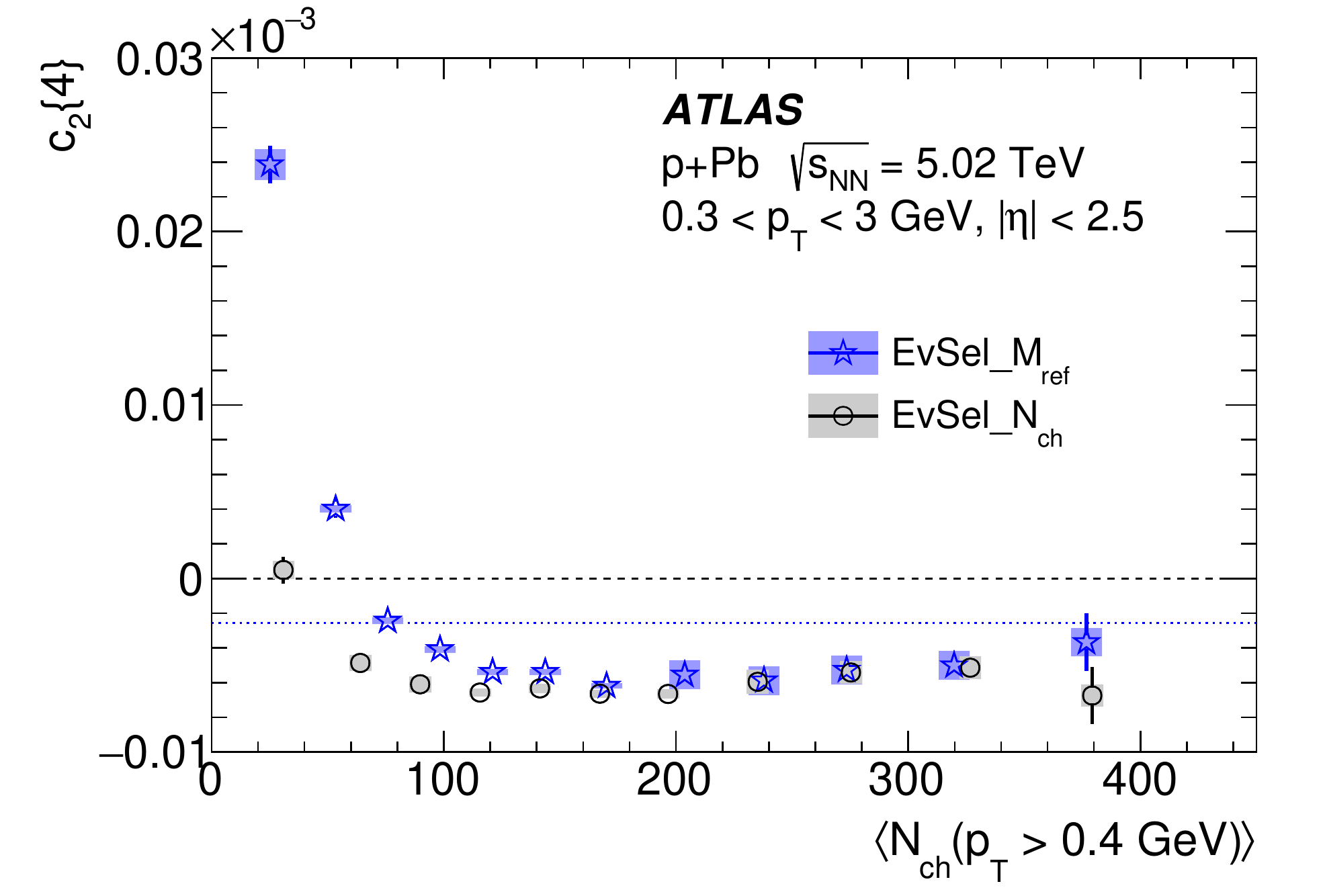}
\includegraphics[width=75mm]{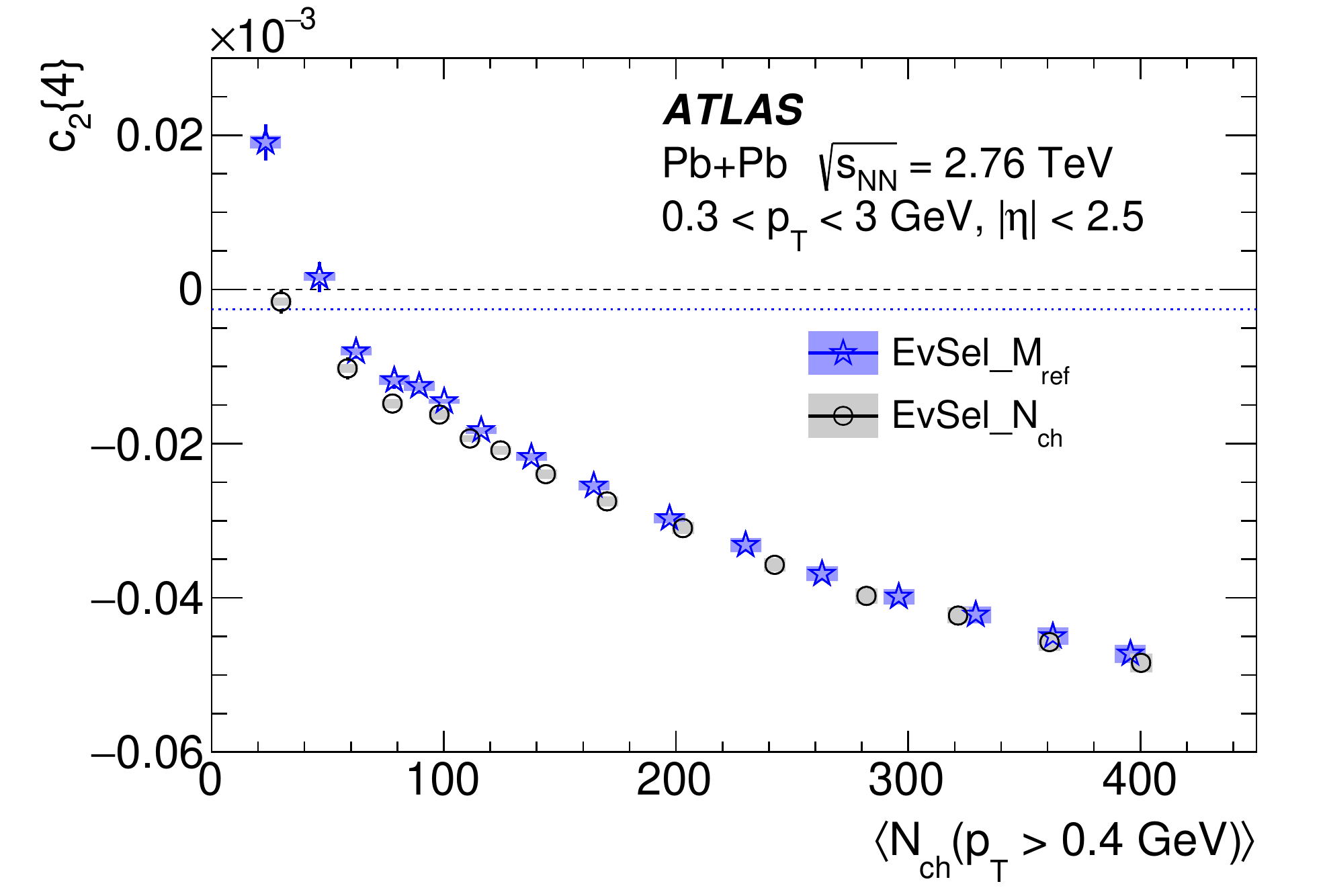}
\caption{Comparison of $c_2\{4\}$ cumulants for reference particles with $0.3< \pT < 3.0$~GeV obtained with two different event selections: events selected according to $M_{\mathrm{ref}}$ (EvSel\_$M_{\mathrm{ref}}$) and according to $N_{\mathrm{ch}}(\pT > 0.4$~GeV) (EvSel\_$N_{\mathrm{ch}}$) for \pp collisions at \sqs= 5.02 and 13 TeV, \pPb collisions at \sqn= 5.02 TeV and low-multiplicity \PbPb collisions at \sqn= 2.76 TeV.  The vertical scale in the upper plots is cut off at $0.03 \times 10^{-3}$ in order to clearly show differences in the region around $c_2\{4\} = 0$ .The error bars  and shaded boxes denote statistical and systematic uncertainties, respectively. Dotted lines indicate the value of $c_2\{4\}$ corresponding to $\mathrm{v}_2\{4\}= 0.04$. }
\label{fig:multFluct} 
\end{center}
\end{figure} 
The two event selections differ in their sensitivity to event-by-event multiplicity fluctuations and are biased in a different manner by contributions from non-flow correlations. In the selection based on $M_{\mathrm{ref}}$, by construction, multiplicity fluctuations are eliminated. This is not the case for the selection using $N_{\mathrm{ch}} (\pT >$ 0.4~GeV): there are strong event-level fluctuations in $M_{\mathrm{ref}}$($ 0.3 < \pT <3$~GeV)  for events selected with fixed values of $N_{\mathrm{ch}}(\pT > 0.4$~GeV). In order to illustrate how multiplicity fluctuations affect the determination of cumulants, the comparison of $c_2\{4\}$ cumulants obtained with two alternative ways of selecting events is shown in Figure~\ref{fig:multFluct} for reference particles with $ 0.3 < \pT <3$~GeV.  In \pp collisions, the cumulants obtained using events with fixed $N_{\mathrm{ch}}(\pT > 0.4$~GeV), thus susceptible to  fluctuations in $M_{\mathrm{ref}}$, are systematically smaller than those obtained using events selected according to $M_{\mathrm{ref}}$. This indicates that non-flow correlations associated with multiplicity fluctuations give negative contributions to the measured $c_2\{4\}$ and, in the case of a small positive $c_2\{4\}$ signal, can mimic the collective effects.  
For \pPb and \PbPb collisions, similar effects are seen at small event multiplicities, where biases from non-flow correlations are most significant. For large multiplicities, the non-flow correlations related to multiplicity fluctuations do not play a dominant role and the two event selections give consistent results. In this paper, the EvSel\_$M_{\mathrm{ref}}$, the event selection based on $M_{\mathrm{ref}}$ that is free of multiplicity fluctuations, is used as the default event selection. 

\begin{figure}[ht!]
\begin{center}
\includegraphics[width=75mm]{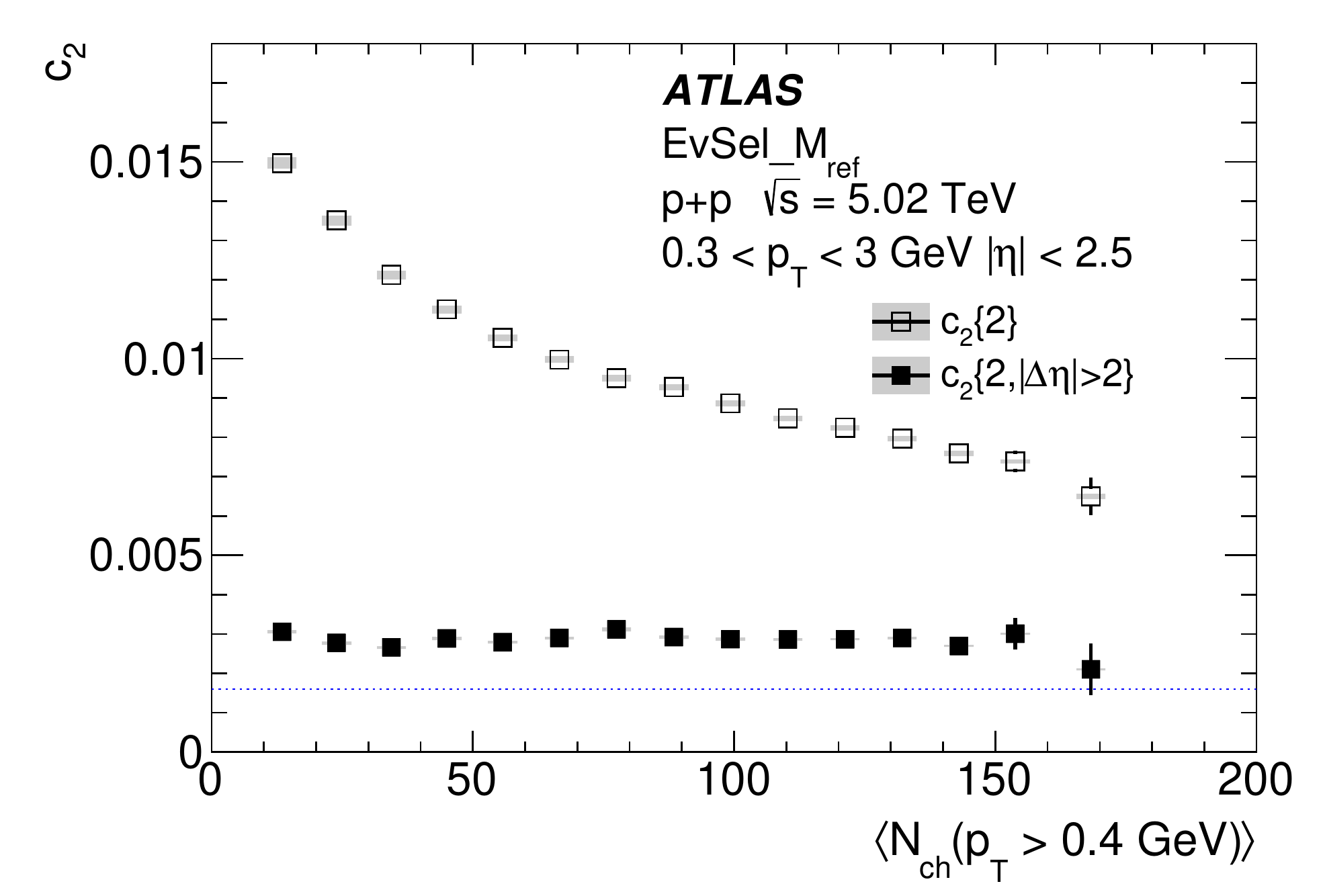}
\includegraphics[width=75mm]{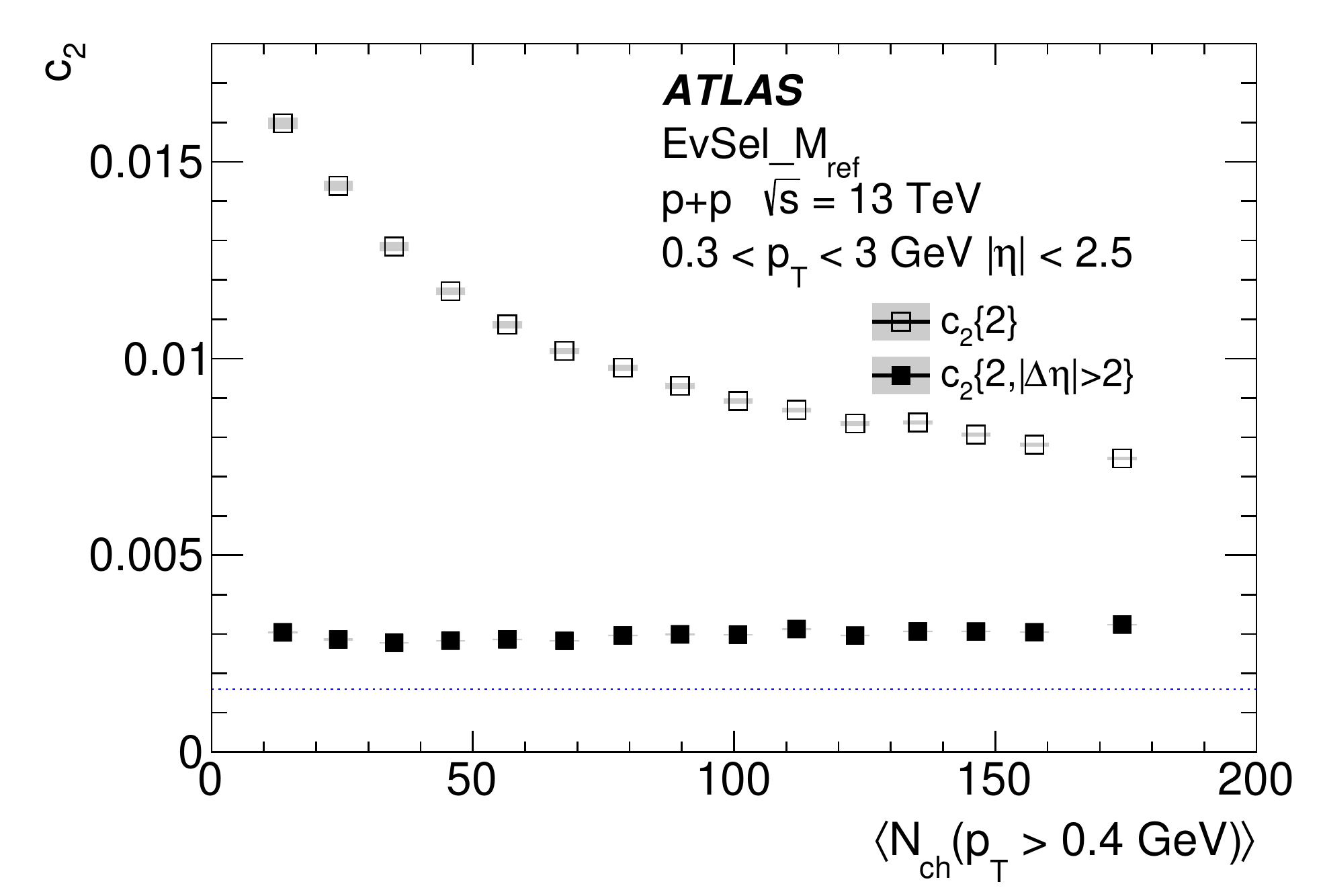}
\includegraphics[width=75mm]{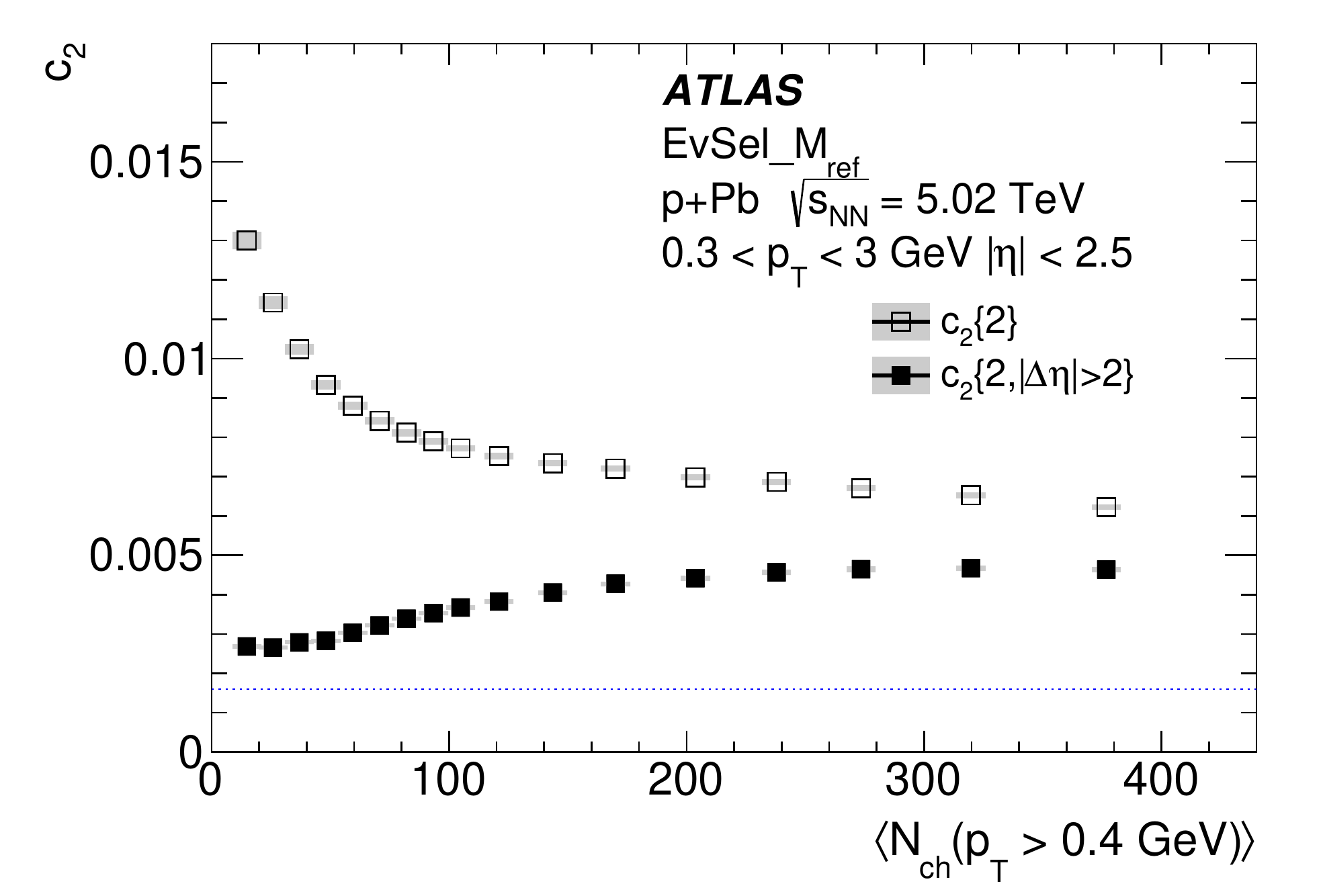}
\includegraphics[width=75mm]{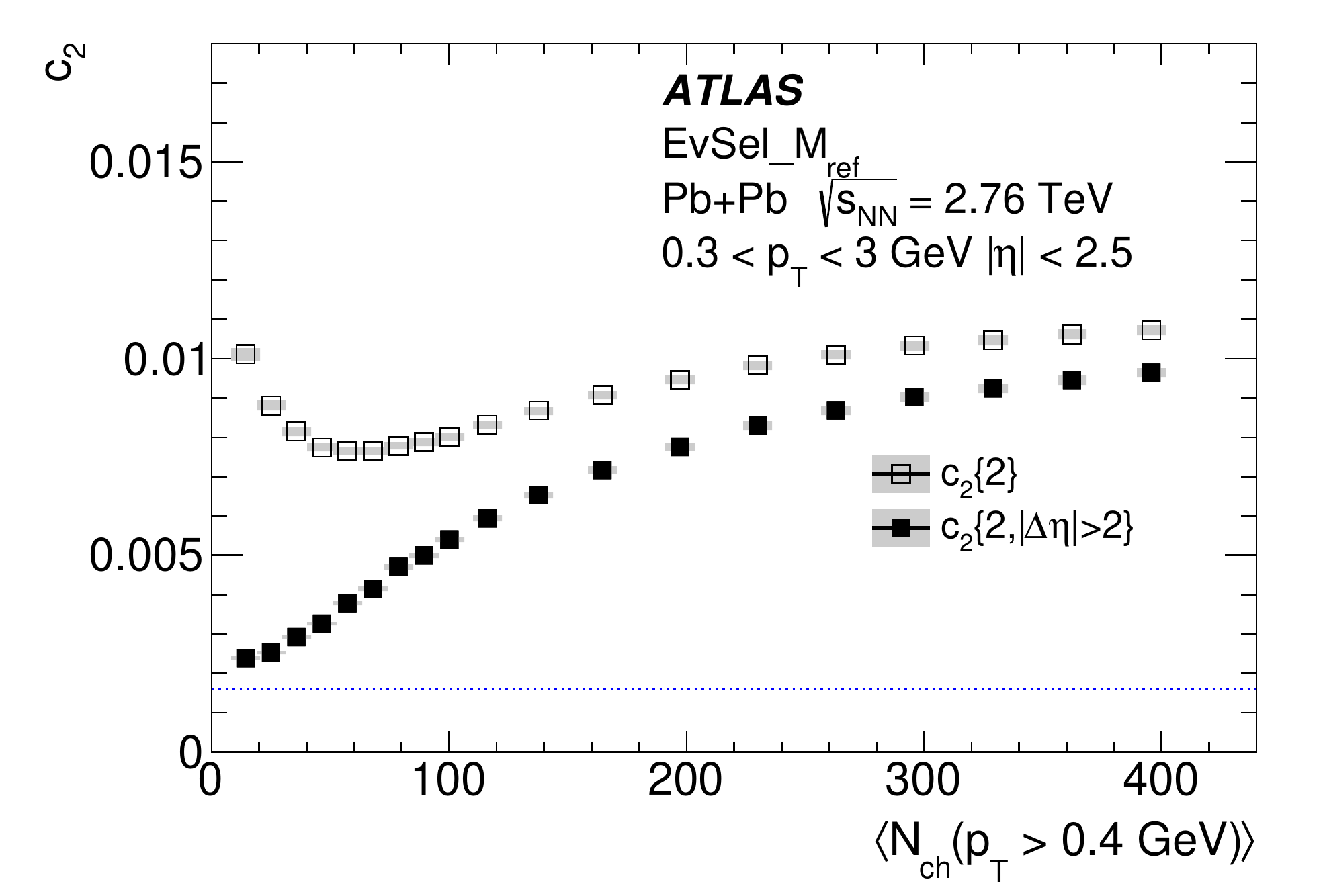}
\caption{Comparison of $c_2\{2\}$ (open symbols) and  $c_2\{2,|\Delta\eta|>2\}$ (filled symbols) for reference particles with $0.3< \pT < 3.0$~GeV for \pp collisions at \sqs= 5.02 and 13 TeV, \pPb collisions at \sqn= 5.02 TeV and low-multiplicity \PbPb collisions at \sqn= 2.76 TeV.   The error bars  and shaded boxes denote statistical and systematic uncertainties, respectively.  Dotted lines indicate the value of $c_2$ corresponding to $\mathrm{v}_2\{2\}= 0.04$.}
\label{fig:EtaGap} 
\end{center}
\end{figure} 

Even when using an event selection free of multiplicity fluctuations, the cumulants calculated with a small number of particles can be contaminated by non-flow correlations. For two-particle cumulants, $c_n\{2\}$, the non-flow correlations can be reduced by requiring a large separation in pseudorapidity between particles forming a pair. As in the analysis of two-particle correlations \cite{pPbatlas1,pPbatlas3,ppatlas1,ppatlas2}, the requirement of $|\Delta\eta|>2$ is implemented in calculating the cumulants $c_n\{2,|\Delta\eta|>2\}$. A comparison of $c_2\{2\}$ calculated without the $|\Delta\eta|>2$  requirement and $c_2\{2,|\Delta\eta|>2\}$ is shown in Figure~\ref{fig:EtaGap} for all collision systems. A strong reduction of the cumulant values can be seen after requiring $|\Delta\eta|>2$, which is the most significant at low multiplicities and for \pp collisions, where the short-range two-particle non-flow correlations dominate.  Unfortunately, such a requirement on $|\Delta\eta|$ cannot be applied in the calculation of cumulants of more than two particles in the standard cumulant approach applied in this analysis. This has to be taken into account when interpreting the results obtained for $c_n\{4\}$. It is known (from  \PYTHIA \cite{jia} and \textsc{Hijing} simulations) that jet and dijet production can generate correlations between four particles, especially in collision systems (e.g. \pp) where collective flow effects are  expected to be small.

Measurements of multi-particle cumulants and the corresponding flow harmonics require very large event samples, especially when considering cumulants and correlations between more than two particles. This analysis uses the two-particle cumulants with a rapidity gap of $|\Delta\eta|>2$ to determine $c_n\{2,|\Delta\eta|>2\}$ for $n= 2$, 3 and 4 for all collision systems.
Four-particle cumulants can be reliably determined for all collision systems only for  $c_2\{4\}$. A statistically significant measurement of higher-order cumulants and harmonics, $n=3,4$, with more than two-particle correlations is not possible with the current data sets.
Statistical limitations are particularly severe for six- and eight-particle cumulants measured in \pp collisions. The statistical uncertainty of the \pp data sets used in this analysis is significantly larger than the expected magnitude of the six- and eight-particle cumulants, preventing reliable measurements of these observables. Therefore, the measurements of $c_2\{6\}$ and $c_2\{8\}$ and the corresponding Fourier harmonics are reported only for \pPb and \PbPb collisions. 

\section{Systematic uncertainties and cross-checks}
\label{sec:syst}
\noindent
The systematic uncertainties are estimated for $c_n\{2,|\Delta\eta|>2\}$ ($n$= 2, 3 and 4) and $c_2\{4\}$, for all collision systems, and for $c_2\{6\}$ and $c_2\{8\}$ only for \pPb and \PbPb data. The two ranges in \pT of reference particles are considered: $0.3 < \pT < 3$~GeV and $0.5 < \pT < 5$~GeV. The $c_n$ uncertainties are then propagated to the corresponding $\mathrm{v}_n$.  Details on the contributions to systematic uncertainties from different sources are collected in tables included in the Appendix.

The following systematic uncertainties are considered: 

\noindent

\noindent
\textbf{Track-quality selections}: The systematic uncertainties resulting from different track selection requirements are estimated as differences between the nominal results and the results obtained with modified track selection criteria. For \pp data, the requirements on the impact parameters are varied from the nominal value of $|d_0| < 1.5$~mm and $|z_0 \sin \theta| < 1.5$~mm, to the tight selection, $|d_0| < 1$~mm and $|z_0 \sin \theta| < 1$~mm,  and to the loose selection, $|d_0| < 2$~mm and $|z_0 \sin \theta| < 2$~mm. For \pPb and \PbPb collisions the nominal selection requirements defined by the cuts on the impact parameters and the cuts on the significance of impact parameters ($|d_0| < 1.5$~mm, $|z_0 \sin \theta| < 1.5$~mm, $|d_0/\sigma_{d_0}|<3$ and $|z_0\sin(\theta)/\sigma_z|<3$) are changed to the loose ones: $|d_0| < 2$~mm, $|z_0 \sin \theta| < 2$~mm, $|d_0/\sigma_{d_0}|<4$ and $|z_0\sin(\theta)/\sigma_z|<4$. The tight selection requirements are: $|d_0| < 1$~mm, $|z_0 \sin \theta| < 1$~mm, $|d_0/\sigma_{d_0}|<2$ and $|z_0\sin(\theta)/\sigma_z|<2$. 

For each collision system, the track reconstruction efficiency is recalculated with the loose and tight track selections. The differences are obtained as averages over three ranges in $N_{\mathrm{ch}} (\pT > 0.4$~GeV). The following ranges are defined: (<50), (50, 100) and (>100) for \pp collisions at 5 and 13 TeV; (<100), (100, 200) and (>200) for \pPb and \PbPb collisions. As a systematic uncertainty the largest difference,
$c_n\{2k\}^{\mathrm{base}} - c_n\{2k\}^{\mathrm{loose}}$ or $c_n\{2k\}^{\mathrm{base}} - c_n\{2k\}^{\mathrm{tight}}$, is taken.

\noindent
\textbf{Tracking efficiency}: Systematic uncertainty in the track reconstruction efficiency results from an imperfect detector geometry description in the simulations. It affects the particle weights determined using the MC-derived tracking efficiency, $\epsilon(\eta,\pT)$.  For \pp collisions, the efficiency uncertainty depends on $\eta$ and \pT, as derived from the studies with the varied detector material budget \cite{STDM-2015-02}. It is found to vary between 1\% and 4\%, depending on $\eta$ and \pT. For \pPb and \PbPb collisions, the efficiency uncertainty is assumed to vary with \pT up to 4\%, independently of $\eta$. The systematic uncertainty of the multi-particle cumulants is estimated by repeating the analysis with the tracking efficiency varied up and down by its corresponding uncertainty. The systematic uncertainty is taken as the largest deviation of the nominal result from the result obtained assuming a higher or lower efficiency. It is estimated for each bin in the charged-particle multiplicity. 

\noindent
\textbf{Pile-up}: The pile-up effects may be important for the analysis of \pp data. The pile-up is significantly reduced by removing events with a second vertex reconstructed from at least four tracks.  Furthermore, in the analysis the $M_{\mathrm{ref}}$ and cumulants are always calculated using the tracks associated with the primary vertex. As a result the pile-up effects should not play a significant role. The exception might be due to events where the pile-up vertex is so close to the primary vertex that the two are merged.  To assess the pile-up effect on the cumulants calculated for 13 TeV \pp data, the results for the low-$\mu$ June data ($\mu < 0.03$) and the moderate-$\mu$ August data ($\mu \sim 0.6$) are compared and the differences are found to be negligible. 

However, such pile-up studies for \pp collisions are strongly affected by statistical fluctuations, which arise due to the small number of data events with low or high pile-up as well as to the smallness of the measured signal. This is particularly true for four-particle cumulants as well as higher-order cumulants  $c_3\{2, |\Delta\eta|>2\}$ and  $c_4\{2, |\Delta\eta|>2\}$, for \pp collisions. Therefore, an alternative approach is also considered, where different criteria are used to reduce the pile-up. In the nominal approach, all events with a second vertex containing at least four tracks are removed. Here, the removal of events with a second vertex reconstructed from at least two or six tracks is also considered and the results for these two selections of events are compared to the nominal results. The maximum difference between the nominal measurement and the cumulants obtained from the data set with higher pile-up or lower pile-up is taken as a systematic uncertainty.

For \pPb results, the pile-up effects are studied by comparing the nominal results, for which events with the second vertex with $\sum \pT > 5$~GeV are removed, to the results obtained without removing the pile-up events. The maximum difference between the nominal measurement and the cumulants obtained without removing the pile-up events is taken as a systematic uncertainty.

For low-multiplicity \PbPb collisions the pile-up is negligibly small ($\mu \approx 10^{-4}$) and not considered to contribute to the systematic uncertainty. 

\noindent
\textbf{Comparison of results for $p$+Pb and Pb+$p$}: For \pPb data the comparison is made between the results obtained during two running configurations with reversed beams directions, \pPb  and Pb+$p$.  The results obtained from two running periods are consistent and give a negligible contribution to the systematic uncertainty.

The systematic uncertainty of the measured cumulants across all systems and the two \pT ranges of reference particles is not dominated by a single source. However, in most
cases the largest contribution is from the track selection uncertainty, which mostly dominates uncertainties for higher-order harmonic cumulants. A sizeable contribution to the total uncertainty is also due to
the tracking efficiency uncertainty, and this uncertainty is the largest for low multiplicities.
The pile-up effects also give sizeable contributions to uncertainties in 5.02 TeV \pp cumulants. The total
systematic uncertainty is obtained by adding all individual contributions in quadrature.
Table~\ref{tab:TotalSyst} lists the total systematic uncertainties of the measured cumulants in different collision systems for reference particles with $0.3 < \pT <3$~GeV. The listed systematic uncertainties are averaged over the $N_{\mathrm{ch}}$ range. For reference particles in the higher transverse momentum range, $0.5 < \pT <5$~GeV, the total systematic uncertainties are included in Table~\ref{tab:TotalSyst1}. The total systematic uncertainty of the cumulants is then propagated to the systematic uncertainties of the Fourier harmonics according to Eqs.~\eqref{eq:vn2} - \eqref{eq:vn8}. 

\begin{table}[h!]
\begin{center}
\caption{Total systematic uncertainties  of the measured multi-particle cumulants for \pp collisions at \sqs= 5.02 and 13 TeV, \pPb collisions at \sqn= 5.02 TeV and low-multiplicity \PbPb collisions at \sqn= 2.76 TeV, for $M_{\mathrm{ref}}$ with  $0.3 < \pT < 3$~GeV as estimated  in a given  $N_{\mathrm{ch}}$ interval.}
\label{tab:TotalSyst}
\begin{tabular}{llrrr}  
\toprule
 \multicolumn{5}{ c }{Total systematic uncertainties} \\
\cline{1-5}
System	 & Systematic uncertainty & $\qquad$ $N_{\mathrm{ch}}$   & $\qquad$ $N_{\mathrm{ch}}$  & $\qquad$ $N_{\mathrm{ch}}$  \\
\midrule
  &  & <50 &  50--100 & >100 \\
  \pp 5~TeV & $\delta c_2\{2,|\Delta\eta|>2\} \times 10^{4}$ & 0.40 &  0.47 & 0.30 \\
  & $\delta c_2\{4\}\times 10^{6}$ & 4.25 & 0.95 & 0.80 \\
  & $\delta c_3\{2,|\Delta\eta|>2\} \times 10^{4}$  & 0.26 & 0.33 & 0.15   \\
  & $\delta c_4\{2,|\Delta\eta|>2\} \times 10^{4}$  & 0.12 & 0.12 &  -    \\ \\
  &  & <50 &  50--100 & >100 \\
   \pp 13~TeV & $\delta c_2\{2,|\Delta\eta|>2\} \times 10^{4}$ & 0.32 &  0.22 & 0.20 \\
  & $\delta c_2\{4\}\times 10^{6}$ & 3.76 & 0.52 & 0.54 \\
  & $\delta c_3\{2,|\Delta\eta|>2\} \times 10^{4}$  & 0.05 & 0.03 & 0.07   \\
  & $\delta c_4\{2,|\Delta\eta|>2\} \times 10^{4}$  & 0.02 & 0.05 &  -    \\ \\
   &  & <100 &  100--200 & >200 \\ 
    \pPb  & $\delta c_2\{2,|\Delta\eta|>2\} \times 10^{4}$ & 0.59 &  0.59 & 0.70 \\
  & $\delta c_2\{4\}\times 10^{6}$ & 0.88 & 0.17 & 0.83 \\
  & $\delta c_2\{6\}\times 10^{7}$ & 0.62 & 0.22 & 0.09 \\
  & $\delta c_2\{8\}\times 10^{8}$ & 3.20 & 0.11 & 0.02 \\
  & $\delta c_3\{2,|\Delta\eta|>2\} \times 10^{4}$  & 0.24 & 0.24 & 0.19   \\
  & $\delta c_4\{2,|\Delta\eta|>2\} \times 10^{4}$  & 0.22 & 0.22 & 0.11   \\ \\
  &  & <100 &  100--200 & >200 \\ 
     \PbPb  & $\delta c_2\{2,|\Delta\eta|>2\} \times 10^{4}$ & 0.66 &  1.00 & 1.27 \\
  & $\delta c_2\{4\}\times 10^{6}$ & 0.82 & 0.67 & 1.19 \\
  & $\delta c_2\{6\}\times 10^{7}$ & 0.35 & 0.23 & 0.44 \\
  & $\delta c_2\{8\}\times 10^{8}$ & 1.23 & 0.13 & 0.31 \\
  & $\delta c_3\{2,|\Delta\eta|>2\} \times 10^{4}$  & 0.10 & 0.09 & 0.13   \\
  & $\delta c_4\{2,|\Delta\eta|>2\} \times 10^{4}$  & 0.03 & 0.04 & 0.05   \\ 
\bottomrule
\end{tabular}
\end{center}
\end{table}
\begin{table}[h!]
\begin{center}
\caption{Total systematic uncertainties of the measured multi-particle cumulants for \pp collisions at \sqs= 5.02 and 13 TeV, \pPb collisions at \sqn= 5.02 TeV and low-multiplicity \PbPb collisions at \sqn= 2.76 TeV, for $M_{\mathrm{ref}}$ with  $0.5 < \pT < 5$~GeV as estimated  in a given  $N_{\mathrm{ch}}$ interval.  }
\label{tab:TotalSyst1}
\begin{tabular}{llrrr}  
\toprule
 \multicolumn{5}{ c }{Total systematic uncertainties} \\
\cline{1-5}
System	 & Systematic uncertainty & $\qquad$ $N_{\mathrm{ch}}$   & $\qquad$ $N_{\mathrm{ch}}$  & $\qquad$ $N_{\mathrm{ch}}$  \\
\midrule
  &  & <50 &  50--100 & >100 \\
  \pp 5~TeV & $\delta c_2\{2,|\Delta\eta|>2\} \times 10^{4}$ & 0.56 &  0.31 & 0.41 \\
  & $\delta c_2\{4\}\times 10^{6}$ & 7.20 & 1.85 & 2.45 \\
  & $\delta c_3\{2,|\Delta\eta|>2\} \times 10^{4}$  & 0.35 & 0.34 & 0.23   \\
  & $\delta c_4\{2,|\Delta\eta|>2\} \times 10^{4}$  & 0.29 & 0.45 &  -    \\ \\
  &  & <50 &  50--100 & >100 \\
   \pp 13~TeV & $\delta c_2\{2,|\Delta\eta|>2\} \times 10^{4}$ & 0.41 &  0.27 & 0.25 \\
  & $\delta c_2\{4\}\times 10^{6}$ & 6.40 & 1.77 & 0.59 \\
  & $\delta c_3\{2,|\Delta\eta|>2\} \times 10^{4}$  & 0.07 & 0.07 & 0.08   \\
  & $\delta c_4\{2,|\Delta\eta|>2\} \times 10^{4}$  & 0.03 & 0.05 & 0.06   \\ \\
   &  & <100 &  100--200 & >200 \\ 
    \pPb  & $\delta c_2\{2,|\Delta\eta|>2\} \times 10^{4}$ & 0.31 &  0.32 & 0.38 \\
  & $\delta c_2\{4\}\times 10^{6}$ & 0.66 & 0.91 & 1.31 \\
  & $\delta c_2\{6\}\times 10^{7}$ & 1.43 & 0.65 & 0.40 \\
  & $\delta c_2\{8\}\times 10^{8}$ & 3.91 & 0.40 & 0.20 \\
  & $\delta c_3\{2,|\Delta\eta|>2\} \times 10^{4}$  & 0.18 & 0.25 & 0.14   \\
  & $\delta c_4\{2,|\Delta\eta|>2\} \times 10^{4}$  & 0.12 & 0.08 & 0.12   \\ \\
  &  & <100 &  100--200 & >200 \\ 
     \PbPb  & $\delta c_2\{2,|\Delta\eta|>2\} \times 10^{4}$ & 0.56 &  0.63 & 0.56 \\
  & $\delta c_2\{4\}\times 10^{6}$ & 1.84 & 0.82 & 0.72 \\
  & $\delta c_2\{6\}\times 10^{7}$ & 0.93 & 0.44 & 0.40 \\
  & $\delta c_2\{8\}\times 10^{8}$ & 0.86 & 0.54 & 0.51 \\
  & $\delta c_3\{2,|\Delta\eta|>2\} \times 10^{4}$  & 0.06 & 0.09 & 0.07   \\
  & $\delta c_4\{2,|\Delta\eta|>2\} \times 10^{4}$  & 0.13 & 0.02 & 0.05   \\ 
\bottomrule
\end{tabular}
\end{center}
\end{table} 

Several cross-checks are also performed to validate the analysis method, but are not included in the systematic uncertainty. To account for the detector imperfections and to make the analysed azimuthal angle distribution uniform, data-determined weights $w_{\phi}(\eta,\phi)$ are used, as described in Section~\ref{sec:analysis}. To verify the robustness of the weighting procedure,  the nominal results for cumulants are compared with those obtained with all weights $w_{\phi}(\eta,\phi)$ set to 1. The difference between the two measurements relative to the nominal results is found to be negligibly small.

Changing the trigger efficiency from 90\% to 95\% is also found to have negligible impact on the measured cumulants.

The global correlation effects should be independent of the charge sign of the produced particles. However, in reality the non-flow contributions may differ  for same-sign and opposite-sign charged particles. To verify whether the results reported here depend on the charge of particles, the analysis is performed separately for same-sign charged particles only and compared to the results for all charged particles. In all cases, no systematic difference is observed when comparing the cumulants for all charged particles with those obtained using only same-sign charged particles. 
\clearpage

\section{Results}
\label{sec:result}
\subsection{Second-order multi-particle cumulants and Fourier harmonics}
\label{sec:elliptic}
\begin{figure}[ht!]
\begin{center}
\includegraphics[width=75mm]{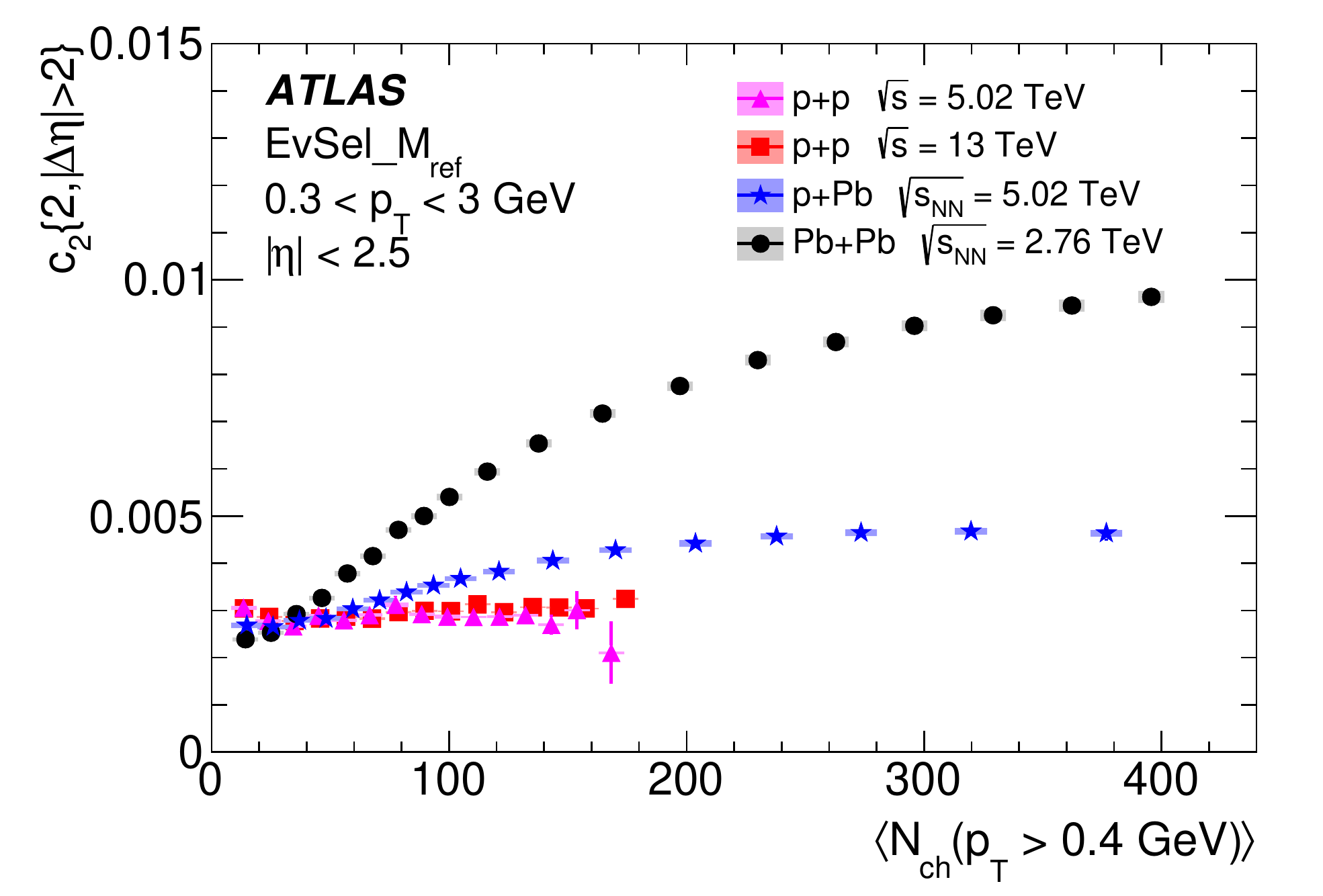}
\includegraphics[width=75mm]{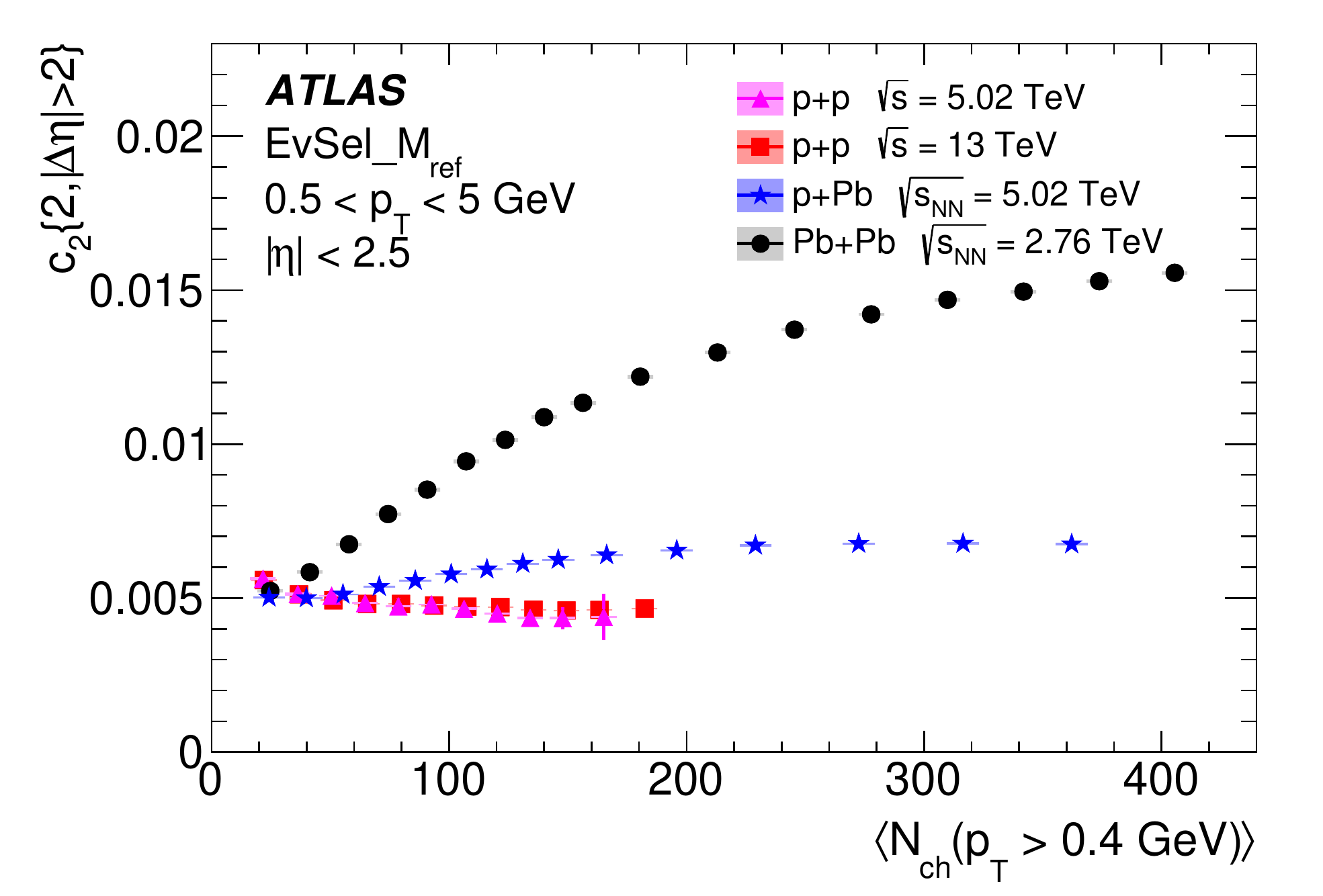}
\caption{The two-particle cumulant $c_2\{2,|\Delta\eta|>2\}$ as a function of $\langle N_{\mathrm{ch}}(\pT > 0.4$~GeV)$\rangle$  for \pp collisions at \sqs= 5.02 and 13 TeV, \pPb collisions at \sqn= 5.02 TeV and low-multiplicity \PbPb collisions at \sqn= 2.76 TeV. The left panel shows the results obtained for $M_{\mathrm{ref}}$ with  $0.3< \pT < 3.0$~GeV while the right panel is for $M_{\mathrm{ref}}$ with  $0.5< \pT < 5.0$~GeV. The error bars and shaded boxes denote statistical and systematic uncertainties, respectively.}
\label{fig:cum22All} 
\end{center}
\end{figure}

The comparison between different collision systems is made for the cumulants calculated in $M_{\mathrm{ref}}$-bins, where the \pT range of reference particles is $0.3< \pT < 3.0$~GeV and $0.5 < \pT < 5.0$~GeV.
A direct comparison of  $c_2\{2,|\Delta\eta|>2\}$ for different collision systems is shown in Figure~\ref{fig:cum22All} as a function of $\langle N_{\mathrm{ch}}(\pT > 0.4$~GeV)$\rangle$.   An ordering in the magnitude of cumulants, with the largest for \PbPb, and then decreasing for smaller collision systems, is observed. Interestingly, for the three systems the $N_{\mathrm{ch}}$-dependence changes from a clear increase for \PbPb, to a weaker increase in \pPb and to no increase or even a decreasing trend  in \pp collisions. There is no dependence on the collision energy for \pp data.

\begin{figure}[ht!]
\begin{center}
\includegraphics[width=75mm]{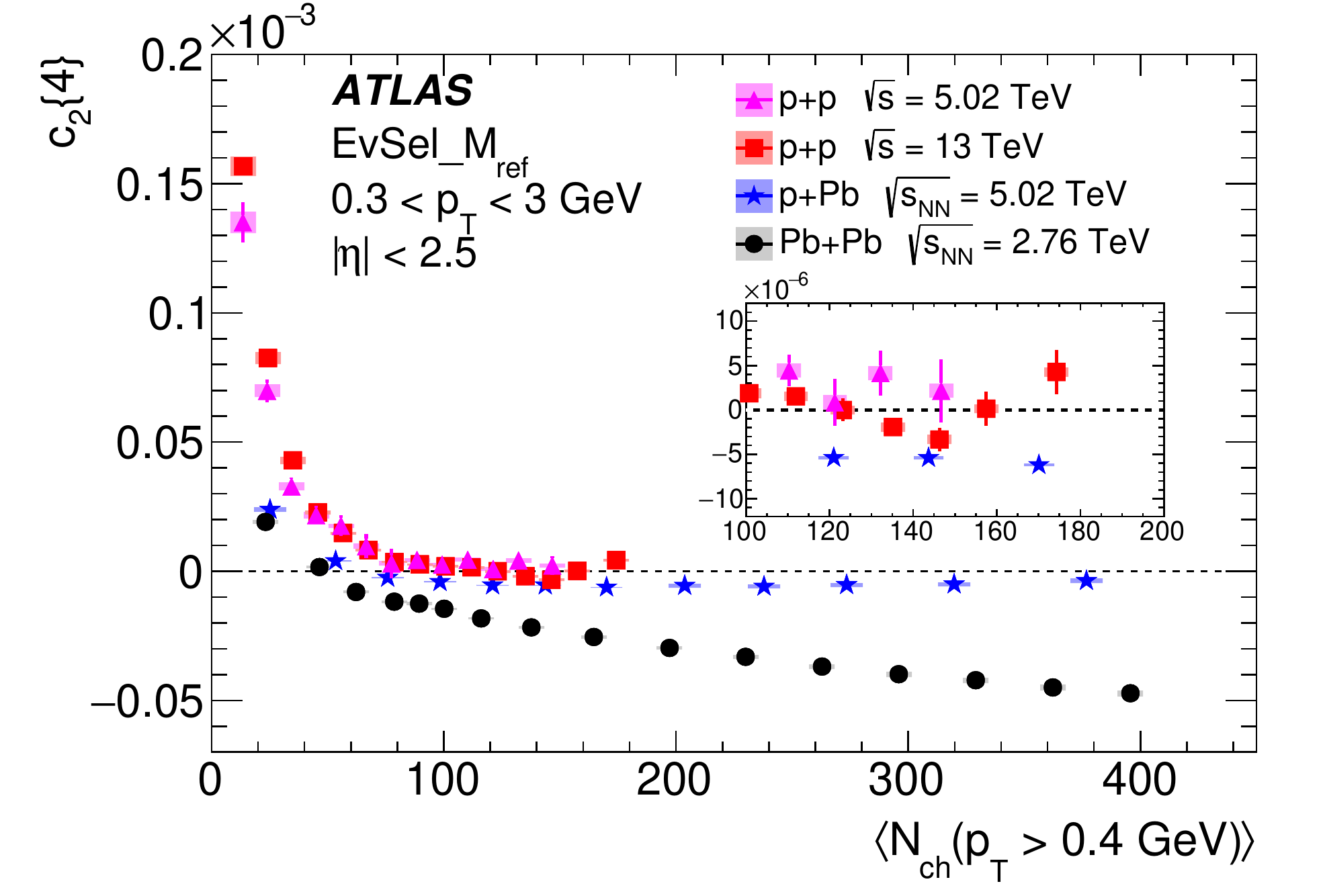}
\includegraphics[width=75mm]{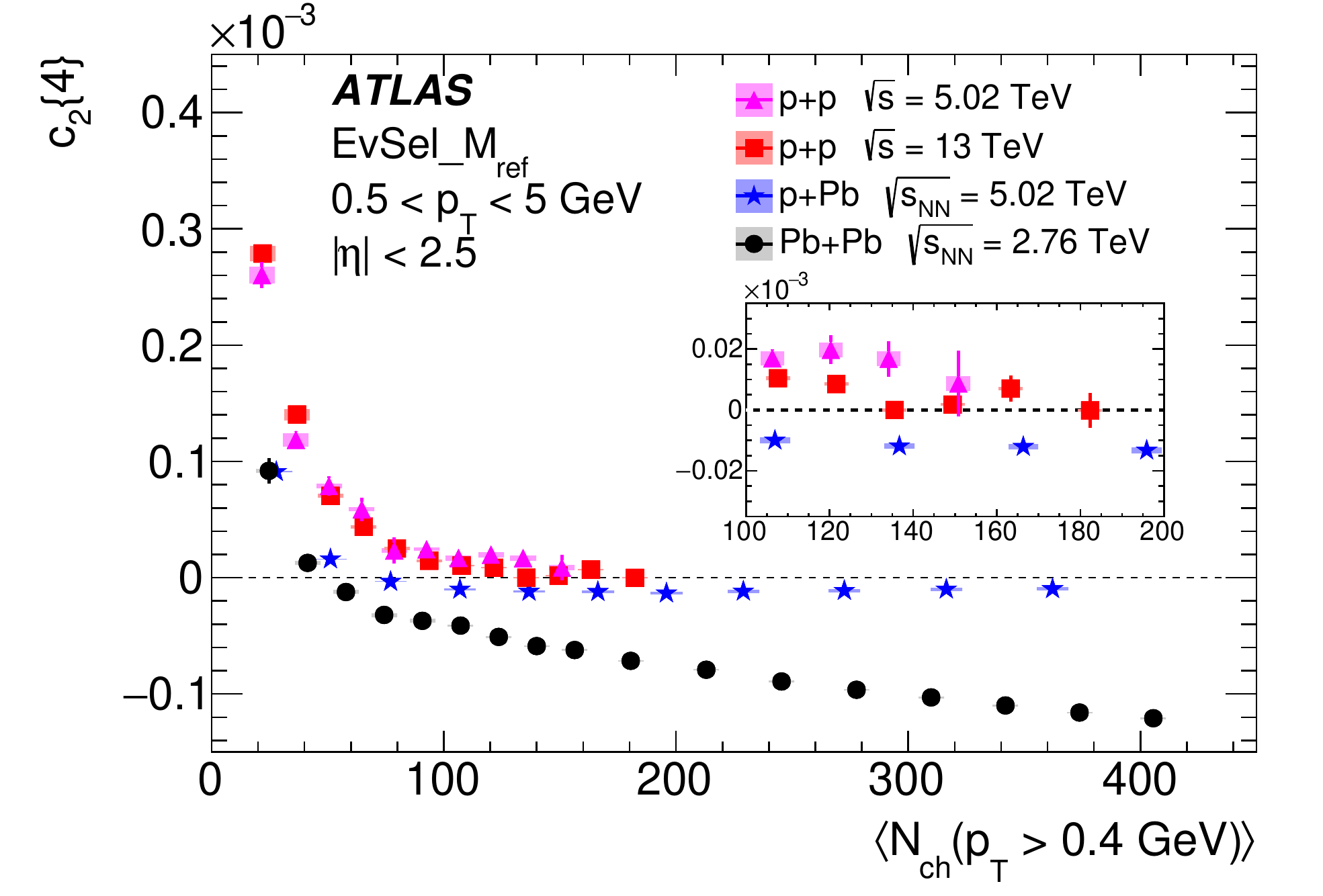}
\caption{The second-order cumulant $c_2\{4\}$ obtained from four-particle correlations as a function of $\langle N_{\mathrm{ch}}(\pT > 0.4$~GeV)$\rangle$  for \pp collisions at \sqs= 5.02 and 13 TeV, \pPb collisions at \sqn= 5.02 TeV and low-multiplicity \PbPb collisions at \sqn= 2.76 TeV. The left panel shows the results obtained for $M_{\mathrm{ref}}$ with  $0.3< \pT < 3.0$~GeV while the right panel is for $M_{\mathrm{ref}}$ with  $0.5< \pT < 5.0$~GeV. The insets zoom in on the region around $c_2\{4\}= 0$. The error bars and shaded boxes denote statistical and systematic uncertainties, respectively.}
\label{fig:cum24All} 
\end{center}
\end{figure}

Four-particle cumulants, as shown in Figure~\ref{fig:cum24All},  follow the ordering $|c_2\{4\}|_{\mathrm{p+Pb}} < |c_2\{4\}|_{\mathrm{Pb+Pb}}$ for $N_{\mathrm{ch}}(\pT > 0.4$~GeV) >100.  The magnitude of $\mathrm{v}_2\{4\}$ derived from $c_2\{4\}$ is larger for \PbPb collisions than for \pPb events with the same $N_{\mathrm{ch}}(\pT > 0.4$~GeV). For \pp collisions, 
the cumulants depend weakly on the collision energy, although systematically larger cumulant values are measured at 13~TeV than at 5.02~TeV at low $N_{\mathrm{ch}}(\pT > 0.4$~GeV). At higher multiplicities, this systematic dependence is reversed.  Over the full range of particle multiplicities, the cumulants are positive or consistent with zero at 5.02~TeV for both \pT ranges and at 13~TeV for  $0.5< \pT < 5.0$~GeV.  For the 13~TeV \pp data, the cumulants for $0.3< \pT < 3.0$~GeV also have positive values over the large range of multiplicities, with the exception of   $N_{\mathrm{ch}}$ from 130 to 150, where $c_2\{4\}$ is negative but less than 1--2 standard deviations from zero. Therefore, these measurements of $c_2\{4\}$ cumulants in \pp collisions, based on the event selection that suppresses the event-by-event fluctuations in the number of reference particles, do not allow determination of the Fourier harmonics. This indicates that the $c_2\{4\}$  obtained with the standard cumulant method used in this paper, even though free of multiplicity fluctuations, may still be biased by non-flow correlations. 
\begin{figure}[H]
\begin{center}
\includegraphics[width=75mm]{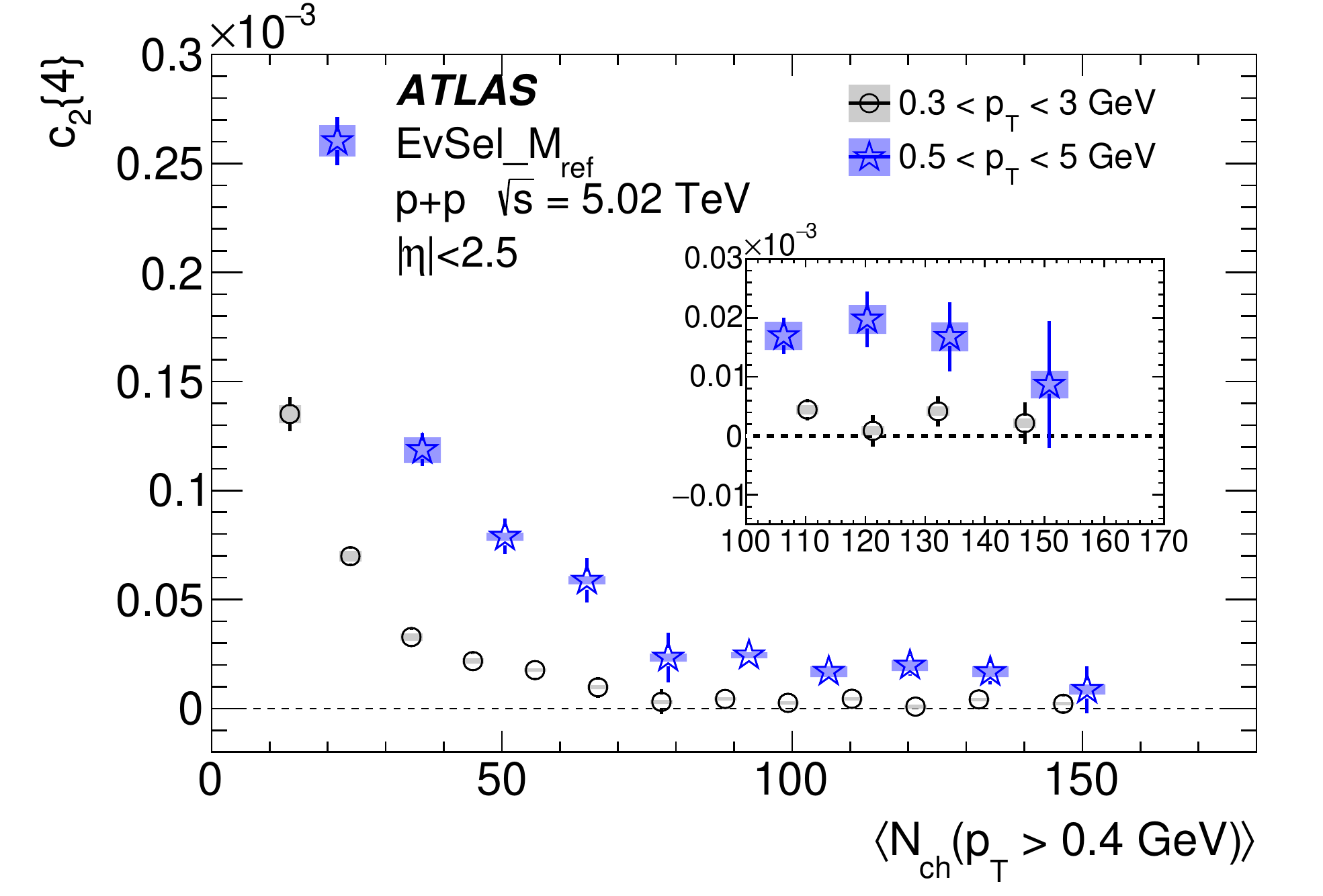}
\includegraphics[width=75mm]{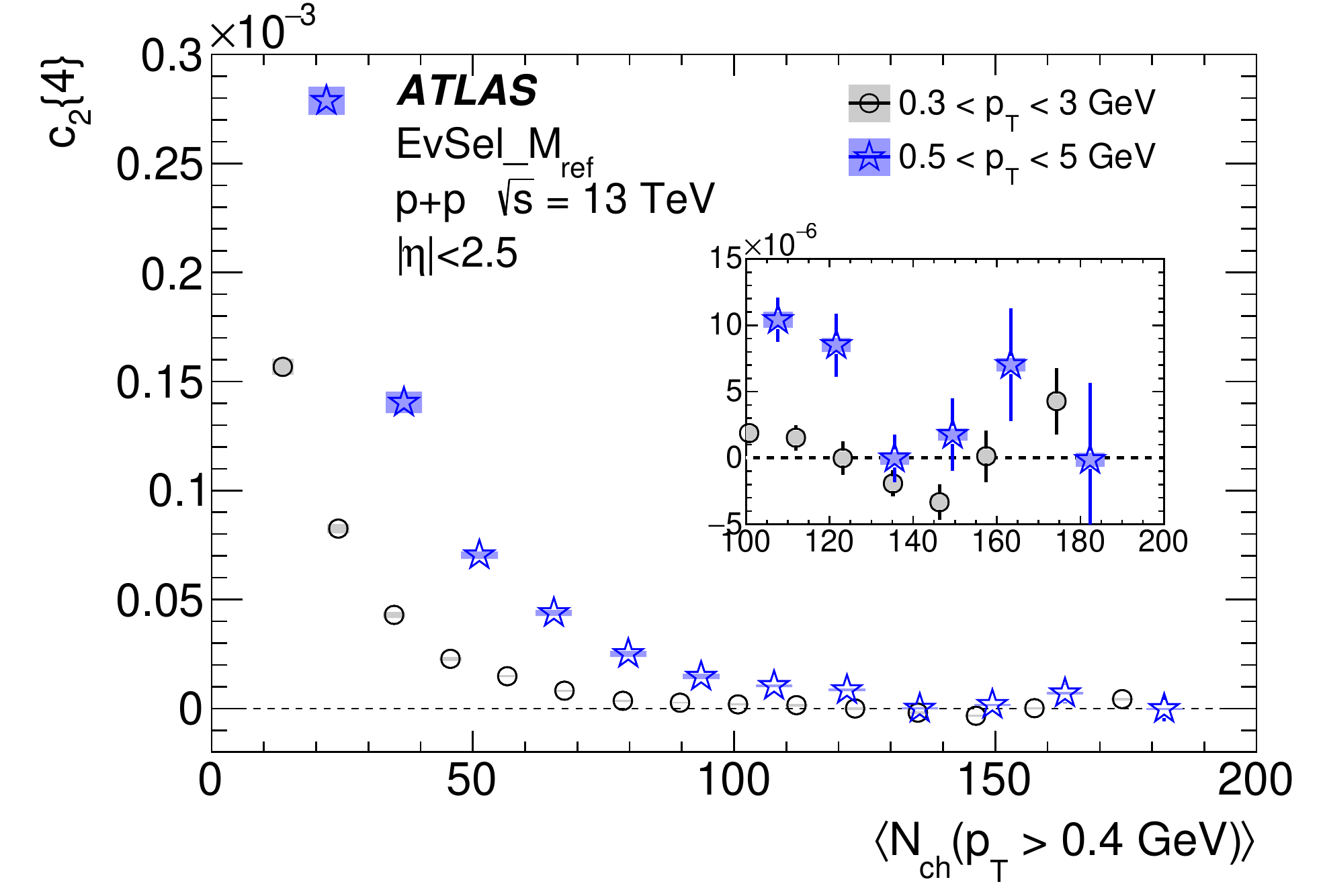}
\includegraphics[width=75mm]{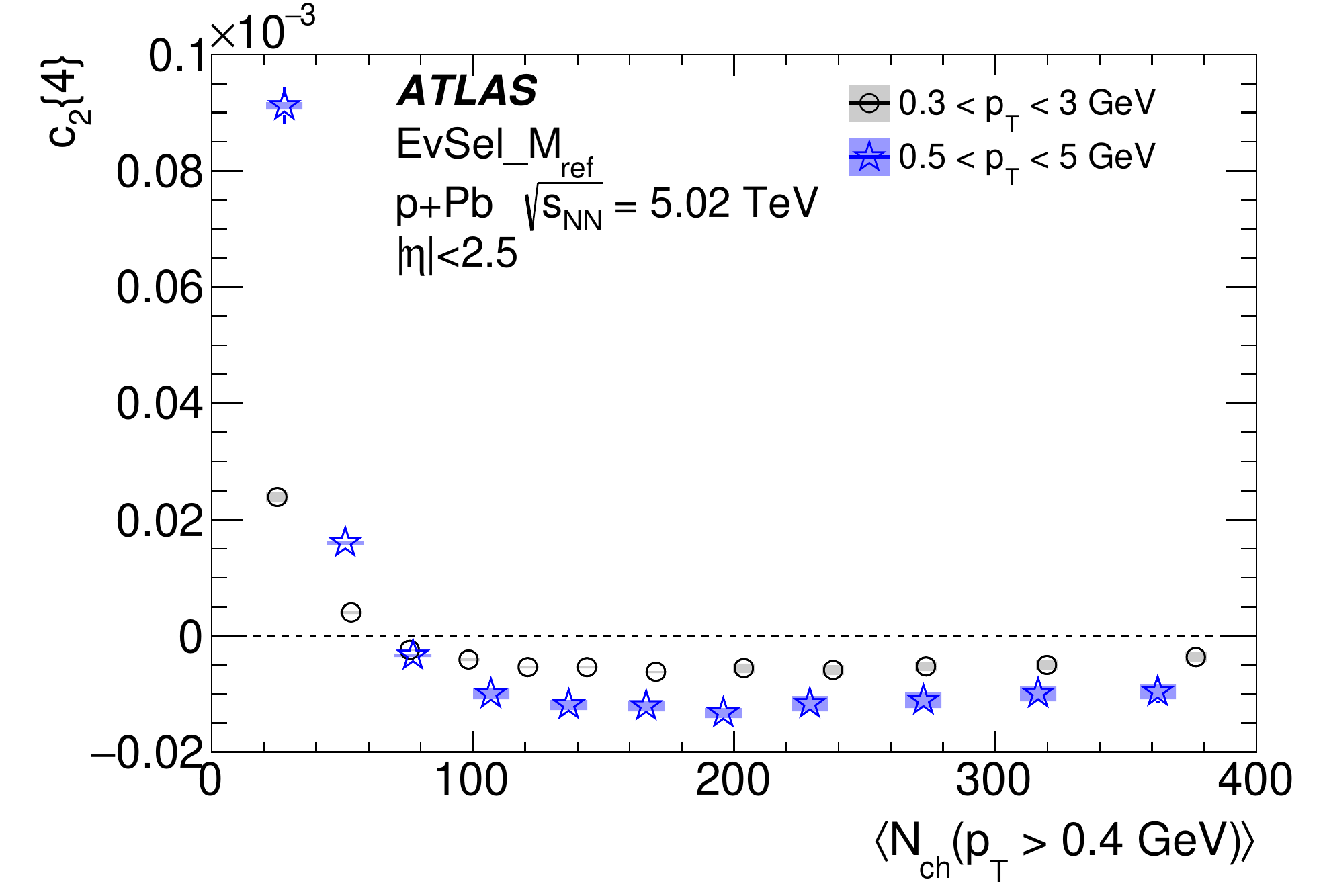}
\includegraphics[width=75mm]{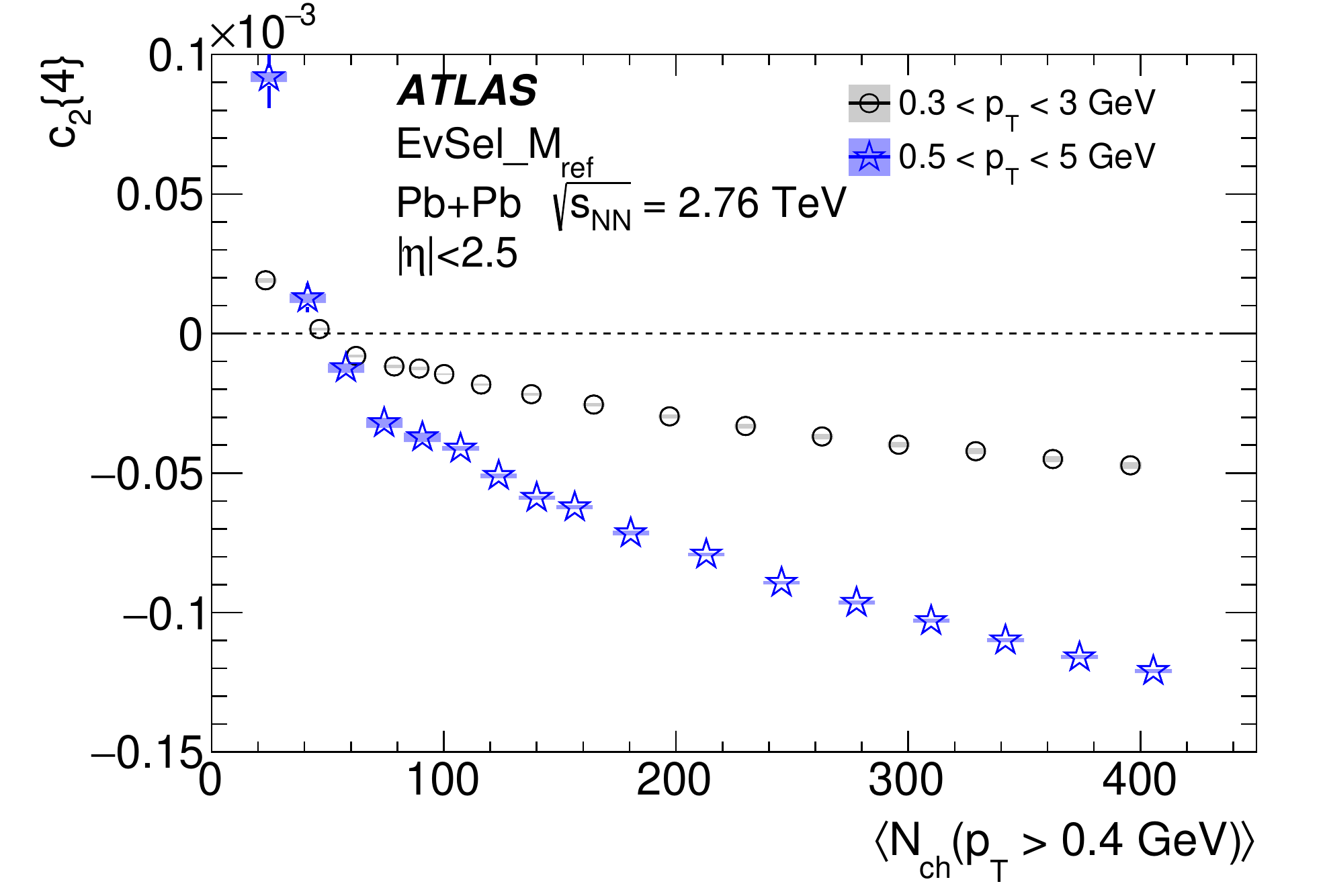}
\caption{Comparison of $c_2\{4\}$ obtained for two \pT ranges of reference tracks as a function of $\langle N_{\mathrm{ch}}(\pT > 0.4$~GeV)$\rangle$  for 5.02 TeV and 13 TeV \pp collisions, and  5.02 TeV \pPb collisions, and 2.76 TeV \PbPb collisions. The insets in the upper panels zoom in on the high-multiplicity data. The error bars and shaded boxes denote statistical and systematic uncertainties, respectively.}
\label{fig:compSystNtrkRef} 
\end{center}
\end{figure} 
A comparison of results for $c_2\{4\}$ obtained with two \pT ranges for reference tracks is shown in Figure~\ref{fig:compSystNtrkRef}. For \pPb and \PbPb collisions, in the region where $c_2\{4\} < 0$, the $|c_2\{4\}|$ is larger for higher-\pT reference particles, as expected due to the rise of $\mathrm{v}_2$ with \pT. For all collision systems, it is observed that for  $c_2\{4\} > 0$,  $c_2\{4\}$ is larger for higher-\pT reference particles. This indicates the influence of non-flow, jet-like correlations.
\begin{figure}[ht!]
\begin{center}
\includegraphics[width=75mm]{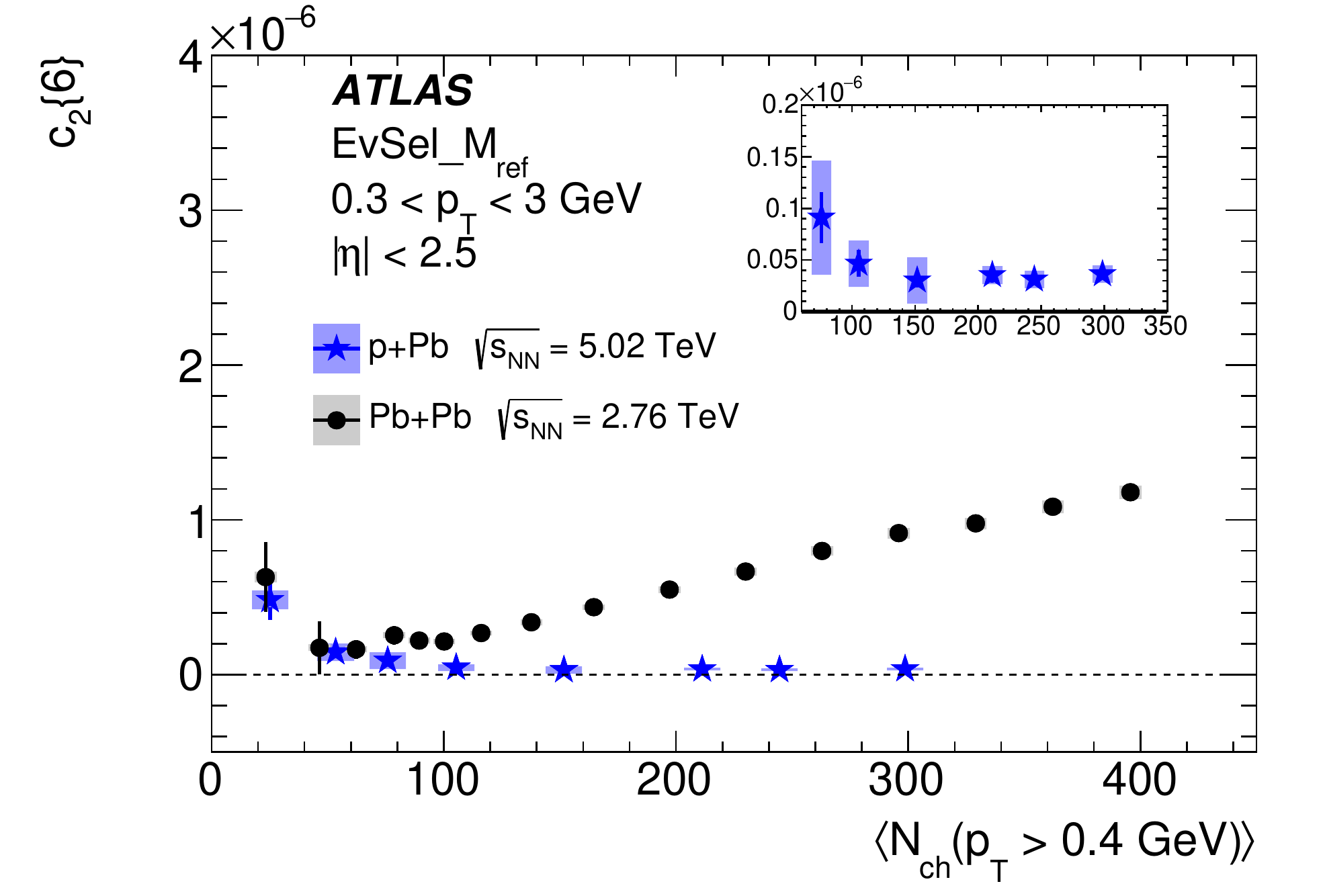}
\includegraphics[width=75mm]{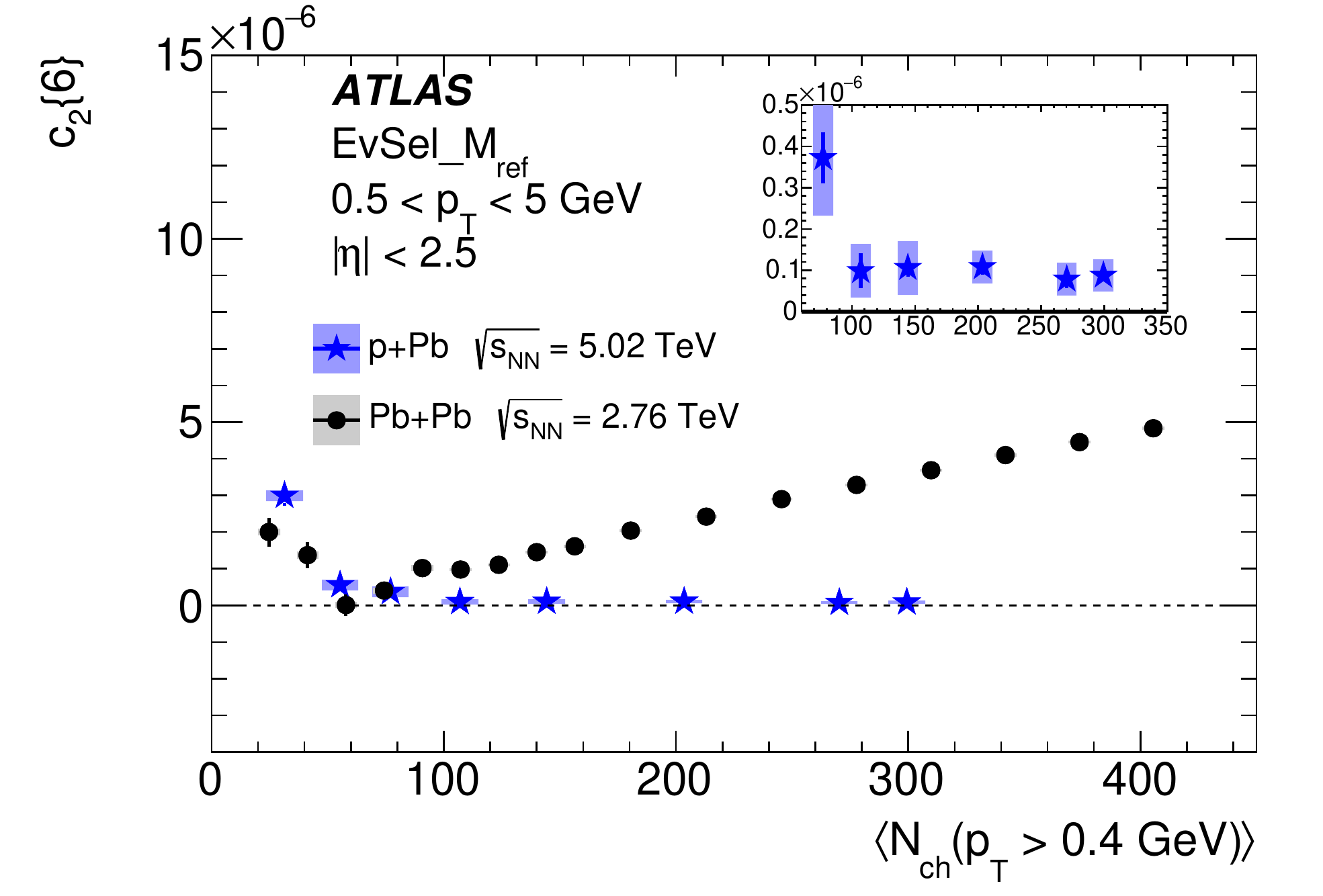}
\includegraphics[width=75mm]{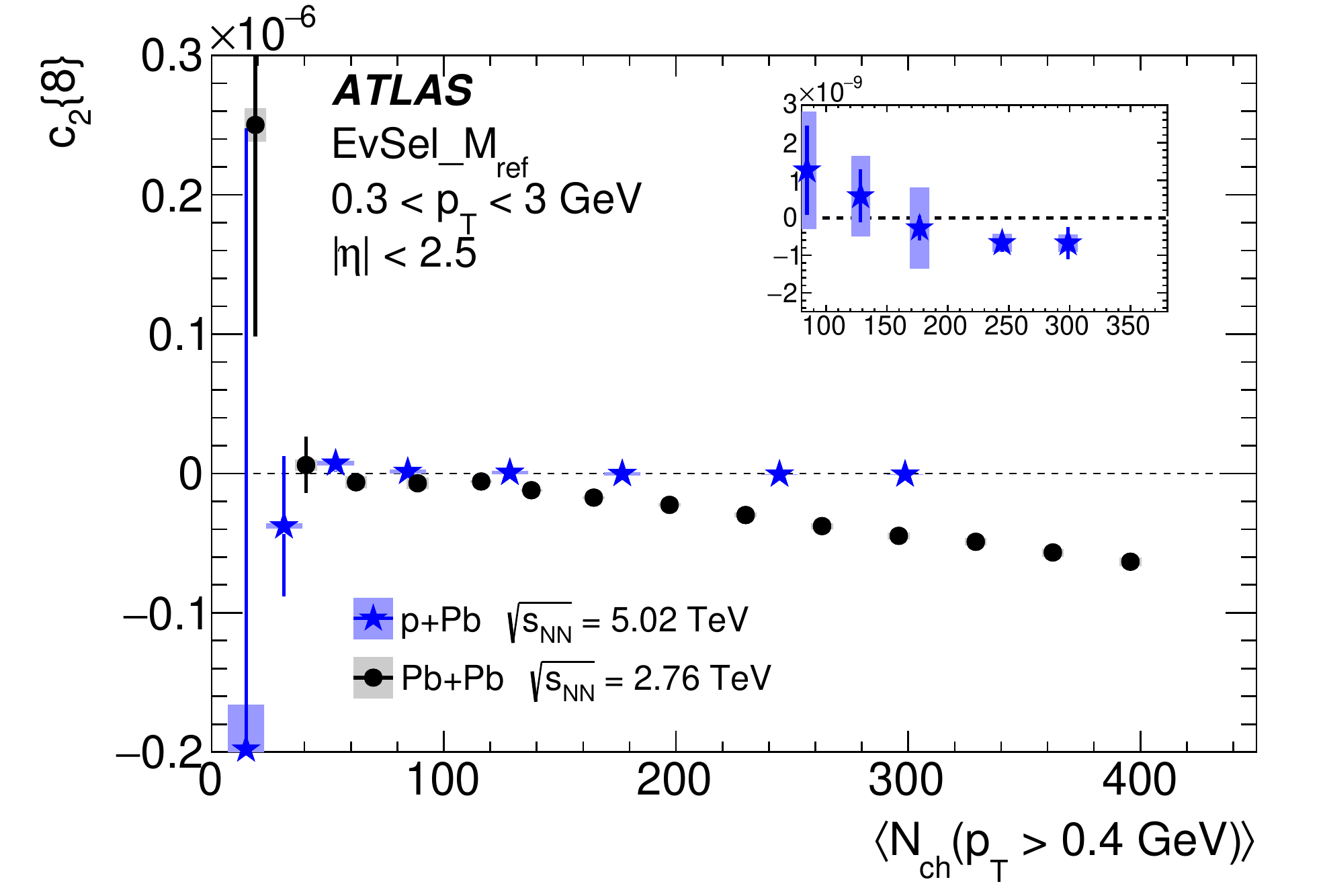}
\includegraphics[width=75mm]{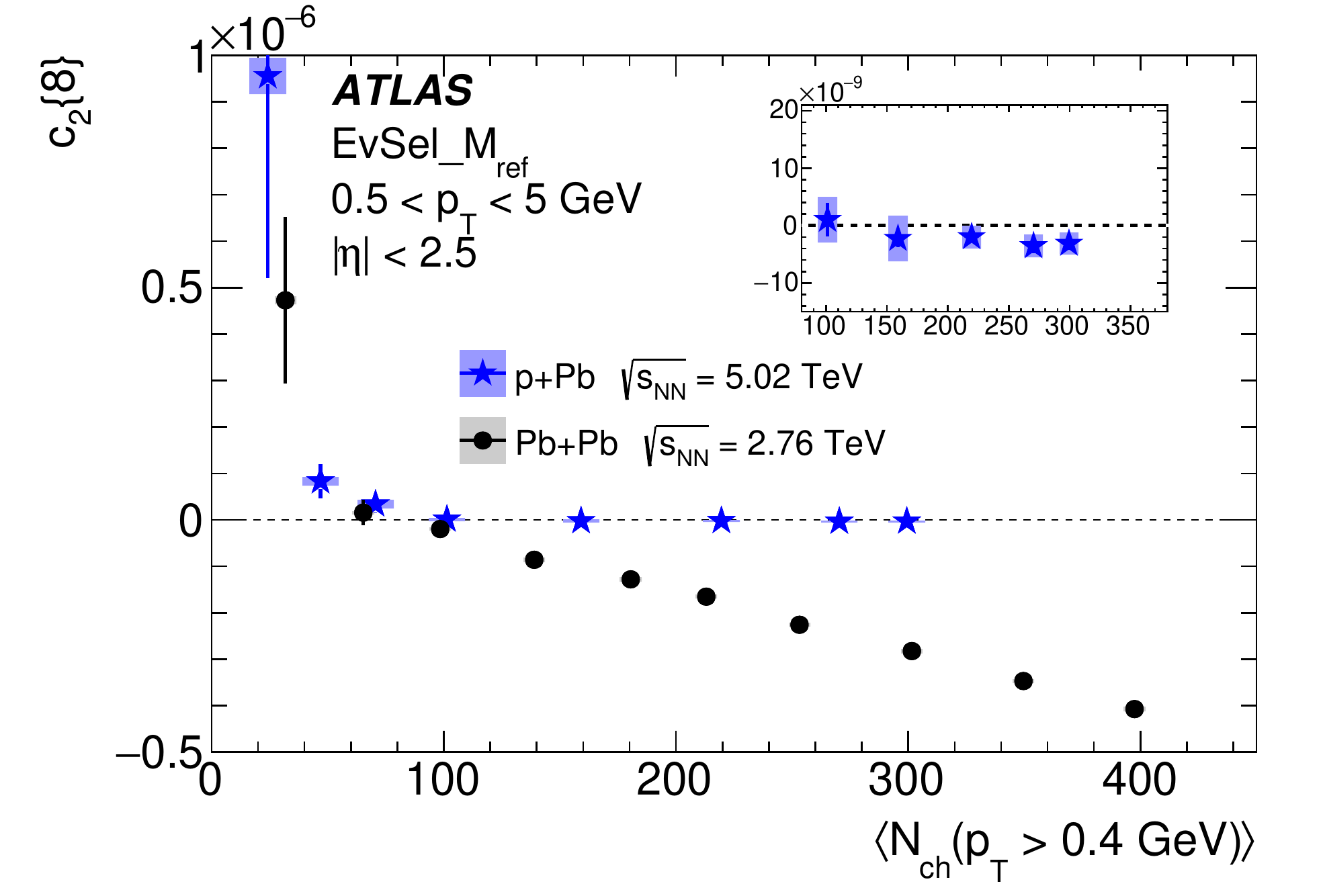}
\caption{Comparison of $c_2\{6\}$ (top) and  $c_2\{8\}$ (bottom) obtained for two \pT ranges of reference tracks as a function of $\langle N_{\mathrm{ch}}(\pT > 0.4$~GeV)$\rangle$  for \pPb collisions at \sqn= 5.02 TeV and low-multiplicity \PbPb collisions at \sqn= 2.76 TeV. The left (right) panels show cumulants calculated for reference particles with $0.3 < \pT <3$~GeV ($0.5 < \pT <5$~GeV). The insets zoom in on the regions around $c_2\{6\}= 0$ and $c_2\{8\}= 0$. The error bars and shaded boxes denote statistical and systematic uncertainties, respectively.}
\label{fig:c268} 
\end{center}
\end{figure} 

The six- and eight-particle $c_2$ cumulants are compared for \pPb and \PbPb collision systems in Figure~\ref{fig:c268}.  The measured $c_2\{6\}$ values are positive for both \pT ranges of reference particles. Positive values of $c_2\{6\}$ allow $\mathrm{v}_2\{6\}$ to be determined (see Eq.~\eqref{eq:vn6}).   For  \PbPb data, the $c_2\{8\}$, obtained for both \pT ranges of reference particles have negative values, and as such permit the evaluation of  $\mathrm{v}_2\{8\}$; however, for \pPb this requirement is only satisfied for a limited range of very high multiplicities. 

\begin{figure}[ht!]
\begin{center}
\includegraphics[width=75mm]{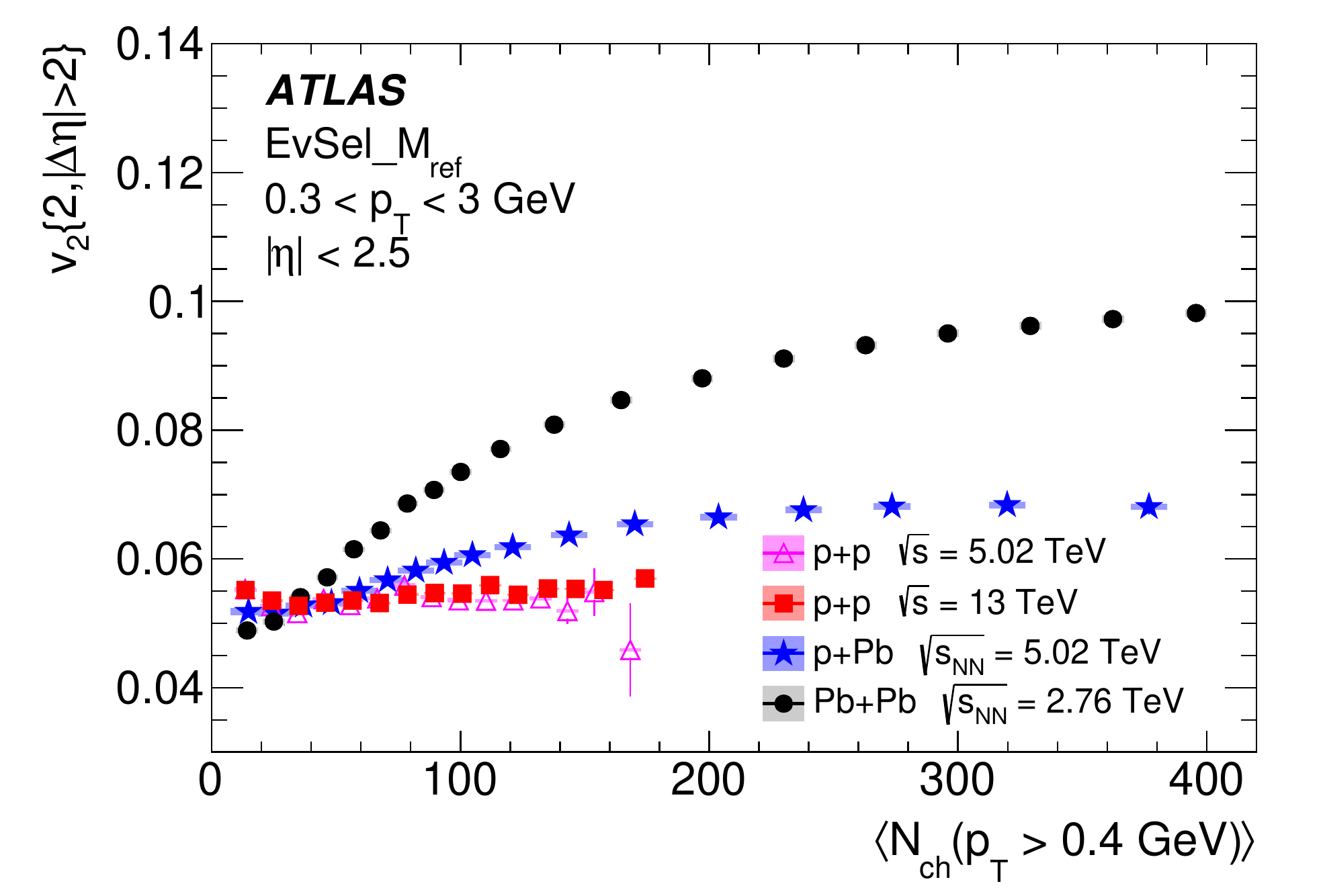}
\includegraphics[width=75mm]{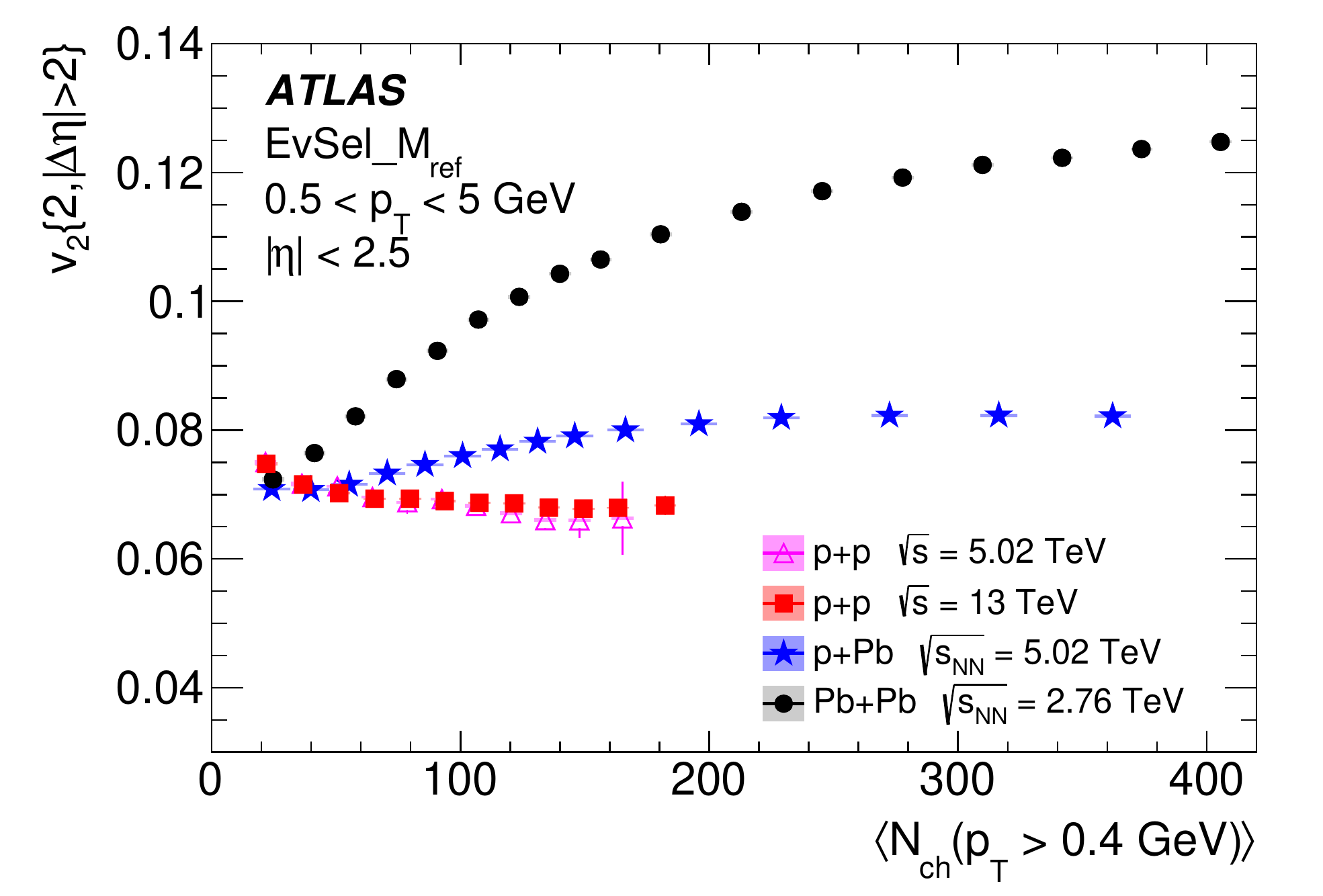}
\caption{Comparison of $\mathrm{v}_2\{2,|\Delta\eta| > 2\}$ as a function of $\langle N_{\mathrm{ch}}(\pT > 0.4$~GeV)$\rangle$ for \pp collisions at \sqs= 5.02 and 13 TeV, \pPb collisions at \sqn= 5.02 TeV and low-multiplicity \PbPb collisions at \sqn= 2.76 TeV, and for two \pT ranges of reference particles. The error bars and shaded boxes denote statistical and systematic uncertainties, respectively.}
\label{fig:Allv22gap} 
\end{center}
\end{figure}

The second-order Fourier harmonics, $\mathrm{v}_2$, is obtained from $c_2$, following Eqs.~\eqref{eq:vn2} - \eqref{eq:vn8}. Real values of $\mathrm{v}_2$ can only be obtained when the values of $c_2\{4\}$ and $c_2\{8\}$ ($c_2\{2,|\Delta\eta|>2\}$ and $c_2\{6\}$) are negative (positive). 
Results for the $\mathrm{v}_2$ harmonic can only be compared for four analysed collision systems for $\mathrm{v}_2\{2,|\Delta\eta| > 2\}$, derived from $c_2\{2,|\Delta\eta| > 2\}$. Such a comparison is shown in Figure~\ref{fig:Allv22gap}.  A number of distinct differences can be observed: (i) for the same $N_{\mathrm{ch}}(\pT > 0.4$~GeV), the largest values of the second-order Fourier harmonic are observed for \PbPb collisions and at the highest multiplicities $\mathrm{v}_2\{2,|\Delta\eta| > 2\}$ for \PbPb is almost twice as large as for \pPb collisions; (ii) the smallest $\mathrm{v}_2$ values are observed for \pp data, which show no dependence on collision energy. For \pp collisions,  the $\mathrm{v}_2\{2,|\Delta\eta| > 2\}$ is weakly dependent on multiplicity, showing a slight decrease for reference particles with higher transverse momenta.  For \pPb and \PbPb collisions, $\mathrm{v}_2\{2,|\Delta\eta| > 2\}$ increases with increasing multiplicity up to $N_{\mathrm{ch}}(\pT > 0.4$~GeV) $\simeq 250$. At higher multiplicities the increase gets weaker for \PbPb collisions, while for \pPb data the second-order flow harmonics are observed to be independent of the multiplicity. Larger $\mathrm{v}_2\{2,|\Delta\eta| > 2\}$ values are observed for reference particles with higher transverse momenta. 
\begin{figure}[H]
\begin{center}
\includegraphics[width=75mm]{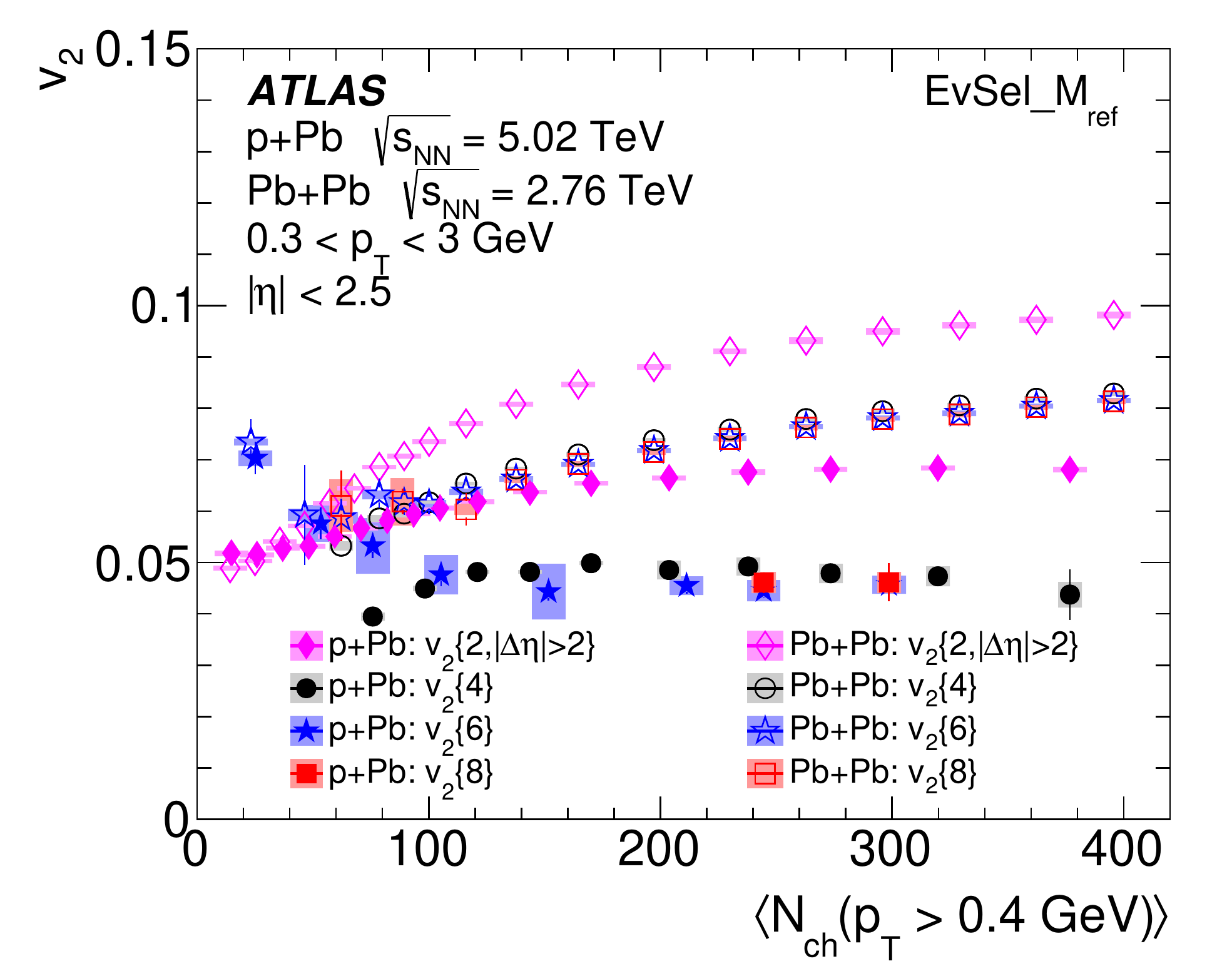}
\includegraphics[width=75mm]{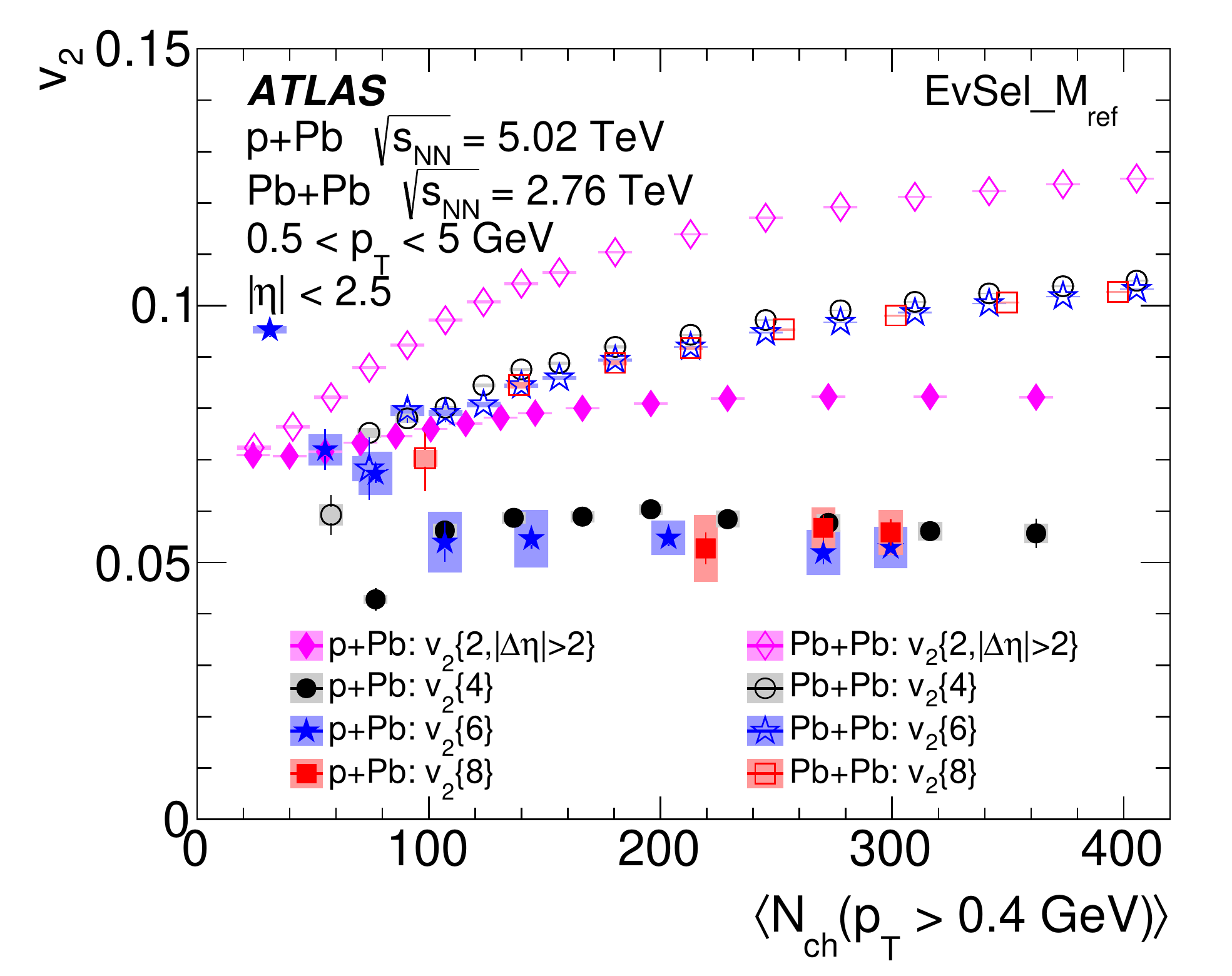}
\caption{Comparison of $\mathrm{v}_2\{2,|\Delta\eta| > 2\}$,
$\mathrm{v}_2\{4\}$, $\mathrm{v}_2\{6\}$ and $\mathrm{v}_2\{8\}$ as a function of $\langle N_{\mathrm{ch}}(\pT > 0.4$~GeV)$\rangle$ for \pPb collisions at \sqn= 5.02 TeV and low-multiplicity \PbPb collisions at \sqn= 2.76 TeV. The results are presented for two \pT ranges of the reference particles as indicated in the legend. The error bars and shaded boxes denote statistical and systematic uncertainties, respectively.}
\label{fig:v2468} 
\end{center}
\end{figure} 

\begin{figure}[ht!]
\begin{center}
\includegraphics[width=75mm]{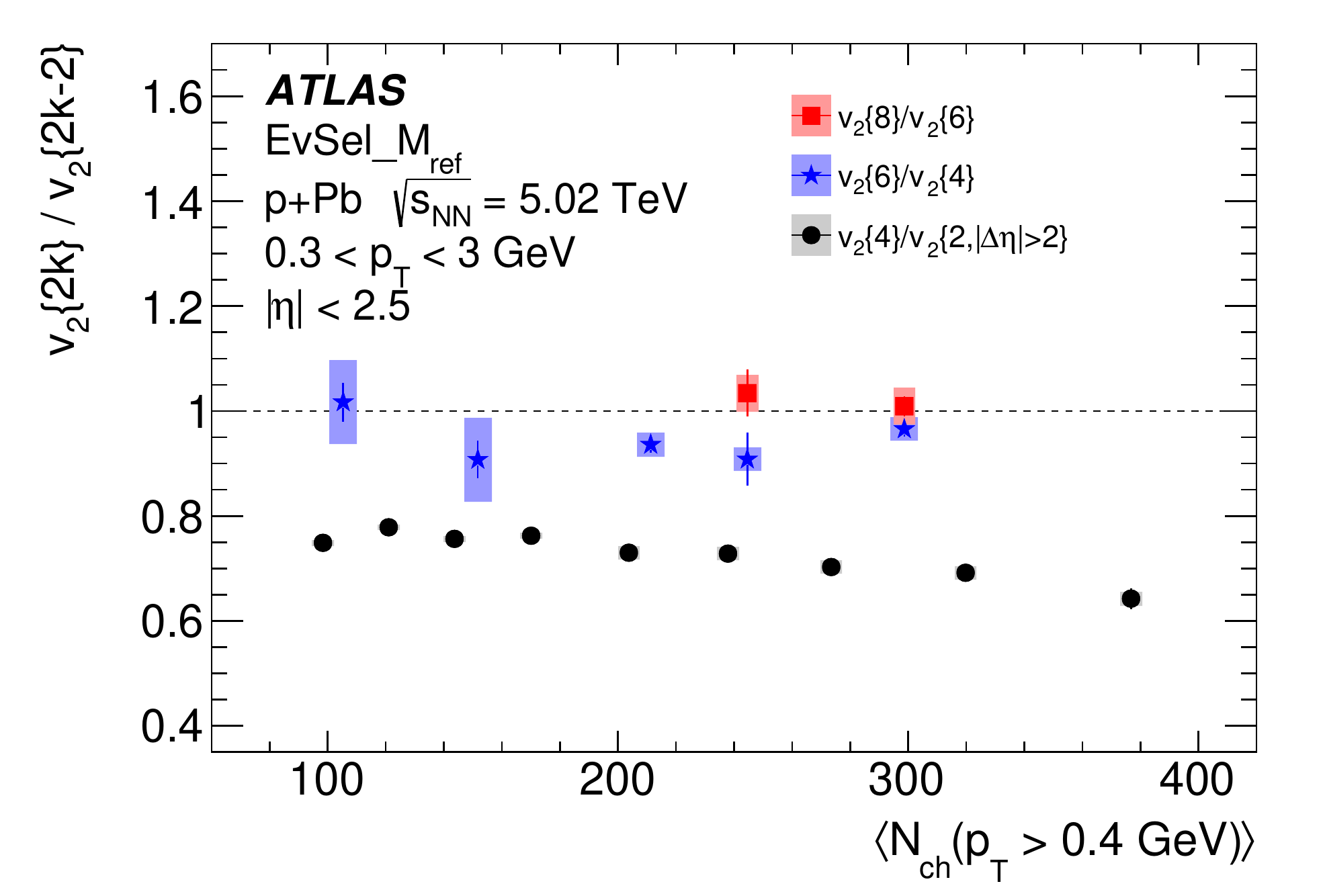}
\includegraphics[width=75mm]{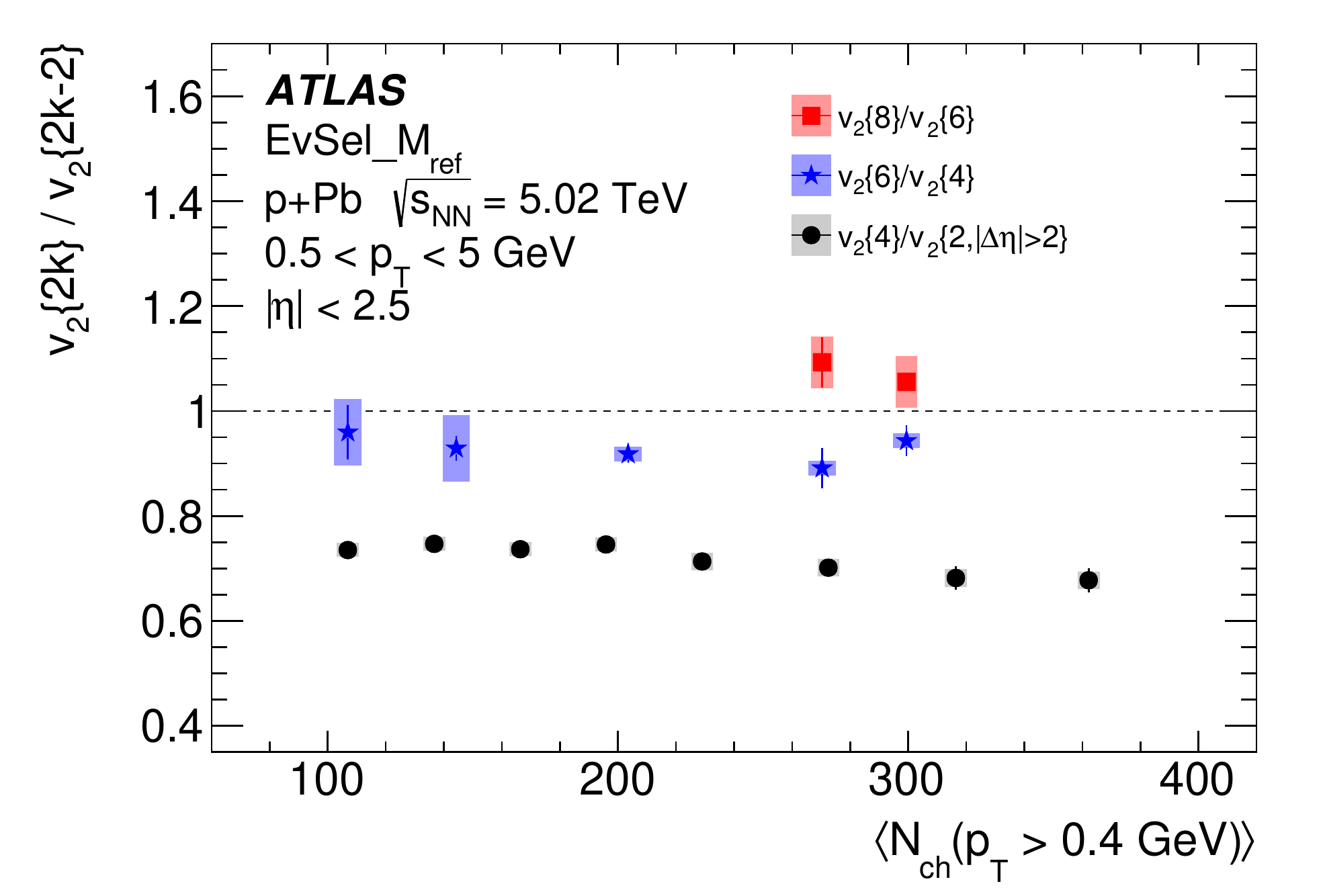}
\includegraphics[width=75mm]{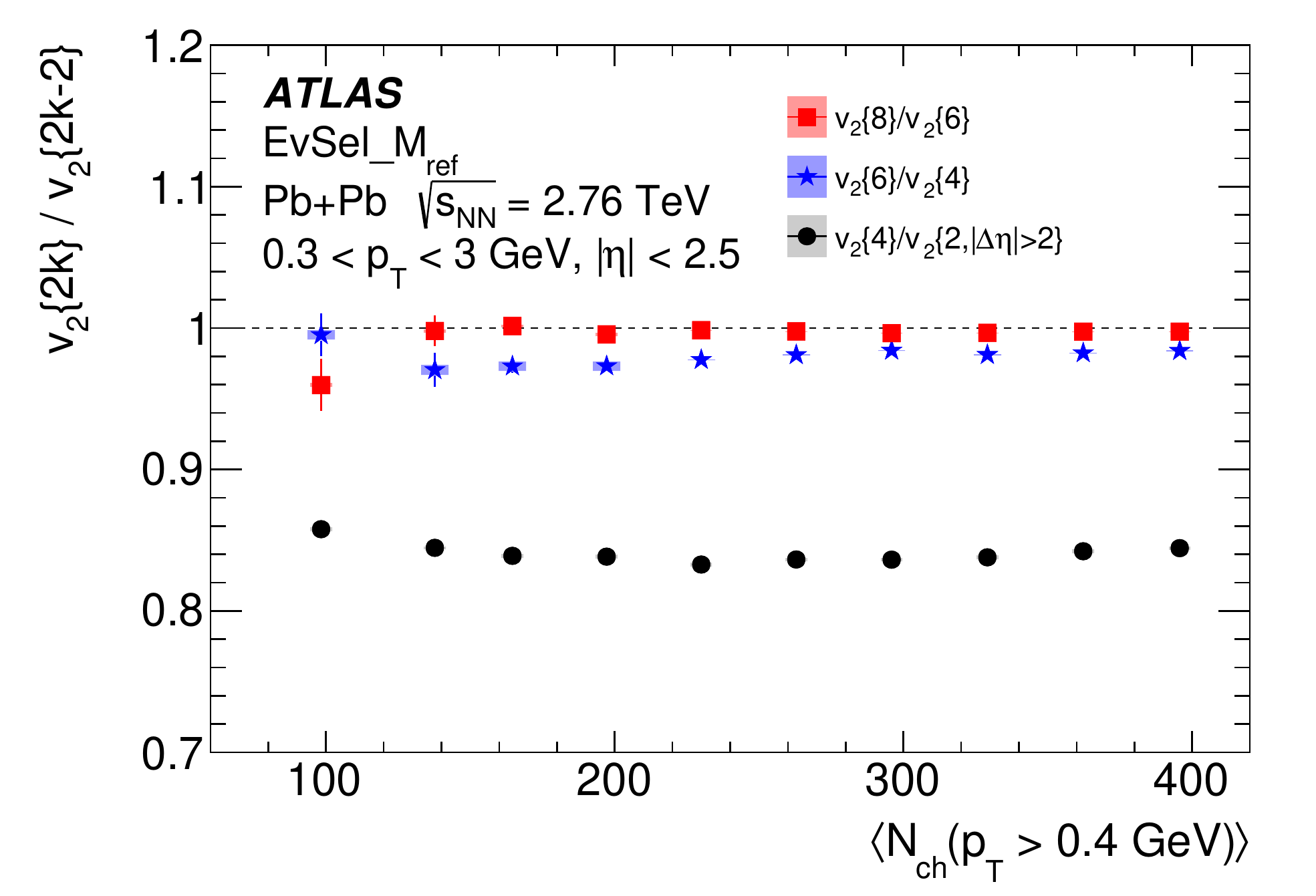}
\includegraphics[width=75mm]{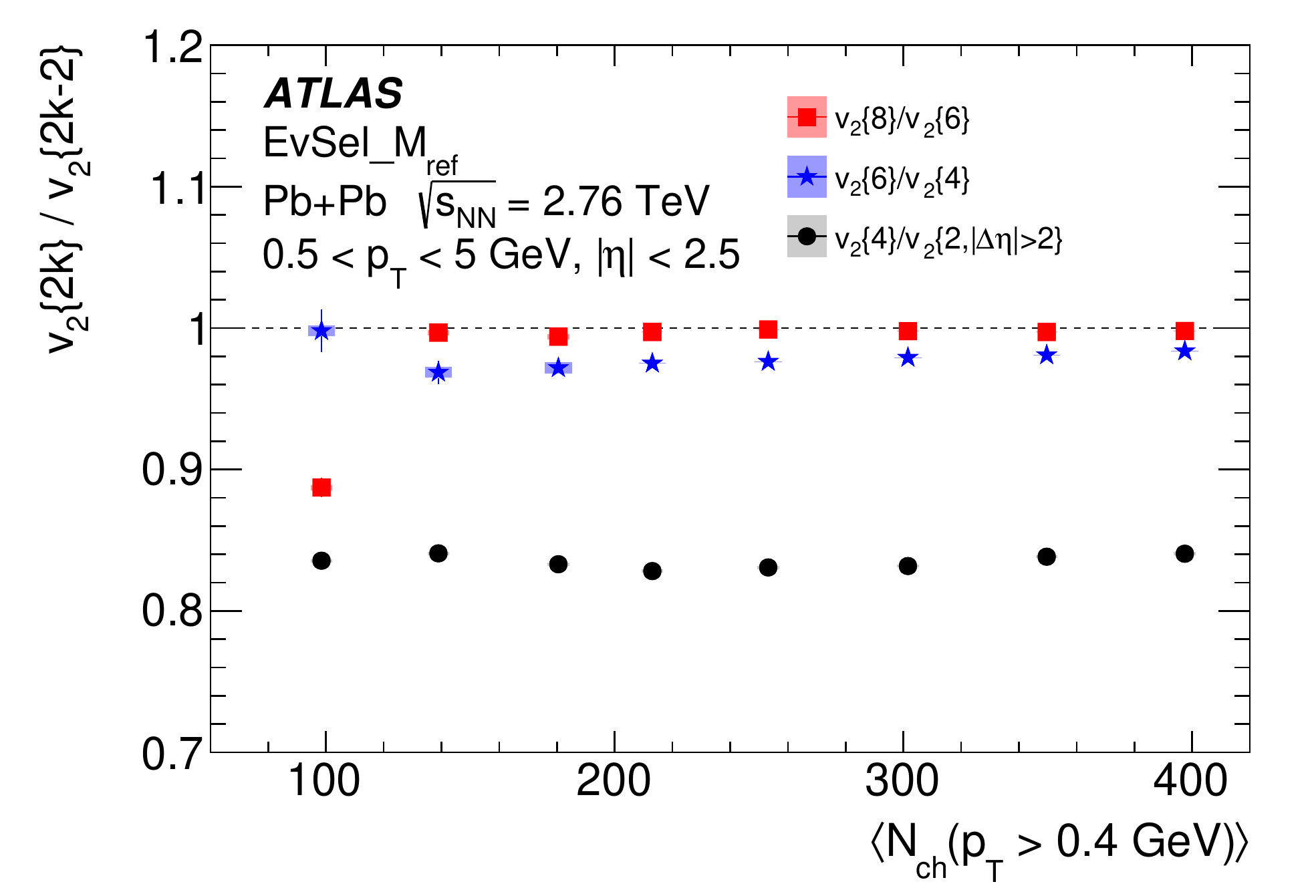}
\caption{The ratios $\mathrm{v}_2\{4\}/\mathrm{v}_2\{2,|\Delta\eta| > 2\}$, $\mathrm{v}_2\{6\}/\mathrm{v}_2\{4\}$ and $\mathrm{v}_2\{8\}/\mathrm{v}_2\{6\}$ 
as a function of $\langle N_{\mathrm{ch}}(\pT > 0.4$~GeV)$\rangle$ for \pPb collisions at \sqn= 5.02 TeV (top) and low-multiplicity \PbPb collisions at \sqn= 2.76 TeV (bottom). Left (right) panels show cumulants calculated for reference particles with $0.3 < \pT <3~\mathrm{GeV}$ ($0.5 < \pT <5~\mathrm{GeV}$). The error bars and shaded boxes denote statistical and systematic uncertainties, respectively.  }
\label{fig:v2ratios} 
\end{center}
\end{figure}

A comparison of the $\mathrm{v}_2$ harmonic obtained with different cumulants, $\mathrm{v}_2\{2,|\Delta\eta| > 2\}$,
$\mathrm{v}_2\{4\}$, $\mathrm{v}_2\{6\}$ and $\mathrm{v}_2\{8\}$, is shown in Figure~\ref{fig:v2468} for \pPb and low-multiplicity \PbPb collisions for the two \pT ranges of reference particles. All derived $\mathrm{v}_2$ harmonics in \PbPb collisions have magnitudes larger than those in \pPb collisions with the same multiplicity. For both systems, $\mathrm{v}_2\{2k\}$ are similar for $k= 2, 3$ and 4 while $\mathrm{v}_2\{2,|\Delta\eta| > 2\}$ are systematically larger. 
However, compared to almost degenerate values of $\mathrm{v}_2\{2k\}, k>1$, a larger $\mathrm{v}_2$ derived from two-particle cumulants is also predicted by models assuming fluctuation-driven initial-state anisotropies in small collision systems, either in the context of hydrodynamics as in Ref.~\cite{v468Ollitrault} or in the effective theory of quantum chromodynamics in the regime of weak coupling \cite{bzdak1,bzdak2}. Figure~\ref{fig:v2ratios} shows the ratio $\mathrm{v}_2\{2k\}/\mathrm{v}_2\{2k-2\}$ for \pPb and low-multiplicity \PbPb collisions as a function of charged-particle multiplicity. Interestingly, for \PbPb collisions all three ratios are independent of $N_{\mathrm{ch}}(\pT > 0.4$~GeV)  beyond 120, independent of the \pT range of reference particles. The $\mathrm{v}_2\{4\}/\mathrm{v}_2\{2,|\Delta\eta| > 2\}$ ratios stay constant at the value of 0.85, while 
$\mathrm{v}_2\{6\}/\mathrm{v}_2\{4\}$ and $\mathrm{v}_2\{8\}/\mathrm{v}_2\{6\}$ ratios are almost degenerate at a value close to one, yet systematically $\mathrm{v}_2\{8\}/\mathrm{v}_2\{6\} > \mathrm{v}_2\{6\}/\mathrm{v}_2\{4\}$. For \pPb collisions, similar universal behaviour of $\mathrm{v}_2\{2k\}/\mathrm{v}_2\{2k-2\}$ ratios is seen, although within much larger uncertainties. The $\mathrm{v}_2\{4\}/\mathrm{v}_2\{2,|\Delta\eta| > 2\}$ ratio has a value of about 0.7, thus smaller than in \PbPb collisions, and shows a tendency to decrease weakly with increasing multiplicity. These observations are qualitatively consistent with the predictions of the model of fluctuating initial geometry from Ref.~\cite{v468Ollitrault}, and provide further constraints on the initial state. 

\subsection{Higher-order multi-particle cumulants and Fourier harmonics}
\label{sec:v3v4}
\begin{figure}[ht!]
\begin{center}
\includegraphics[width=75mm]{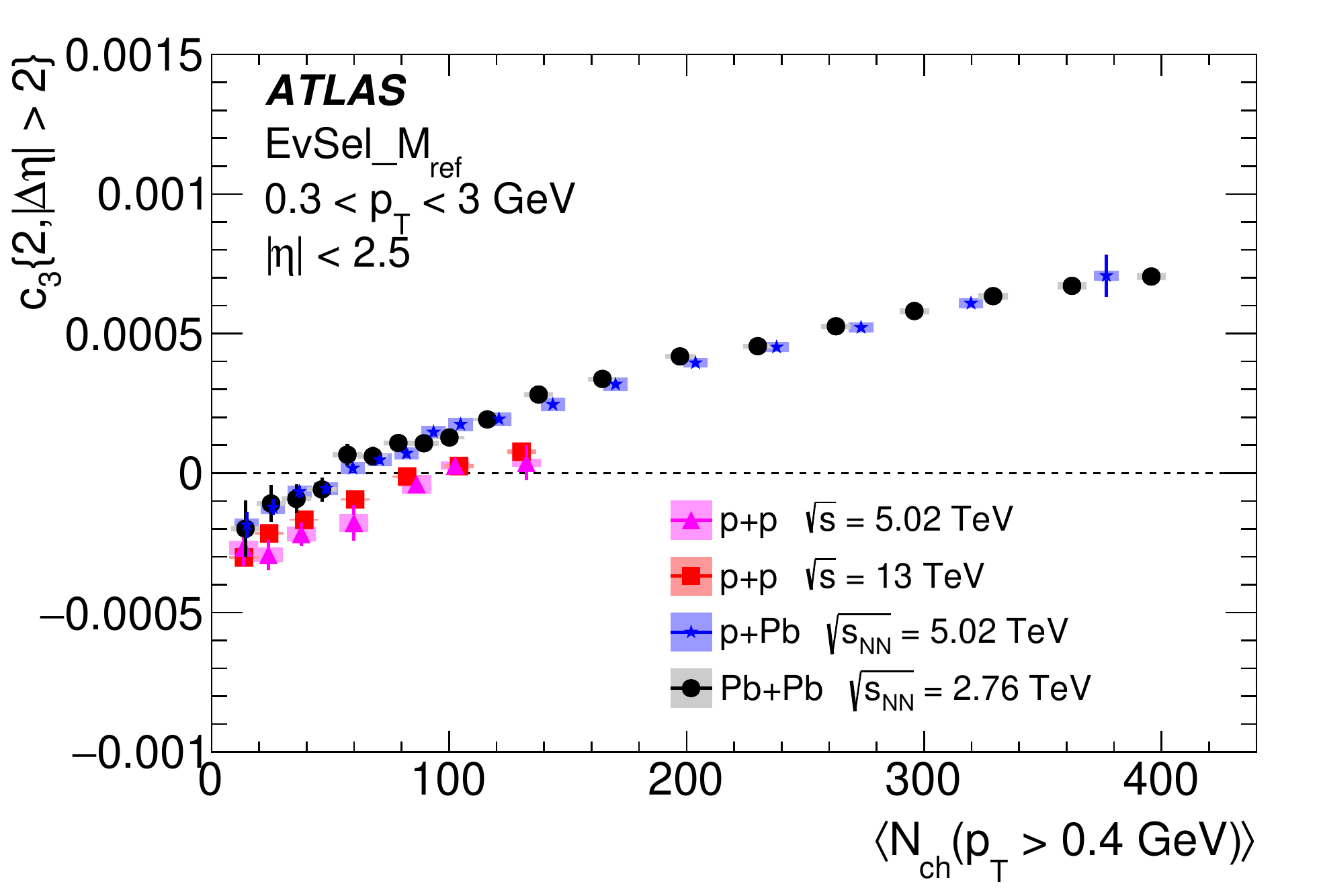}
\includegraphics[width=75mm]{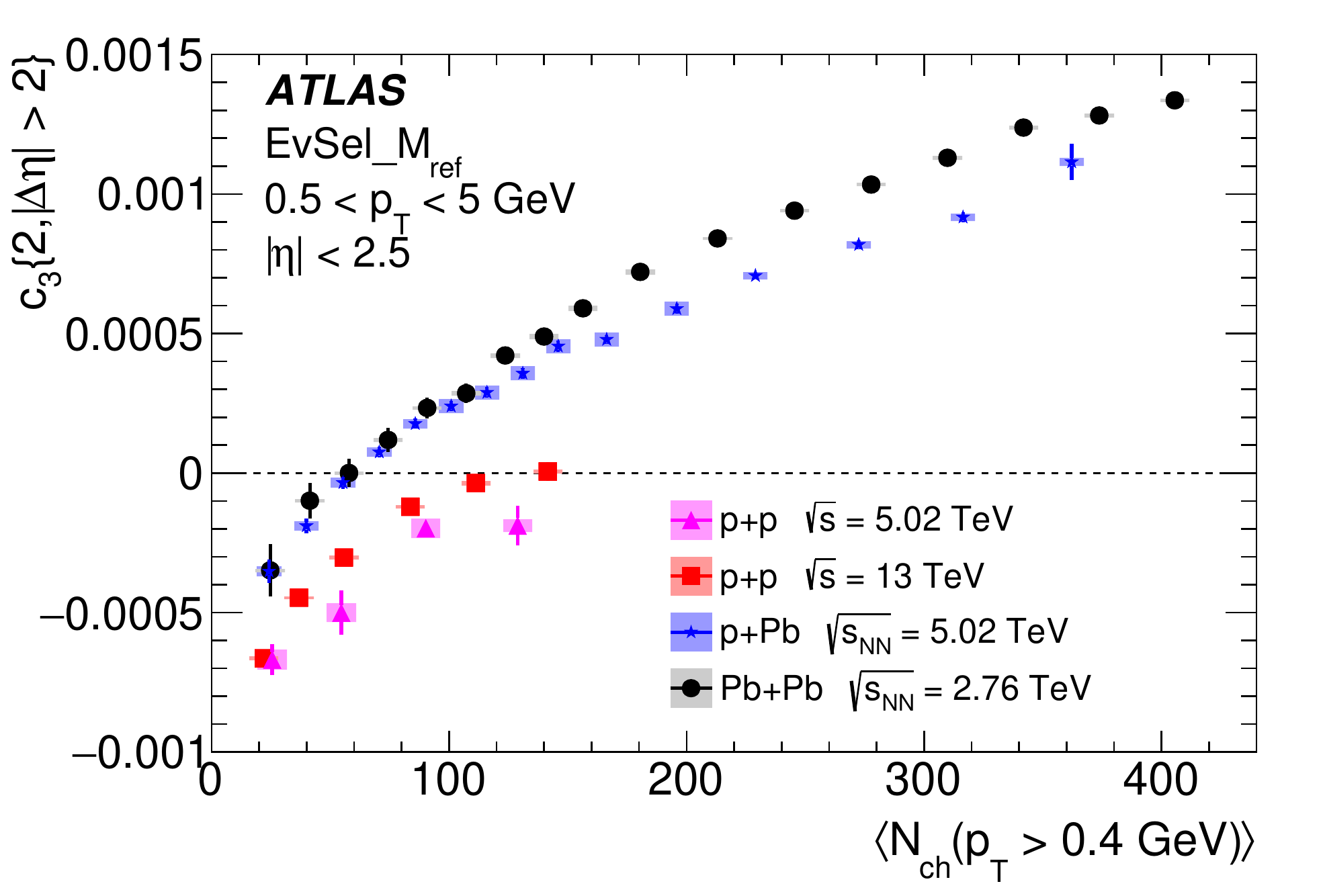}
\includegraphics[width=75mm]{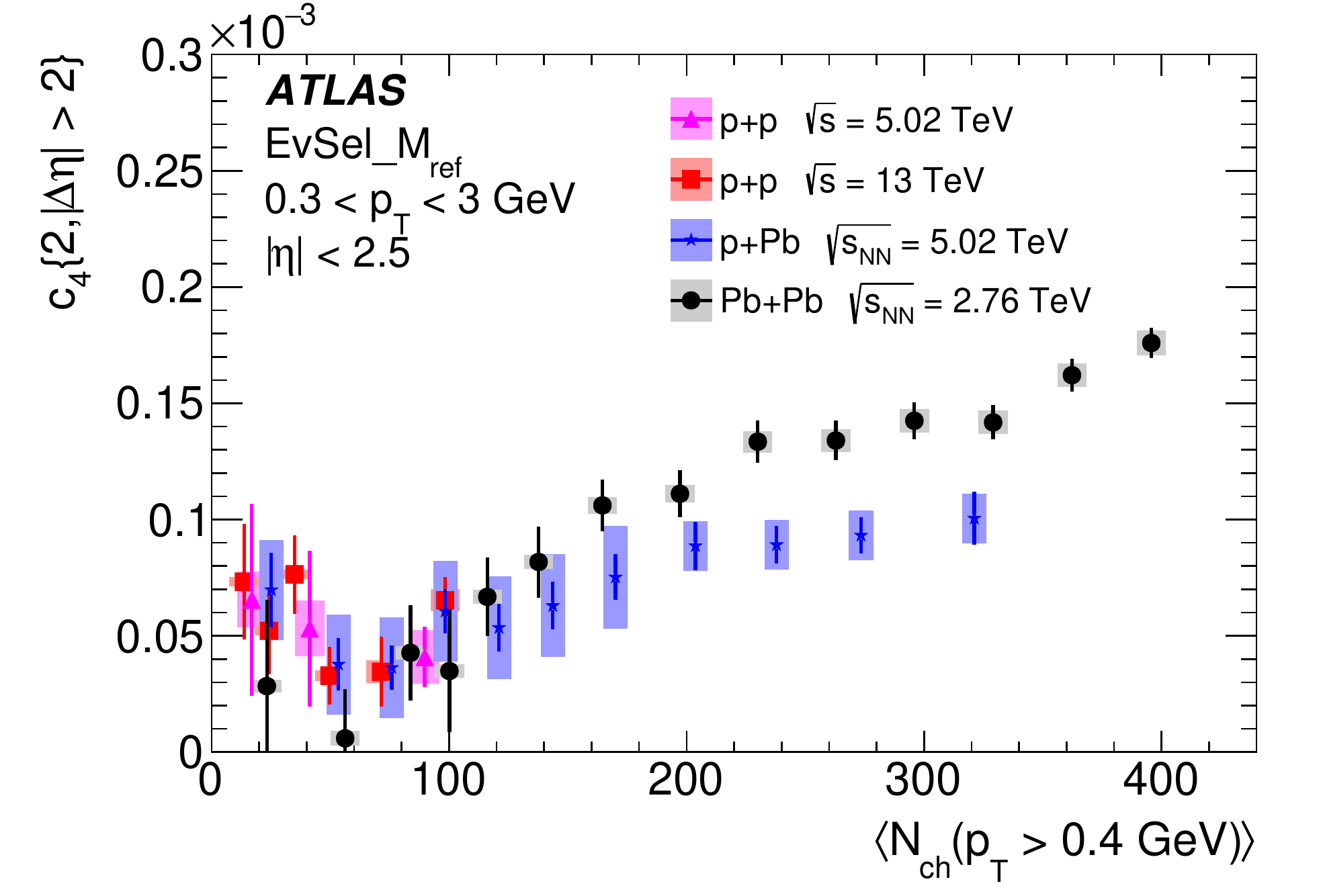}
\includegraphics[width=75mm]{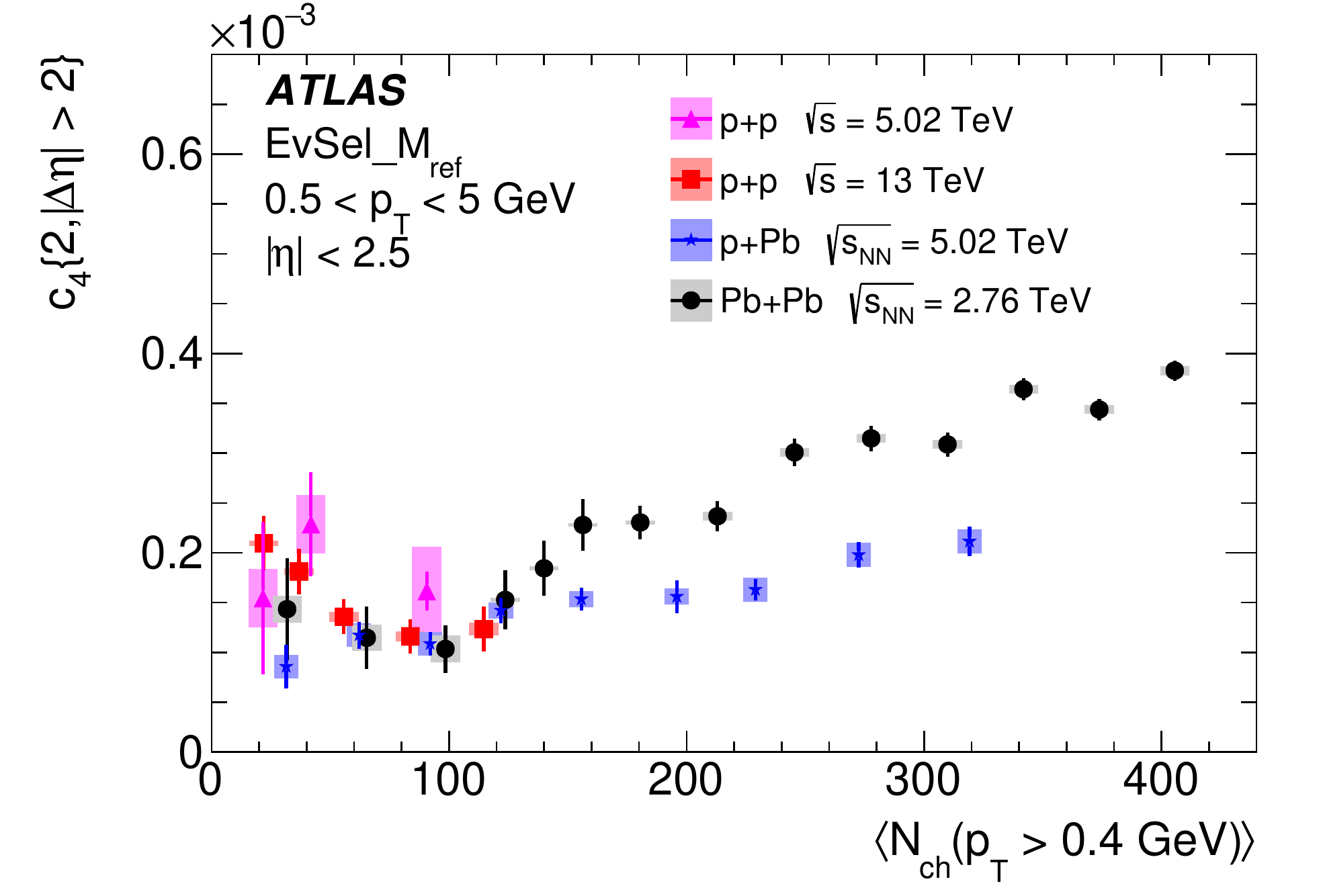}
\caption{The two-particle cumulant $c_3$ (top) and $c_4$ (bottom) calculated with the $|\Delta\eta|>2$ requirement as a function of $\langle N_{\mathrm{ch}}(\pT > 0.4$~GeV)$\rangle$  for \pp collisions at \sqs= 5.02 and 13 TeV, \pPb collisions at \sqn= 5.02 TeV and low-multiplicity \PbPb collisions at \sqn= 2.76 TeV for two \pT ranges of reference particles. The error bars and shaded boxes denote statistical and systematic uncertainties, respectively.}
\label{fig:c3c4} 
\end{center}
\end{figure}

Calculations of $c_3$ and $c_4$ multi-particle cumulants are statistics-limited and statistically significant results can only be obtained using two-particle cumulants with the superimposed $|\Delta\eta|>2$ gap. Figure~\ref{fig:c3c4} shows a comparison between different collision systems for $c_3\{2,|\Delta\eta|>2\}$ and $c_4\{2,|\Delta\eta|>2\}$ cumulants calculated for $M_{\mathrm{ref}}$, where the \pT range of reference particles is either ${0.3 < \pT < 3.0~\mathrm{GeV}}$ or ${0.5 < \pT < 5.0~\mathrm{GeV}}$.   

For \pp data, the $c_3\{2,|\Delta\eta|>2\}$ values are either negative or consistent with zero over the whole range of $N_{\mathrm{ch}}(\pT > 0.4$~GeV), except for the two highest multiplicities measured for \pp collisions at 13~TeV. Therefore, for  $N_{\mathrm{ch}}(\pT > 0.4$~GeV) < 100, the $\mathrm{v}_3$ signal in \pp collisions is undefined or zero within the errors. A positive $c_3$ signal is obtained from \pPb and \PbPb data, except for the charged-particle multiplicities below about 50. The magnitude of $c_3$ is comparable for \PbPb and \pPb collisions when reference particles with  $0.3< \pT < 3.0$~GeV are selected, and systematically slightly larger for \PbPb than for \pPb for reference particles with $0.5< \pT < 5.0$~GeV.  The fourth-order cumulants, $c_4$, have positive values of $c_4\{2,|\Delta\eta|>2\}$ even for the \pp data, and their  magnitude  is comparable to that for \pPb and \PbPb collisions in the overlapping range of  $N_{\mathrm{ch}}$. For  $N_{\mathrm{ch}}(\pT > 0.4$~GeV) > 120, where only the measurements for \pPb and \PbPb are accessible, the $c_4$ cumulants measured at the same charged-particle multiplicity are larger for \PbPb than for \pPb.  
\begin{figure}[H]
\begin{center}
\includegraphics[width=75mm]{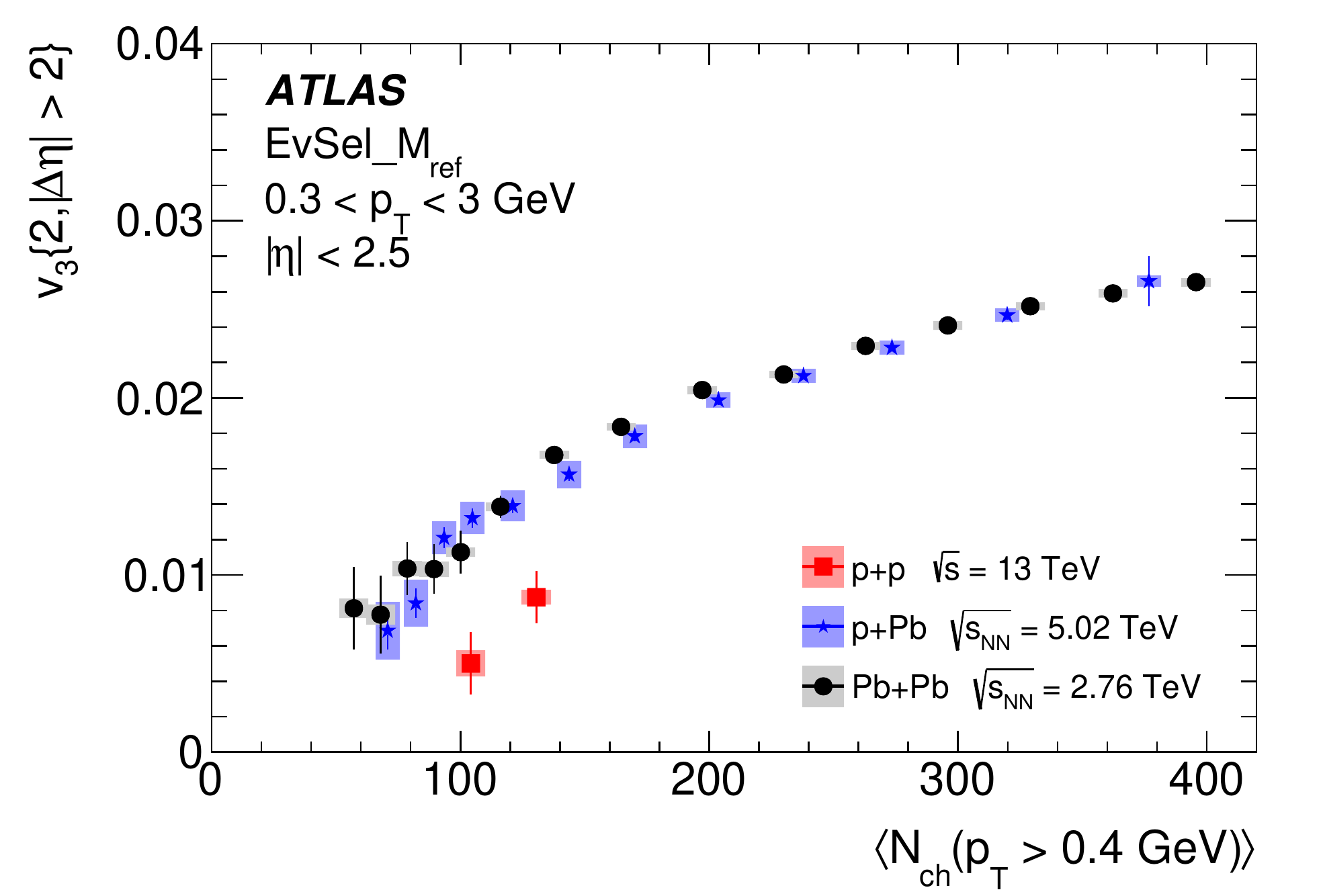}
\includegraphics[width=75mm]{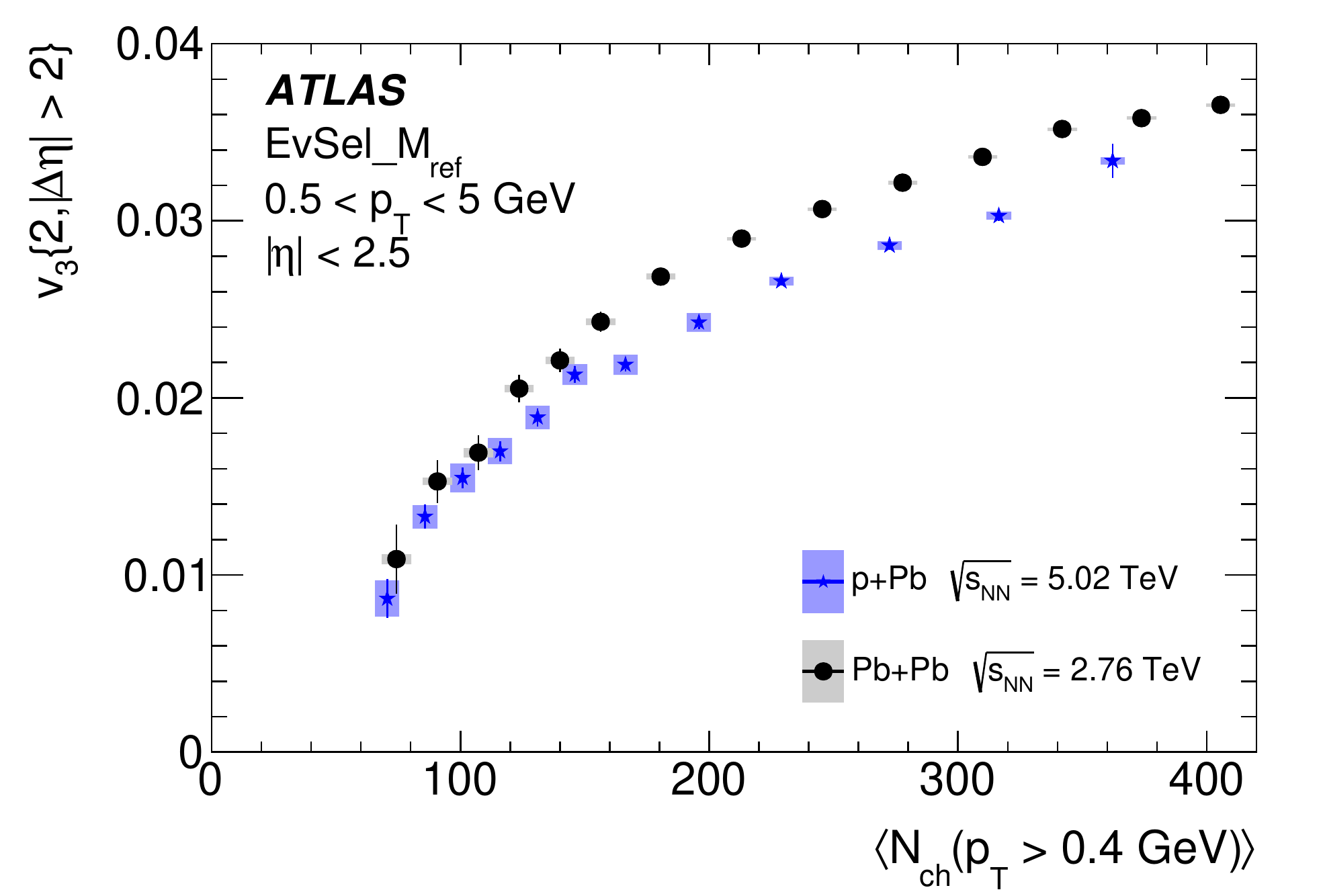}
\includegraphics[width=75mm]{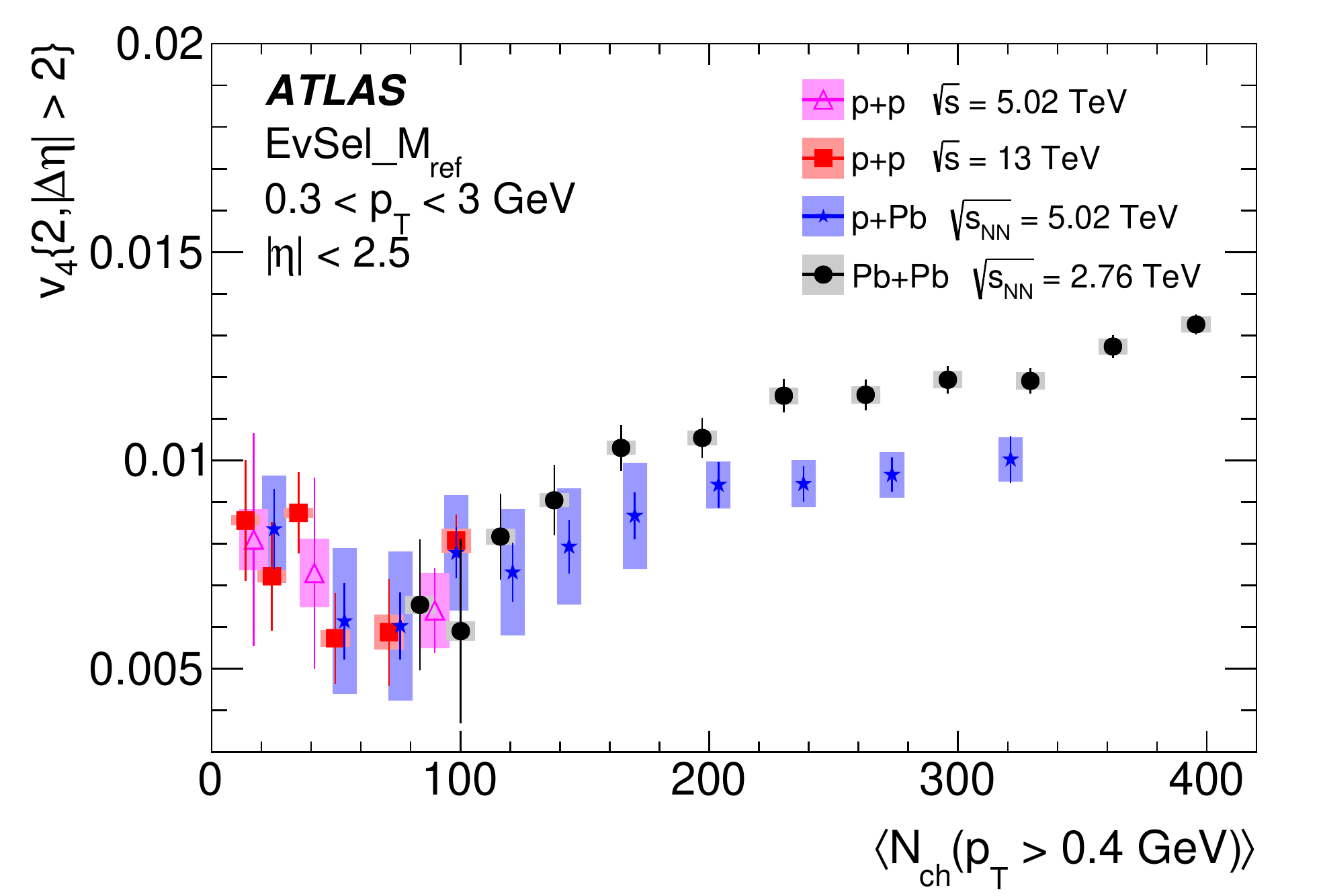}
\includegraphics[width=75mm]{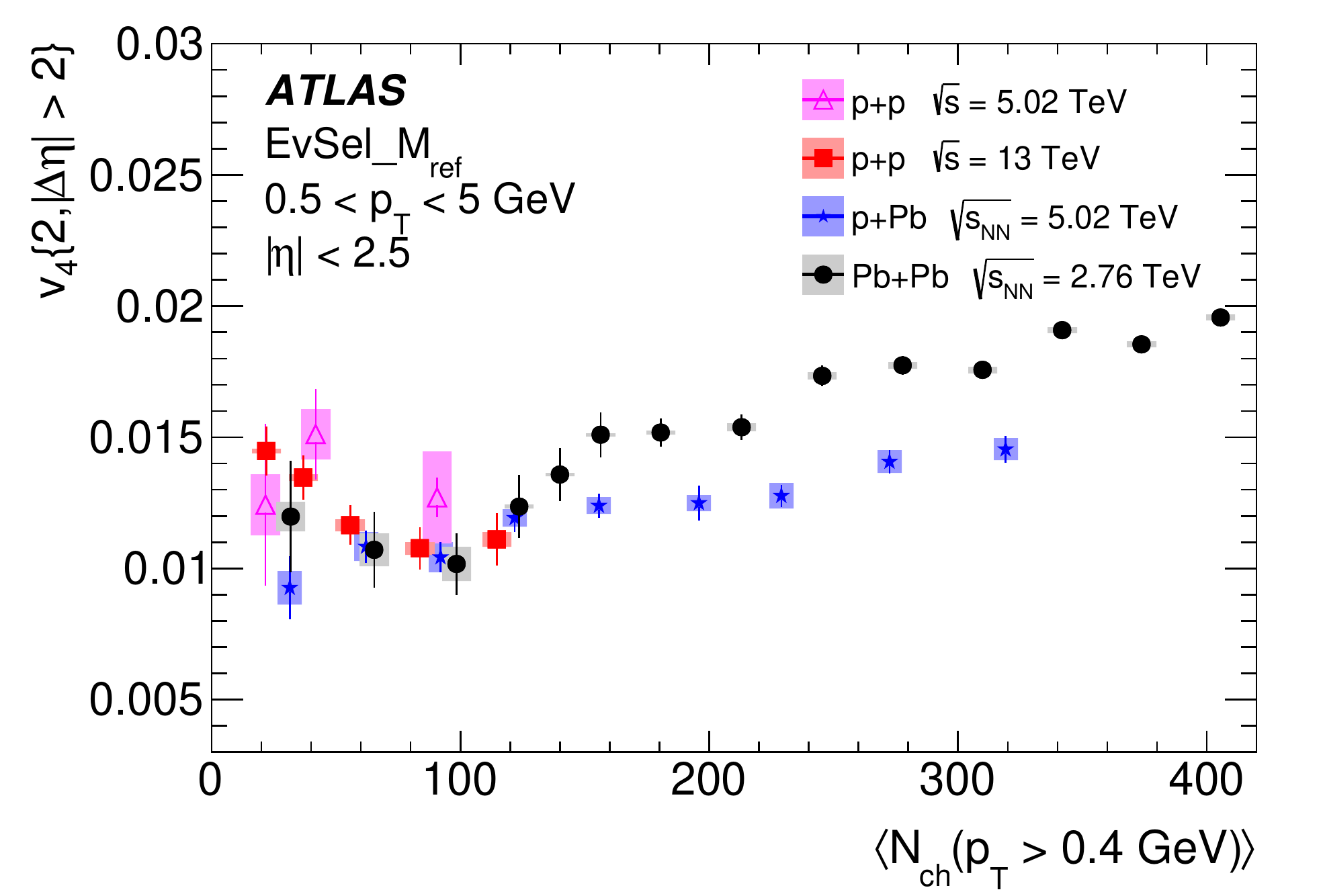}
\caption{The $\mathrm{v}_3\{2,|\Delta\eta|>2\}$ (top) and $\mathrm{v}_4\{2,|\Delta\eta|>2\}$ (bottom) as a function of $\langle N_{\mathrm{ch}}(\pT > 0.4$~GeV)$\rangle$ for \pp collisions at \sqs= 5.02 and 13 TeV, \pPb collisions at \sqn= 5.02 TeV and low-multiplicity \PbPb collisions at \sqn= 2.76 TeV, and for two \pT ranges of the reference particles. The error bars and shaded boxes denote statistical and systematic uncertainties, respectively.}
\label{fig:comp_pPb_PbPb_v34} 
\end{center}
\end{figure} 

The third- and fourth-order flow harmonics, $\mathrm{v}_3$ and $\mathrm{v}_4$, calculated with two-particle cumulants  with the $|\Delta\eta|>2$ requirement are shown in Figure~\ref{fig:comp_pPb_PbPb_v34}.   For \pPb and \PbPb collisions the $\mathrm{v}_3\{2,|\Delta\eta|>2\}$ values are similar for reference particles with $0.3< \pT < 3.0$~GeV, and much larger than for the 13~TeV \pp data. For higher-\pT reference particles, the  \PbPb $\mathrm{v}_3$ is systematically larger than $\mathrm{v}_3$ in \pPb collisions with  the same multiplicity. The $\mathrm{v}_3$ increases with increasing multiplicity. A weaker increase is seen for  $\mathrm{v}_4\{2,|\Delta\eta|>2\}$, but at high multiplicities the values observed in \PbPb collisions are systematically larger than in \pPb, for two \pT ranges of reference particles. For multiplicities below 100, where the $\mathrm{v}_4\{2,|\Delta\eta|>2\}$ can also be obtained from \pp collisions, no system dependence is seen. 

\subsection{Comparison to other results}
\label{sec:comp}
\begin{figure}[t!]
\begin{center}
\includegraphics[width=70mm]{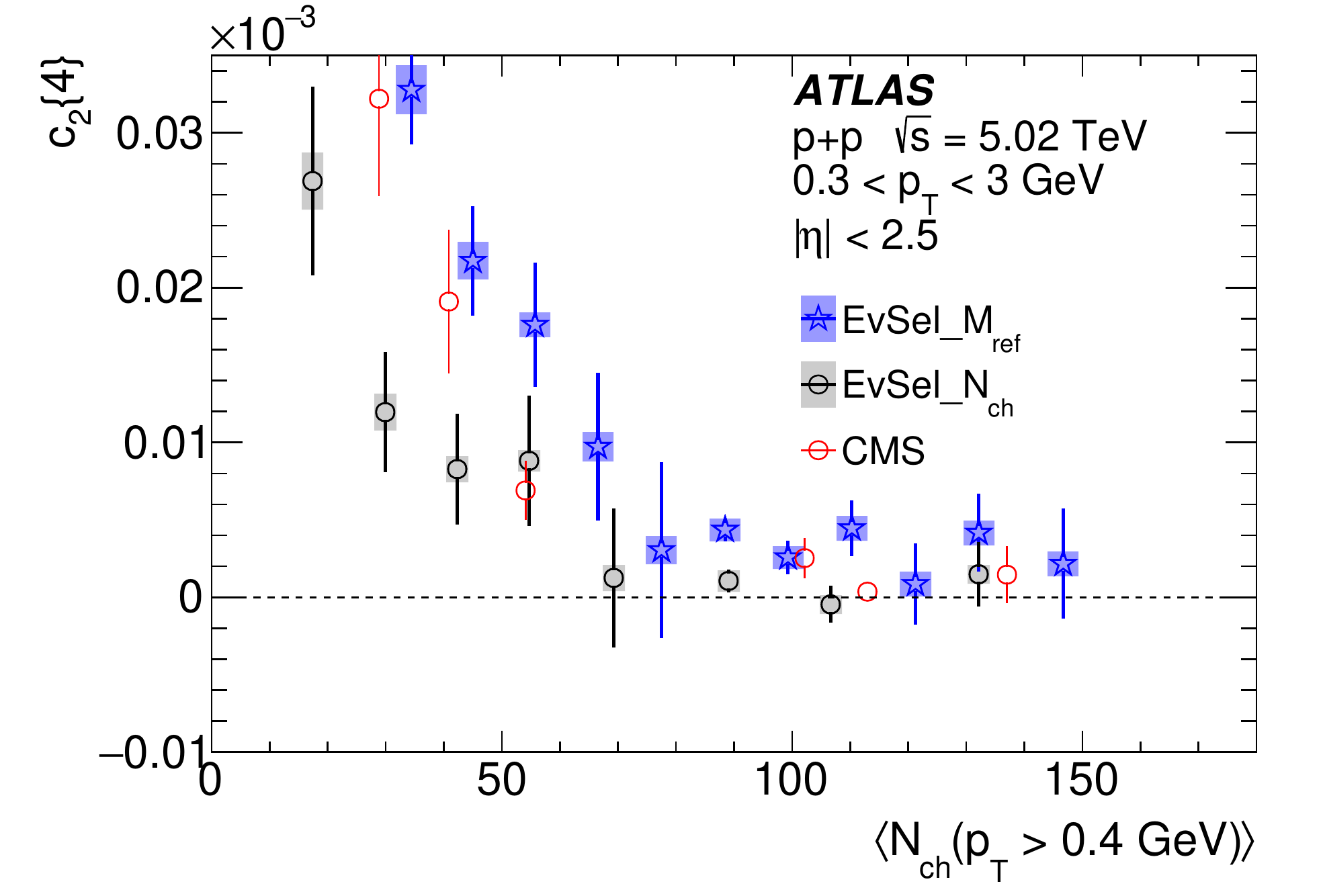}
\includegraphics[width=70mm]{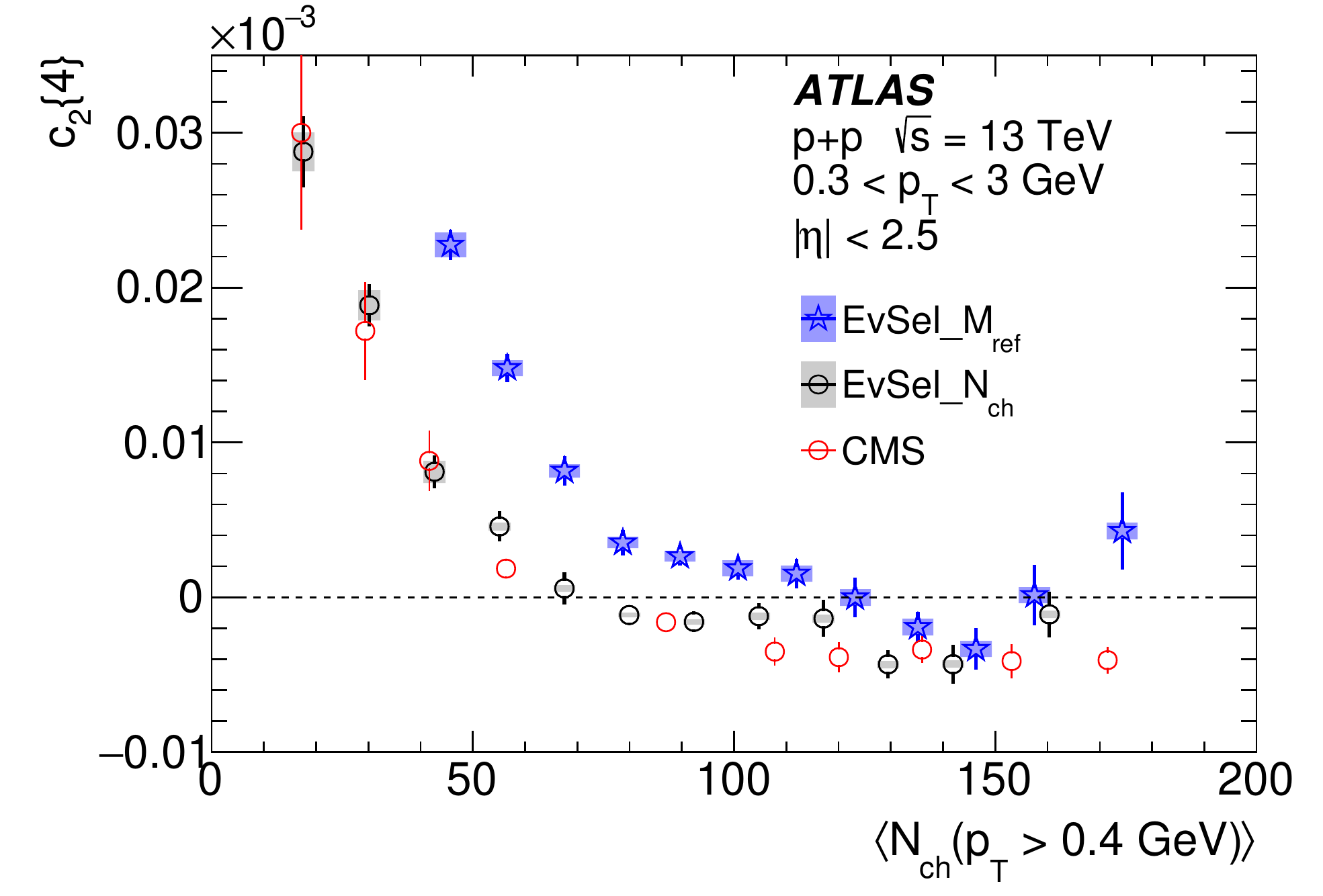}
\caption{Comparison of  the ATLAS and CMS \cite{ppcms3} results for $c_2\{4\}$ cumulants in \pp collisions at 5.02~TeV (left) and 13~TeV (right)  shown as a function of $\langle N_{\mathrm{ch}}(\pT > 0.4$~GeV)$\rangle$. The ATLAS results are shown for two event selections: EvSel\_$M_{\mathrm{ref}}$ and EvSel\_$N_{\mathrm{ch}}$ with the error bars and shaded boxes denoting statistical and systematic uncertainties, respectively. For the CMS results, the error bars indicate statistical and systematic uncertainties added in quadrature.}
\label{fig:comp_ppcms} 
\end{center}
\end{figure}
\begin{figure}[htb!]
\begin{center}
\includegraphics[width=70mm]{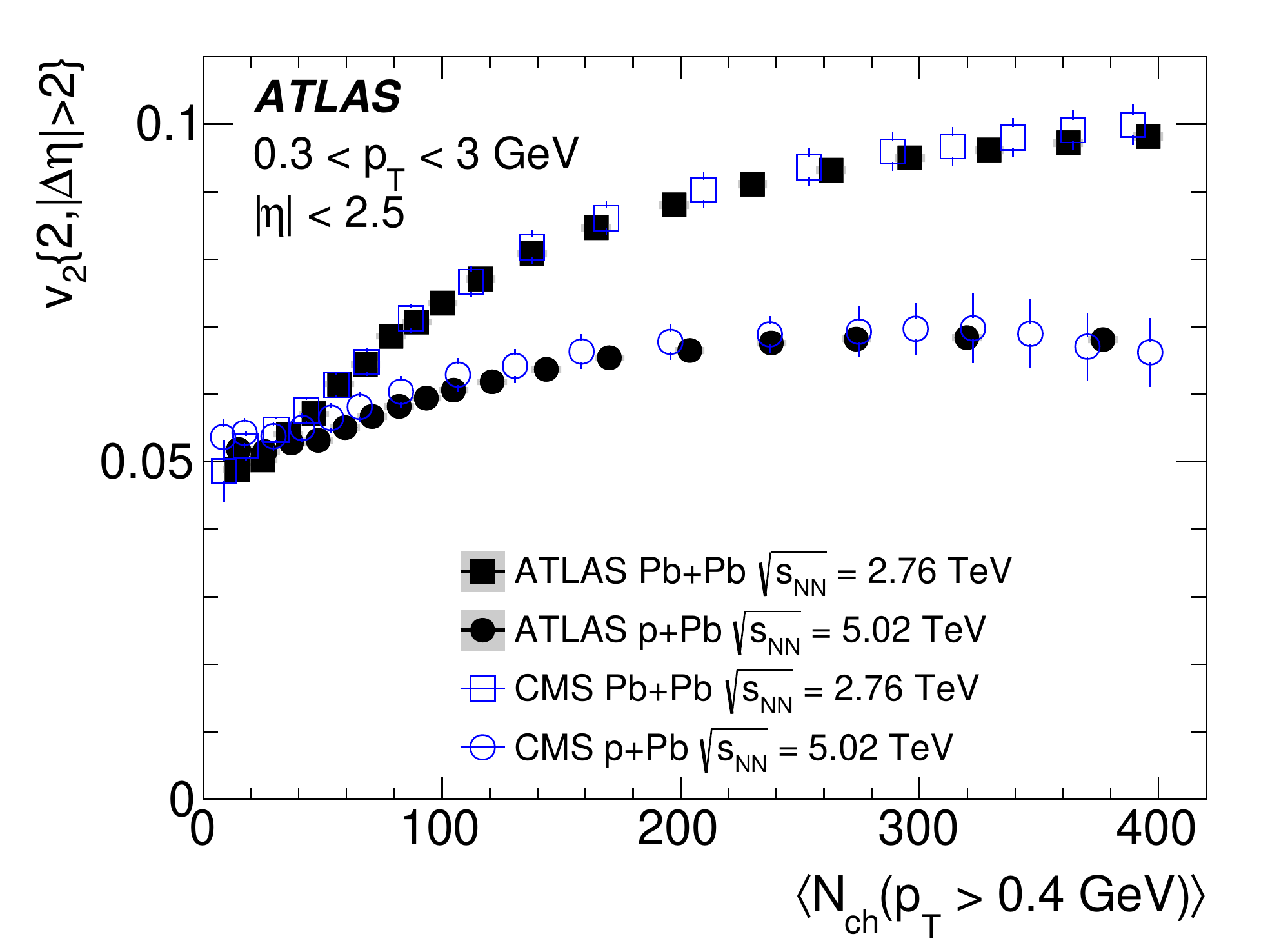}
\includegraphics[width=70mm]{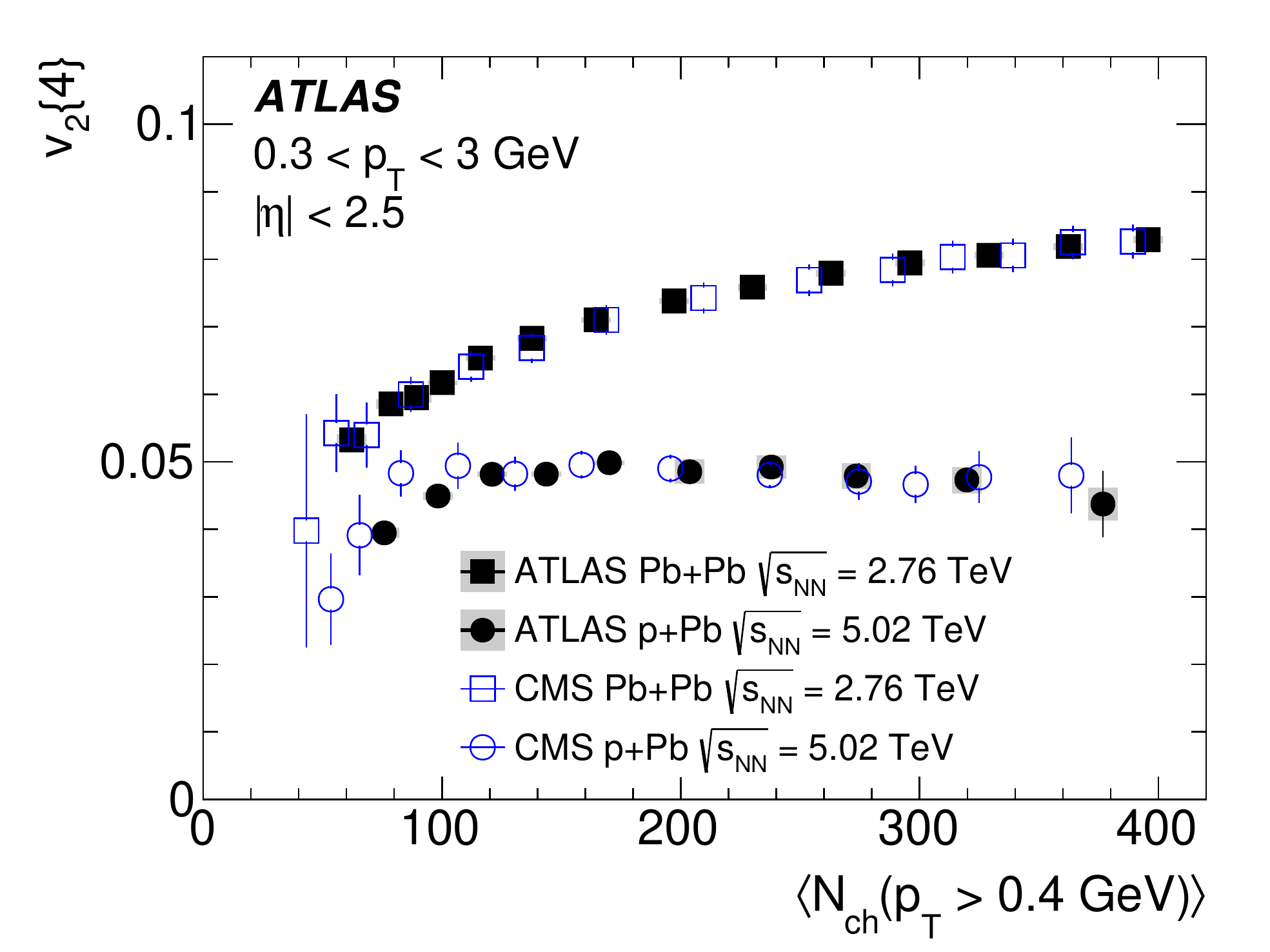}
\includegraphics[width=70mm]{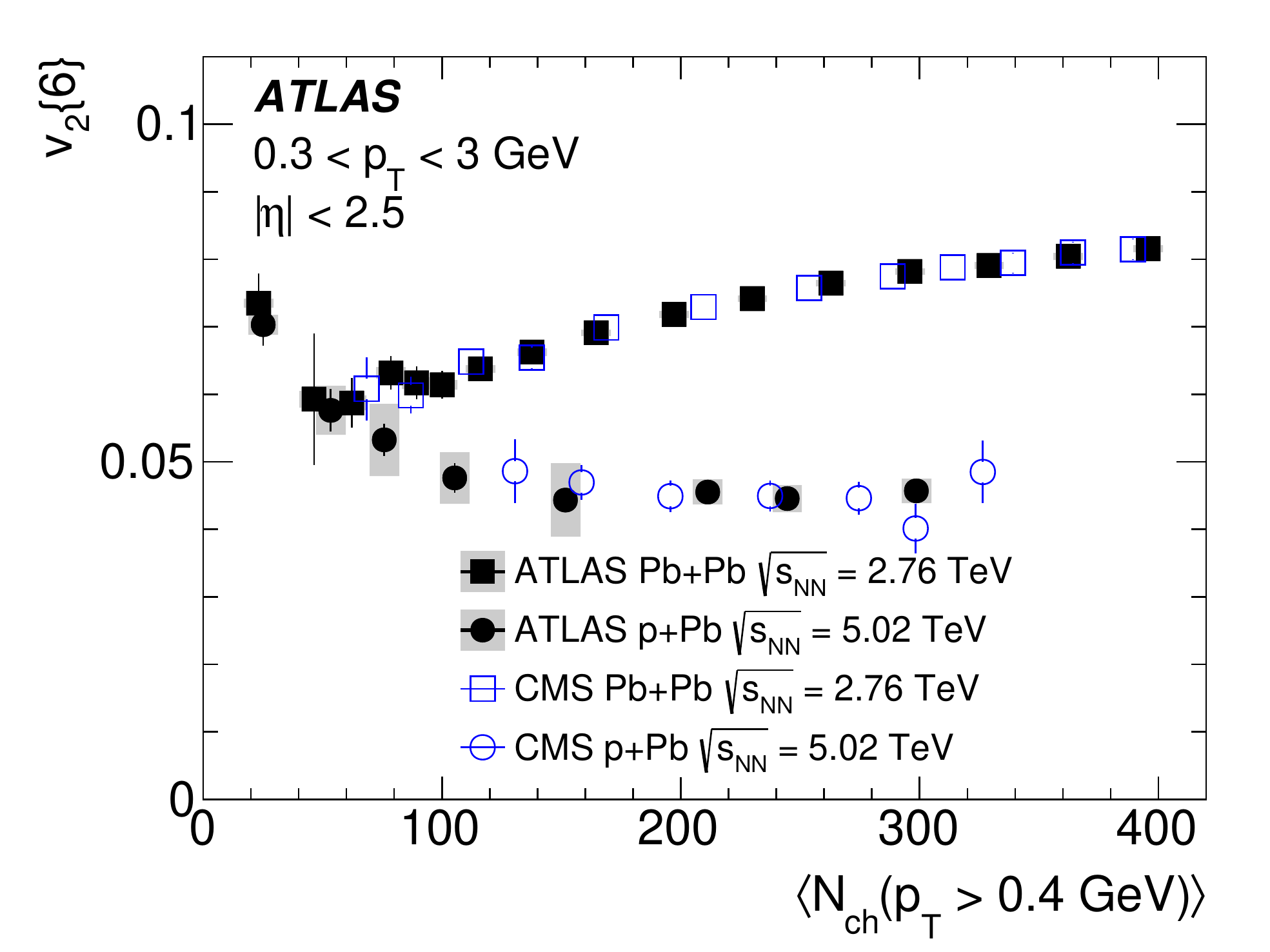}
\includegraphics[width=70mm]{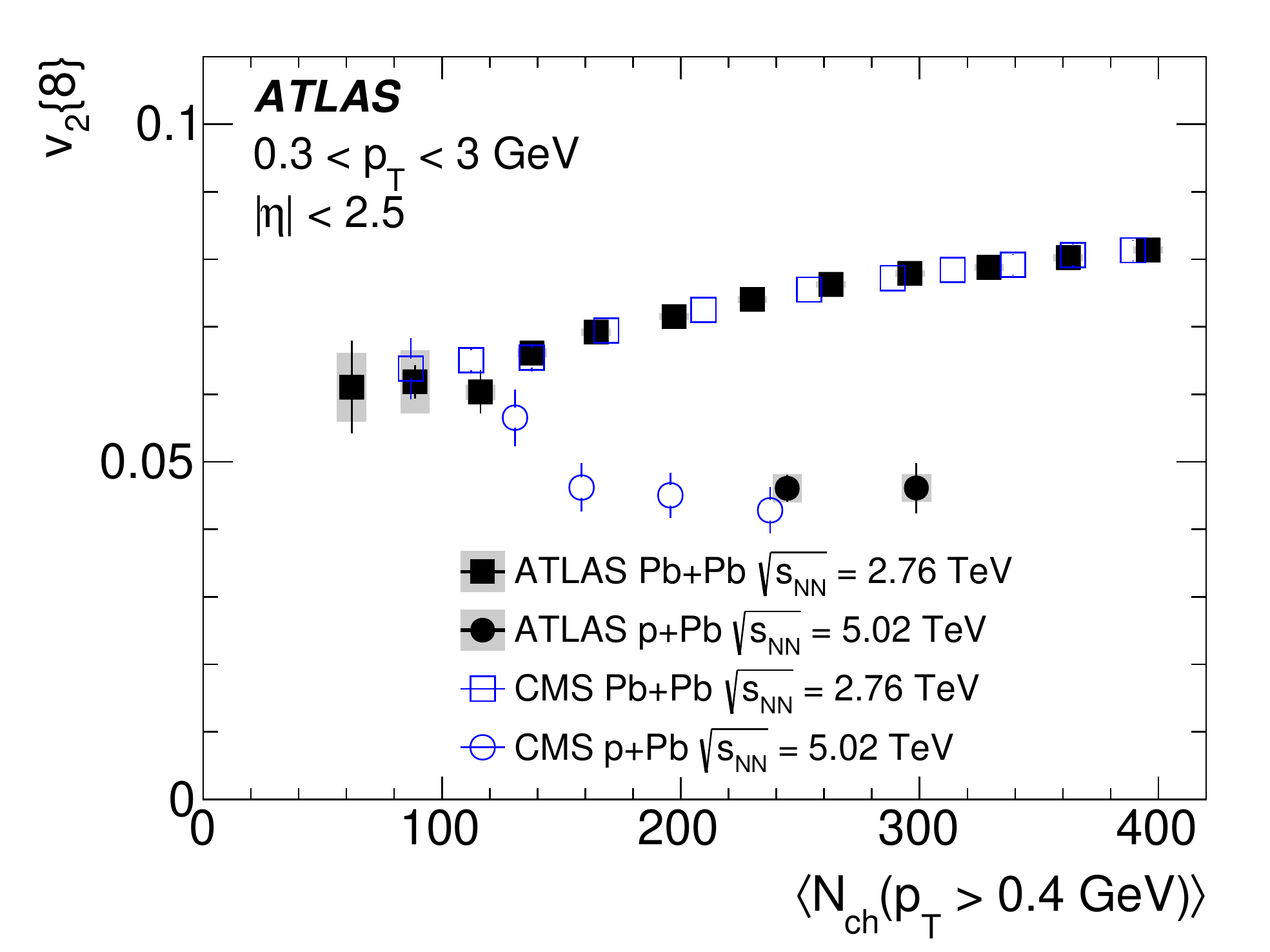}
\caption{Comparison of  the ATLAS  (EvSel\_$M_{\mathrm{ref}}$) and CMS \cite{pPbcms3} results for $\mathrm{v}_2$ harmonics obtained with multi-particle cumulants in \pPb collisions at 5.02 TeV and \PbPb collisions at 2.76 TeV  shown as a function of $\langle N_{\mathrm{ch}}(\pT > 0.4$~GeV)$\rangle$. The ATLAS results are shown with the error bars and shaded boxes denoting statistical and systematic uncertainties, respectively. For the CMS results, the error bars indicate statistical and systematic uncertainties added in quadrature.}
\label{fig:comp_Pbcms} 
\end{center}
\end{figure}
ATLAS results for  $c_2\{4\}$ cumulants measured for \pp data at 5.02~TeV and 13 TeV are compared to similar results obtained by CMS \cite{ppcms3} in Figure~\ref{fig:comp_ppcms}. The ATLAS results are shown for two event selections: EvSel\_$M_{\mathrm{ref}}$ and EvSel\_$N_{\mathrm{ch}}$  (see Section~\ref{sec:analysis}). For the nominal event selection (EvSel\_$M_{\mathrm{ref}}$), the $c_2\{4\}$
cumulants at 5.02 TeV agree with the CMS measurement at high multiplicities, while at low multiplicities
the CMS cumulants are systematically smaller in magnitude than those measured by ATLAS. This discrepancy is more pronounced at 13 TeV, and extends over the full range of collision multiplicities. At
high multiplicities, CMS reported a clear signal of negative $c_2\{4\}$ in  contrast to our default method based on EvSel\_$M_{\mathrm{ref}}$, but is roughly consistent with our measurements based on selecting events according to EvSel\_$N_{\mathrm{ch}}$. 

\begin{figure}[ht!]
\begin{center}
\includegraphics[width=70mm]{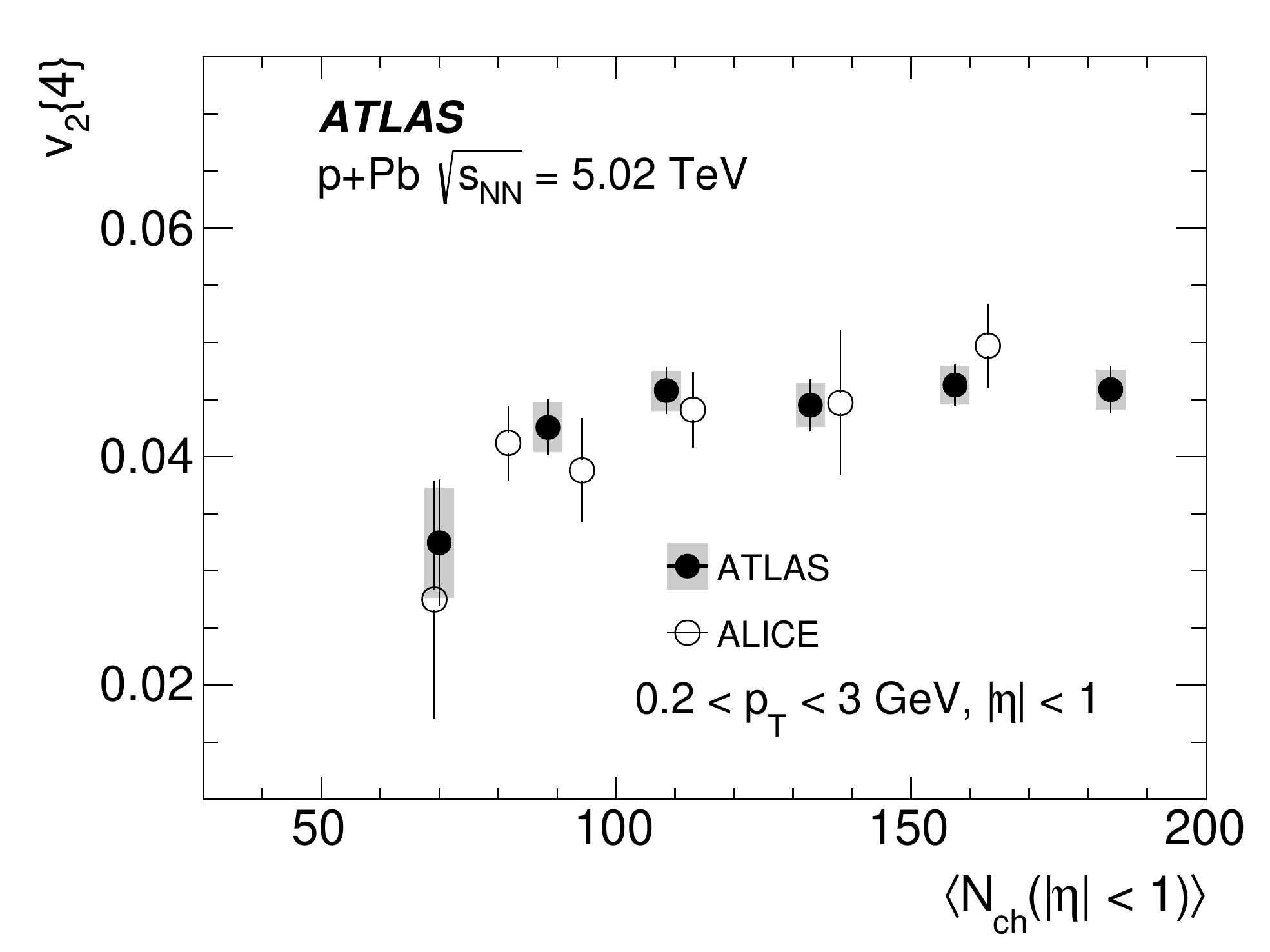}
\includegraphics[width=70mm]{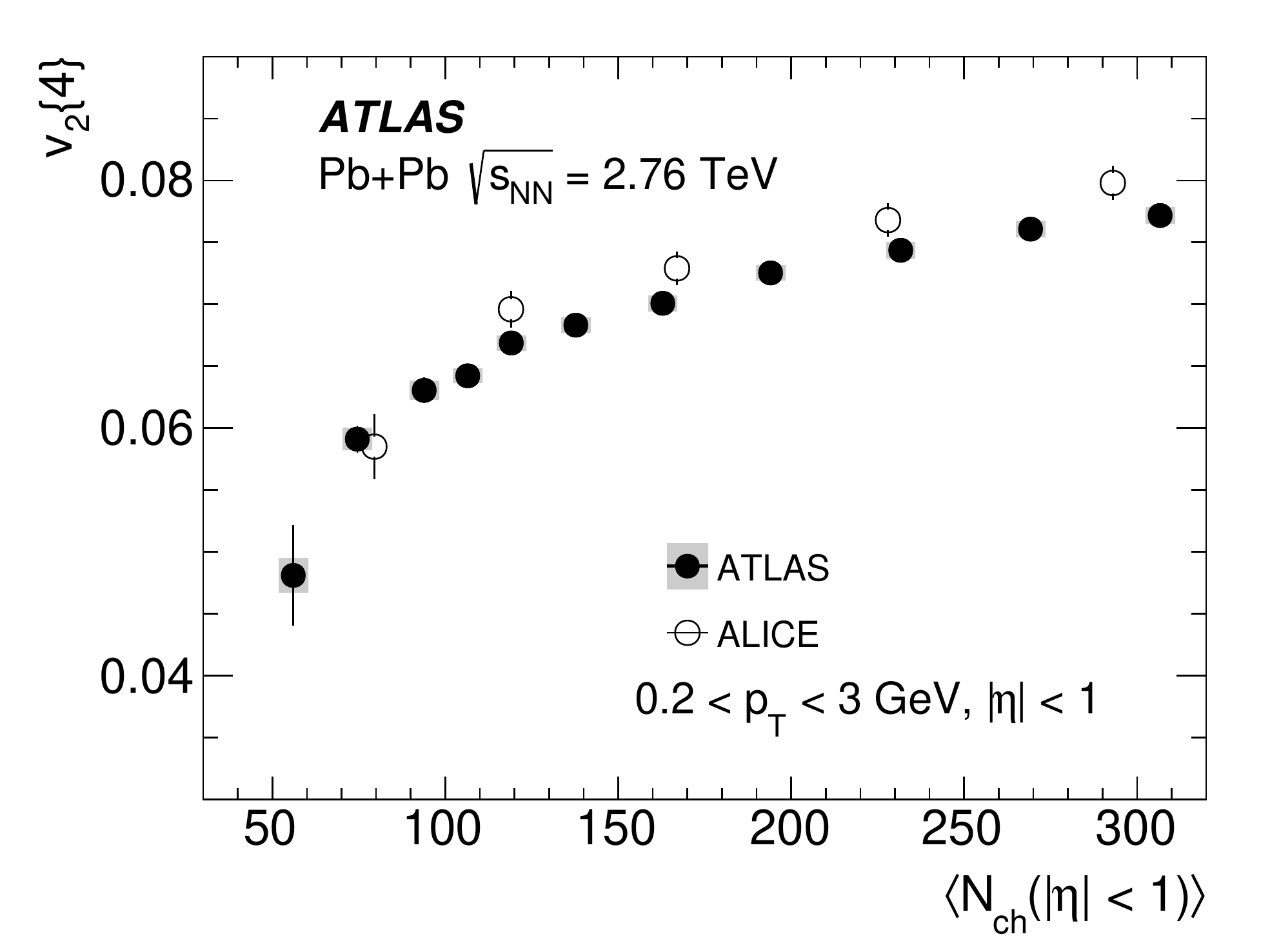}
\caption{Comparison of  the ATLAS  (EvSel\_$M_{\mathrm{ref}}$) and ALICE \cite{Alice9} results for $\mathrm{v}_2\{4\}$ harmonics obtained with four-particle cumulants in \pPb collisions at 5.02 TeV (left) and \PbPb collisions at 2.76 TeV  (right)  shown as a function of $\langle N_{\mathrm{ch}}(|\eta|<1) \rangle$. The ATLAS results are shown with the error bars and shaded boxes denoting statistical and systematic uncertainties, respectively. For the ALICE results, the error bars indicate statistical and systematic uncertainties added in quadrature.}
\label{fig:comp_Pbalice} 
\end{center}
\end{figure}

For \pPb and \PbPb collisions, the results for  $\mathrm{v}_2$ harmonics obtained with multi-particle cumulants agree very well with the CMS data \cite{pPbcms3}, as shown in Figure~\ref{fig:comp_Pbcms}.  Figure~\ref{fig:comp_Pbalice} shows a similar compability of ATLAS and ALICE \cite{Alice9} measurements of $\mathrm{v}_2\{4\}$ in \pPb collisions. For \PbPb collisions, the ALICE results on $\mathrm{v}_2\{4\}$  are slightly above those measured by ATLAS.
\begin{figure}[ht!]
\begin{center}
\includegraphics[width=70mm]{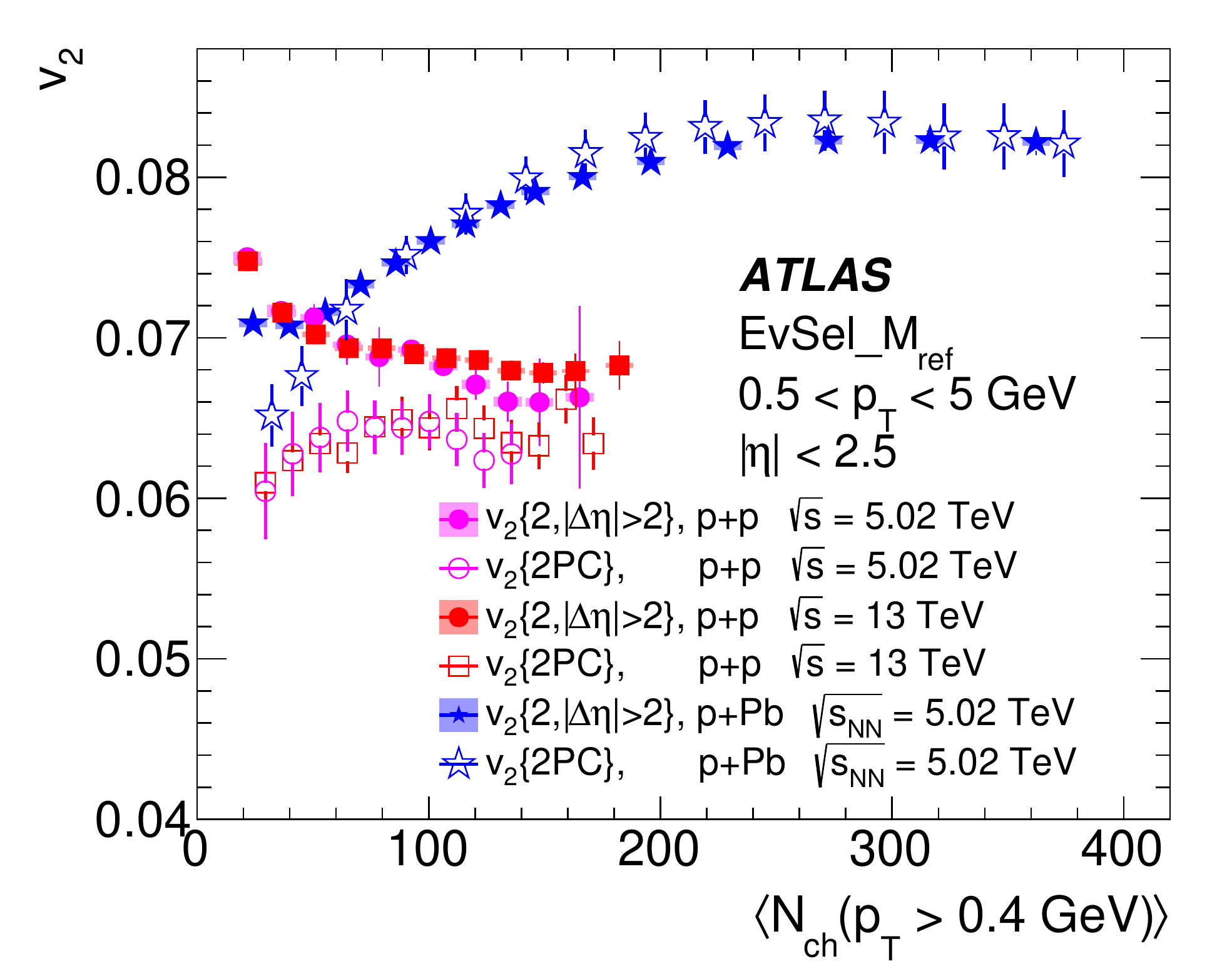}
\includegraphics[width=70mm]{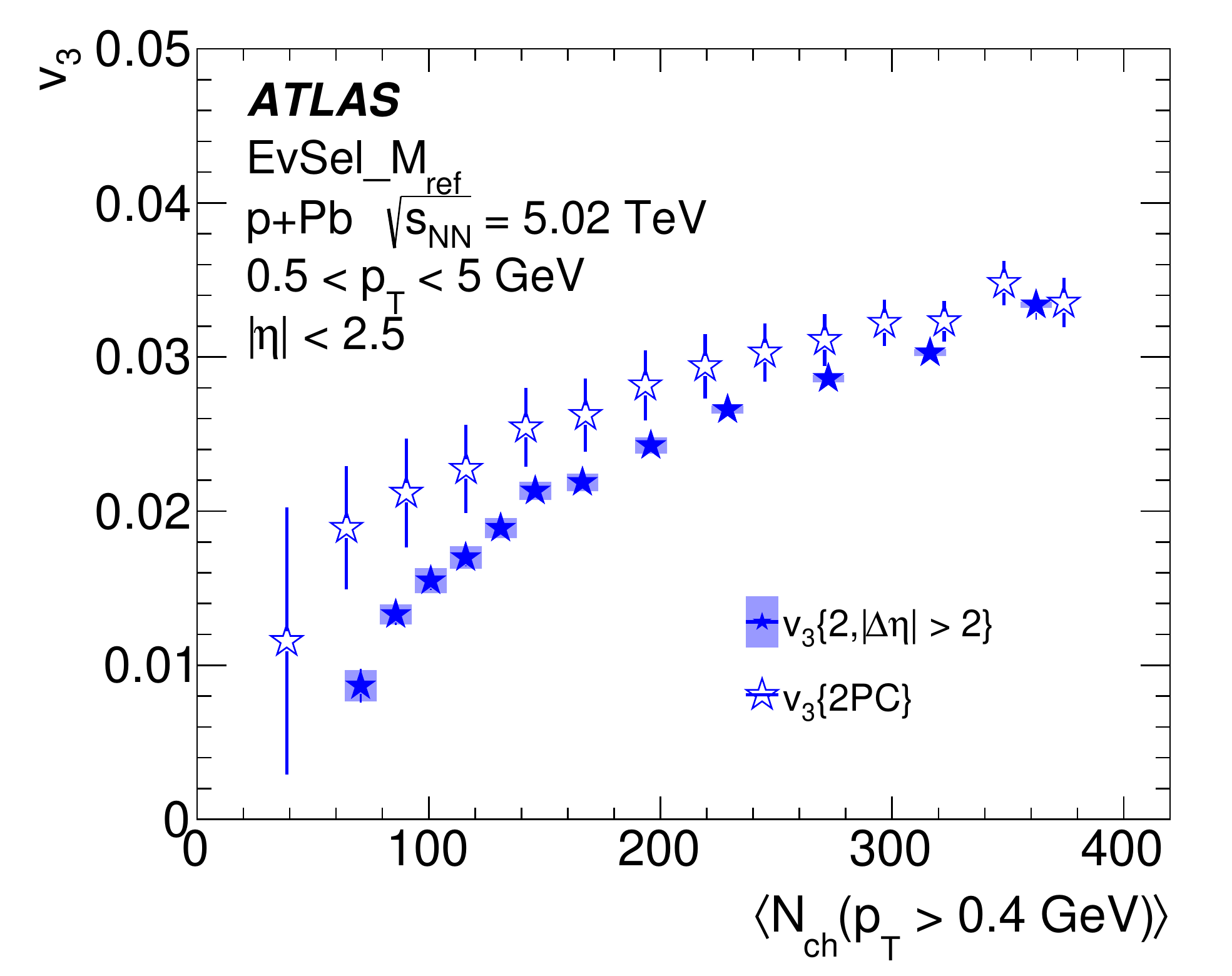}
\includegraphics[width=70mm]{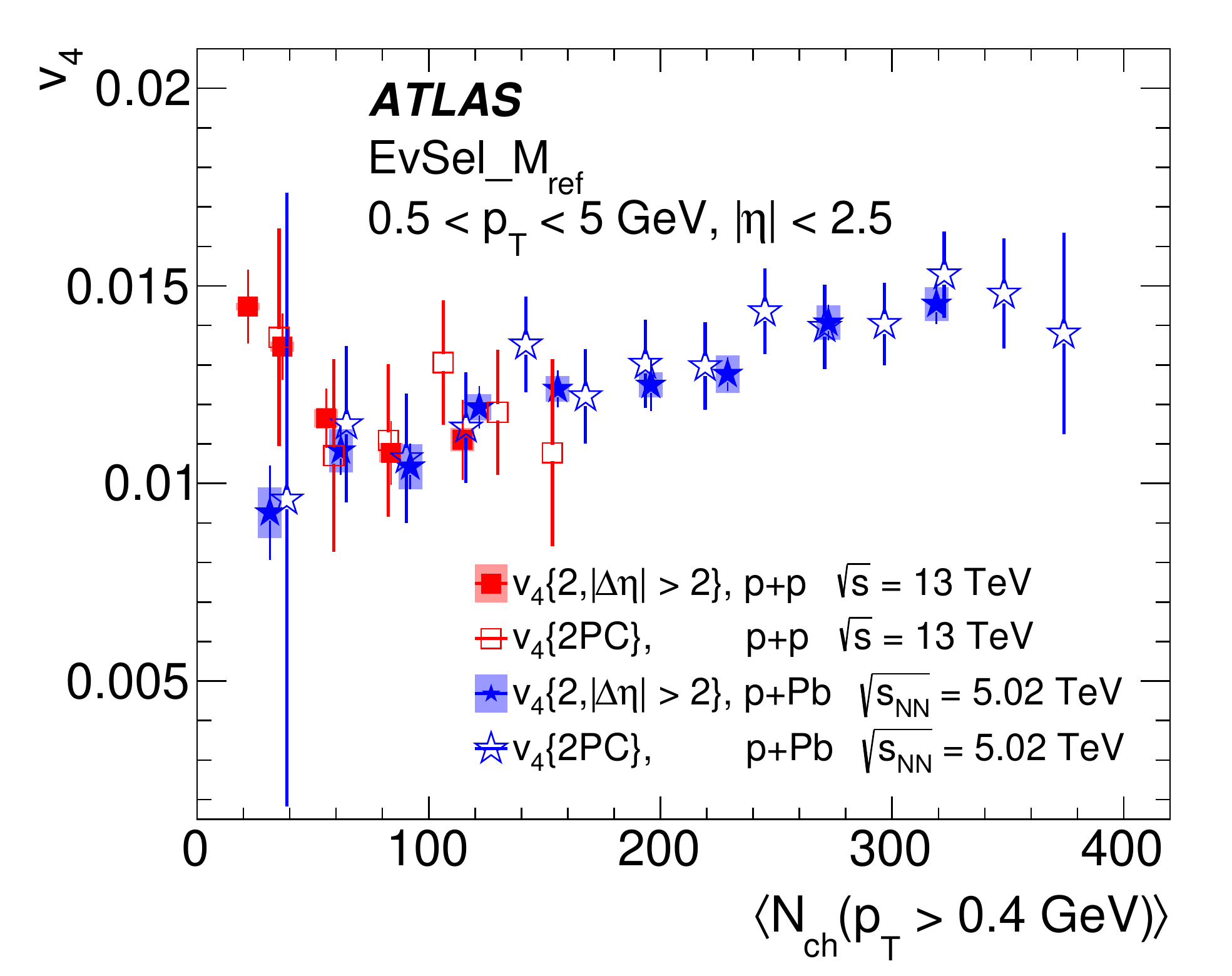}
\caption{Comparison of  the ATLAS  (EvSel\_$M_{\mathrm{ref}}$) measurements of $\mathrm{v}_2$ (top left), $\mathrm{v}_3$ (top right) and $\mathrm{v}_4$ (bottom) harmonics obtained with two-particle cumulants (filled symbols) and two-particle correlation function method (open symbols) for \pp collisions at \sqs= 5.02 and 13 TeV, \pPb collisions at \sqn= 5.02 TeV. The error bars and shaded boxes for the cumulant measurements denote statistical and systematic uncertainties, respectively. For the two-particle correlation function results, the error bars indicate statistical and systematic uncertainties added in quadrature. }
\label{fig:comp_2pc} 
\end{center}
\end{figure}

A comparison of flow harmonics measured with distinct analysis methods, which primarily differ in their sensitivity to non-flow correlations, is also of interest. The method, commonly used to measure flow harmonics in small collision systems, is the two-particle correlation function method (2PC). This method was used by ATLAS  to measure $\mathrm{v}_n$ harmonics in \pp and \pPb collisions \cite{ppatlas2}.  In that measurement,  the non-flow correlations were suppressed by requiring the $|\Delta\eta|>2$ gap, as in this analysis.  However, additional procedures were undertaken in  Ref. \cite{ppatlas2} to also suppress the jet--jet correlations. The ATLAS results for flow harmonics obtained using the two-particle correlation function method, $\mathrm{v}_n\{2\mathrm{PC}\}$, are compared with the results reported here, obtained with two-particle cumulants,  in Figure~\ref{fig:comp_2pc}. For the $\mathrm{v}_2$ harmonic the two measurements give consistent results for \pPb collisions. For \pp collisions the cumulant method gives $\mathrm{v}_2$ values larger than those obtained with the 2PC method, suggesting that the former are contaminated by the non-flow correlations not removed by the $|\Delta\eta|>2$ requirement. The fact that in contrast to $\mathrm{v}_2 \{2\mathrm{PC}\}$ the $\mathrm{v}_2$ harmonic  cannot be determined from the measurement of the $c_2\{4\}$ cumulant reported here for \pp collisions  clearly indicates that this cumulant is biased by non-flow correlations. In the case of the third-order flow harmonic, $\mathrm{v}_3$, the comparison can be made only for \pPb collisions, and here it can be seen that $\mathrm{v}_3\{2,|\Delta\eta|>2\} < \mathrm{v}_3\{2\mathrm{PC}\} $. This difference results from elimination of the jet--jet non-flow correlations by the additional procedure supplementing the $|\Delta\eta|>2$ gap in the 2PC method.  The two methods give consistent results for the $\mathrm{v}_4$ harmonic measured in \pPb collisions at 5.02 TeV as well as in \pp collisions at 13 TeV, indicating that the aforementioned differences between the two analysis methods have a negligible impact on $\mathrm{v}_4$.

\FloatBarrier

\section{Summary and conclusions}
\label{sec:conclusion}
Multi-particle cumulants and corresponding Fourier harmonics are measured by the ATLAS experiment at the LHC for azimuthal angle distributions of charged particles in \pp collisions at \sqs = 5.02 and 13~TeV and in \pPb collisions at \sqn = 5.02~TeV, and compared to the results obtained from low-multiplicity \PbPb collisions at \sqn = 2.76~TeV. The results are presented as a function of charged-particle multiplicity for two ranges of the particles' transverse momenta: $0.3 < \pT < 3$~GeV and $0.5 < \pT < 5$~GeV. For the same charged-particle multiplicity the second-order cumulants and harmonics ($c_2\{2,|\Delta\eta|>2\}$ and $\mathrm{v}_2\{2,|\Delta\eta|>2\}$), derived from two-particle correlations with the $|\Delta\eta|>2$ gap, have larger magnitudes in \PbPb collisions than in \pPb collisions. The smallest signal is observed in \pp collisions. The latter show no energy or multiplicity dependence while the cumulants and the second-order harmonic increase  with increasing multiplicity in  \pPb and \PbPb collisions. 

Four-particle cumulants, $c_2\{4\}$, show that $|c_2\{4\}|$ in \pPb collisions is less than 
$ |c_2\{4\}|$ measured for \PbPb data. For charged-particle multiplicities above 100, the $c_2\{4\}$ cumulants have negative values in  \pPb and \PbPb collisions, confirming the collective nature of multi-particle correlations in these collision systems. The derived magnitude of the $\mathrm{v}_2\{4\}$ harmonic is larger in \PbPb collisions than in \pPb collisions with the same multiplicity. In \pp collisions, over the full range of particle multiplicities, the cumulants are positive or consistent with zero at 5.02~TeV for both \pT ranges. In the 13 TeV \pp data, the cumulants measured for charged particles with lower \pT ($0.3 < \pT < 3$~GeV) also have positive values over the large range of multiplicities. Therefore, these measurements of four-particle cumulants in \pp collisions, based on a method that suppresses the non-flow correlations associated with event-by-event fluctuations in the number of reference particles, generally do not satisfy the requirement of being negative. This indicates that  $c_2\{4\}$ cumulants obtained from the
standard procedure used in this paper may still be biased by residual non-flow correlations. The $c_2\{4\}$ cumulant in 13 TeV \pp collisions measured by CMS is smaller over the full range of collision multiplicities than the $c_2\{4\}$ cumulant obtained by ATLAS with the nominal event selection (EvSel\_$M_{\mathrm{ref}}$) while it is consistent with the ATLAS measurement obtained with the EvSel\_$N_{\mathrm{ch}}$ event selection. 

Six- and eight-particle $c_2$ cumulants can be obtained with sufficient statistical precision only for \pPb and \PbPb collisions.   All derived $\mathrm{v}_2$ harmonics have larger magnitudes for \PbPb collisions than for  \pPb collisions with the same multiplicity. For both systems, $\mathrm{v}_2\{2k\}$ are similar for $k=$ 2, 3 and 4 while $\mathrm{v}_2\{2,|\Delta\eta| > 2\}$ is systematically larger than the second-order Fourier component calculated with cumulants of more than two-particles. Compared to the almost degenerate values of $\mathrm{v}_2\{2k\}, k>1$, a larger $\mathrm{v}_2$ derived from two-particle cumulants is also  predicted by models assuming fluctuation-driven initial-state anisotropies in small collision systems. Interestingly, the ratios $\mathrm{v}_2\{2k\}/\mathrm{v}_2\{2k-2\}$ for \pPb and low-multiplicity \PbPb collisions  are independent of the charged-particle multiplicity  for $N_{\mathrm{ch}}$  > 120, regardless of the \pT range of particles used to calculate the cumulants and Fourier harmonics. 

Higher-order cumulants, $c_3$ and $c_4$, are measured only  using two-particle cumulants with an imposed $|\Delta\eta|>2$ gap. For \pp data $c_3\{2,|\Delta\eta|>2\}$ values are either negative or consistent with zero over almost the full range of $N_{\mathrm{ch}}$ multiplicities, except at the highest multiplicities measured in \pp collisions at 13~TeV. Therefore, the $\mathrm{v}_3$ signal for \pp collisions is undefined or zero within the errors. A positive $c_3$ signal is obtained for \pPb and \PbPb data, except for the charged-particle multiplicities below $\sim$120. The magnitude of $c_3$ and the corresponding $\mathrm{v}_3\{2,|\Delta\eta| > 2\}$ harmonic are comparable for \PbPb and \pPb collisions when particles with  $0.3< \pT < 3.0$~GeV are considered, and systematically slightly larger for \PbPb than for \pPb for particles with $0.5< \pT < 5.0$~GeV.  The fourth-order cumulants, $c_4$, have positive values of $c_4\{2,|\Delta\eta|>2\}$ even for the \pp data, and their  magnitude  is comparable to that for \pPb and \PbPb collisions in the overlapping range of  $N_{\mathrm{ch}}$. For $N_{\mathrm{ch}}$ > 120, where only the measurements for \pPb and \PbPb are accessible, the $c_4$ cumulants measured at the same charged-particle multiplicity are larger for \PbPb than for \pPb.  

The ATLAS results are compared to measurements reported by CMS and ALICE. An agreement across
the experiments is observed for \pPb and \PbPb collisions. The comparison with the ATLAS results obtained for \pp and \pPb collisions with the two-particle correlation method shows some differences, which can be explained by the additional requirements applied in the two-particle correlation method in order to reduce the jet--jet correlations.

The comprehensive data on multi-particle cumulants presented in this paper provide insights into the origin of azimuthal angle anisotropies in small collision systems, and as such can be used to constrain the theoretical modelling of the underlying mechanism. 
\section*{Acknowledgements}

We thank CERN for the very successful operation of the LHC, as well as the
support staff from our institutions without whom ATLAS could not be
operated efficiently.

We acknowledge the support of ANPCyT, Argentina; YerPhI, Armenia; ARC, Australia; BMWFW and FWF, Austria; ANAS, Azerbaijan; SSTC, Belarus; CNPq and FAPESP, Brazil; NSERC, NRC and CFI, Canada; CERN; CONICYT, Chile; CAS, MOST and NSFC, China; COLCIENCIAS, Colombia; MSMT CR, MPO CR and VSC CR, Czech Republic; DNRF and DNSRC, Denmark; IN2P3-CNRS, CEA-DSM/IRFU, France; SRNSF, Georgia; BMBF, HGF, and MPG, Germany; GSRT, Greece; RGC, Hong Kong SAR, China; ISF, I-CORE and Benoziyo Center, Israel; INFN, Italy; MEXT and JSPS, Japan; CNRST, Morocco; NWO, Netherlands; RCN, Norway; MNiSW and NCN, Poland; FCT, Portugal; MNE/IFA, Romania; MES of Russia and NRC KI, Russian Federation; JINR; MESTD, Serbia; MSSR, Slovakia; ARRS and MIZ\v{S}, Slovenia; DST/NRF, South Africa; MINECO, Spain; SRC and Wallenberg Foundation, Sweden; SERI, SNSF and Cantons of Bern and Geneva, Switzerland; MOST, Taiwan; TAEK, Turkey; STFC, United Kingdom; DOE and NSF, United States of America. In addition, individual groups and members have received support from BCKDF, the Canada Council, CANARIE, CRC, Compute Canada, FQRNT, and the Ontario Innovation Trust, Canada; EPLANET, ERC, ERDF, FP7, Horizon 2020 and Marie Sk{\l}odowska-Curie Actions, European Union; Investissements d'Avenir Labex and Idex, ANR, R{\'e}gion Auvergne and Fondation Partager le Savoir, France; DFG and AvH Foundation, Germany; Herakleitos, Thales and Aristeia programmes co-financed by EU-ESF and the Greek NSRF; BSF, GIF and Minerva, Israel; BRF, Norway; CERCA Programme Generalitat de Catalunya, Generalitat Valenciana, Spain; the Royal Society and Leverhulme Trust, United Kingdom.

The crucial computing support from all WLCG partners is acknowledged gratefully, in particular from CERN, the ATLAS Tier-1 facilities at TRIUMF (Canada), NDGF (Denmark, Norway, Sweden), CC-IN2P3 (France), KIT/GridKA (Germany), INFN-CNAF (Italy), NL-T1 (Netherlands), PIC (Spain), ASGC (Taiwan), RAL (UK) and BNL (USA), the Tier-2 facilities worldwide and large non-WLCG resource providers. Major contributors of computing resources are listed in Ref.~\cite{ATL-GEN-PUB-2016-002}.

\clearpage
\appendix
\part*{Appendix}
\addcontentsline{toc}{part}{Appendix}
The following tables list individual contributions to the systematic uncertainty for the following collision systems:   \pp  at 5.02~TeV, \pp at 13~TeV, \pPb at 5.02~TeV and \PbPb at 2.76~TeV. 
 Table~\ref{tab:SystTC} lists contributions to the systematic uncertainty from the track selection requirements, taken as the maximum difference between the base measurement and the results obtained with loose or tight track selections in a given range of $N_{\mathrm{ch}}$, for reference particles with  $0.3 < \pT < 3$~GeV. Table~\ref{tab:SystTC1} includes systematic uncertainties for reference particles with $0.5 < \pT < 5$~GeV.
\begin{table}[h!]
\begin{center}
\caption{Systematic uncertainties related to the track selection requirements for multi-particle cumulants measured in different collision systems for $M_{\mathrm{ref}}$ with  $0.3 < \pT < 3$~GeV.  }
\label{tab:SystTC}
\begin{tabular}{llrrr}  
\toprule
 \multicolumn{5}{ c }{Systematic uncertainties due to track selection requirements} \\
\cline{1-5}
System	 & Systematic uncertainty & $\qquad$ $N_{\mathrm{ch}}$   & $\qquad$ $N_{\mathrm{ch}}$  & $\qquad$ $N_{\mathrm{ch}}$  \\
\midrule
  &  & <50 &  50--100 & >100 \\
  \pp 5~TeV & $\delta c_2\{2,|\Delta\eta|>2\} \times 10^{4}$ & --0.05 &  0.33 & 0.13 \\
  & $\delta c_2\{4\}\times 10^{6}$ & --0.87 & 0.50 & --0.48 \\
  & $\delta c_3\{2,|\Delta\eta|>2\} \times 10^{4}$  & --0.26 & 0.32 & 0.11   \\
  & $\delta c_4\{2,|\Delta\eta|>2\} \times 10^{4}$  & 0.01 & 0.11 & -   \\ \\
  &  & <50 &  50--100 & >100 \\
   \pp 13~TeV & $\delta c_2\{2,|\Delta\eta|>2\} \times 10^{4}$ & --0.07 &  --0.09 & --0.02 \\
  & $\delta c_2\{4\}\times 10^{6}$ & --0.54 & --0.33 & 0.50 \\
  & $\delta c_3\{2,|\Delta\eta|>2\} \times 10^{4}$  & --0.03 & 0.02 & --0.07   \\
  & $\delta c_4\{2,|\Delta\eta|>2\} \times 10^{4}$  & 0.02 & --0.04 & -   \\ \\
   &  & <100 &  100--200 & >200 \\ 
    \pPb  & $\delta c_2\{2,|\Delta\eta|>2\} \times 10^{4}$ & 0.26 &  0.26 & 0.41 \\
  & $\delta c_2\{4\}\times 10^{6}$ & 0.18 & --0.06 & 0.68 \\
  & $\delta c_2\{6\}\times 10^{7}$ & --0.45 & --0.05 & 0.04 \\
  & $\delta c_2\{8\}\times 10^{8}$ & 0.15 & 0.01 & --0.02 \\
  & $\delta c_3\{2,|\Delta\eta|>2\} \times 10^{4}$  & --0.19 & --0.23 & --0.16   \\
  & $\delta c_4\{2,|\Delta\eta|>2\} \times 10^{4}$  & --0.21 & --0.17 & --0.10   \\ \\
  &  & <100 &  100--200 & >200 \\ 
     \PbPb  & $\delta c_2\{2,|\Delta\eta|>2\} \times 10^{4}$ & 0.12 &  0.23 & 0.35 \\
  & $\delta c_2\{4\}\times 10^{6}$ & --0.52 & --0.12 & --0.48 \\
  & $\delta c_2\{6\}\times 10^{7}$ & 0.19 & --0.13 & 0.16 \\
  & $\delta c_2\{8\}\times 10^{8}$ & --0.43 & 0.08 & --0.11 \\
  & $\delta c_3\{2,|\Delta\eta|>2\} \times 10^{4}$  & --0.09 & --0.06 & 0.04   \\
  & $\delta c_4\{2,|\Delta\eta|>2\} \times 10^{4}$  & 0.03 & 0.03 & 0.04   \\ 
\bottomrule
\end{tabular}
\end{center}
\end{table}
\begin{table}[h!]
\begin{center}
\caption{Systematic uncertainties related to the track selection requirements for multi-particle cumulants measured in different collision systems for $M_{\mathrm{ref}}$ with  $0.5 < \pT < 5$~GeV.  }
\label{tab:SystTC1}
\begin{tabular}{llrrr}  
\toprule
 \multicolumn{5}{ c }{Systematic uncertainties due to track selection requirements} \\
\cline{1-5}
System	 & Systematic uncertainty & $\qquad$ $N_{\mathrm{ch}}$   & $\qquad$ $N_{\mathrm{ch}}$  & $\qquad$ $N_{\mathrm{ch}}$  \\
\midrule
  &  & <50 &  50--100 & >100 \\
  \pp 5~TeV & $\delta c_2\{2,|\Delta\eta|>2\} \times 10^{4}$ & --0.08 &  0.10 & 0.19 \\
  & $\delta c_2\{4\}\times 10^{6}$ & --3.73 & --1.09 & 0.88 \\
  & $\delta c_3\{2,|\Delta\eta|>2\} \times 10^{4}$  & 0.01 & --0.20 & 0.21   \\
  & $\delta c_4\{2,|\Delta\eta|>2\} \times 10^{4}$  & --0.12 & --0.15 & -   \\ \\
  &  & <50 &  50--100 & >100 \\
   \pp 13~TeV & $\delta c_2\{2,|\Delta\eta|>2\} \times 10^{4}$ & --0.19 &  --0.12 & 0.01 \\
  & $\delta c_2\{4\}\times 10^{6}$ & --4.02 & --1.11 & --0.34 \\
  & $\delta c_3\{2,|\Delta\eta|>2\} \times 10^{4}$  & --0.03 & --0.06 & 0.08   \\
  & $\delta c_4\{2,|\Delta\eta|>2\} \times 10^{4}$  & --0.02 & --0.05 & --0.05   \\ \\
   &  & <100 &  100--200 & >200 \\ 
    \pPb  & $\delta c_2\{2,|\Delta\eta|>2\} \times 10^{4}$ & 0.26 &  0.29 & 0.35 \\
  & $\delta c_2\{4\}\times 10^{6}$ & 0.16 & 0.09 & --0.59 \\
  & $\delta c_2\{6\}\times 10^{7}$ & 0.13 & --0.19 & 0.31 \\
  & $\delta c_2\{8\}\times 10^{8}$ & 0.89 & --0.01 & --0.19 \\
  & $\delta c_3\{2,|\Delta\eta|>2\} \times 10^{4}$  & --0.06 & --0.15 & --0.08   \\
  & $\delta c_4\{2,|\Delta\eta|>2\} \times 10^{4}$  & --0.08 & --0.06 & 0.04   \\ \\
  &  & <100 &  100--200 & >200 \\ 
     \PbPb  & $\delta c_2\{2,|\Delta\eta|>2\} \times 10^{4}$ & 0.53 &  0.58 & 0.46 \\
  & $\delta c_2\{4\}\times 10^{6}$ & --1.72 & --0.79 & --0.62 \\
  & $\delta c_2\{6\}\times 10^{7}$ & 0.85 & --0.43 & 0.33 \\
  & $\delta c_2\{8\}\times 10^{8}$ & --0.37 & 0.54 & --0.45 \\
  & $\delta c_3\{2,|\Delta\eta|>2\} \times 10^{4}$  & 0.06 & 0.09 & 0.06   \\
  & $\delta c_4\{2,|\Delta\eta|>2\} \times 10^{4}$  & --0.13 & --0.02 & 0.04   \\ 
\bottomrule
\end{tabular}
\end{center}
\end{table}

Table~\ref{tab:SystTE} lists contributions to the systematic uncertainty due to the uncertainty in the tracking efficiency, taken as the largest difference between the base measurement and the results obtained with the efficiency varied up or down for reference particles with  $0.3 < \pT < 3$~GeV. The maximum deviations in a given range of $N_{\mathrm{ch}}$ are listed. Table~\ref{tab:SystTE1} includes systematic uncertainties for reference particles with $0.5 < \pT < 5$~GeV.
\begin{table}[h!]
\begin{center}
\caption{Systematic uncertainties related to the tracking efficiency uncertainty for multi-particle cumulants measured in different collision systems for $M_{\mathrm{ref}}$ with  $0.3 < \pT < 3$~GeV.  The maximum deviations in a given range of $N_{\mathrm{ch}}$ are listed.}
\label{tab:SystTE}
\begin{tabular}{llrrr}  
\toprule
 \multicolumn{5}{ c }{Systematic uncertainties due to the tracking efficiency uncertainty} \\
\cline{1-5}
System	 & Systematic uncertainty & $\qquad$ $N_{\mathrm{ch}}$   & $\qquad$ $N_{\mathrm{ch}}$  & $\qquad$ $N_{\mathrm{ch}}$  \\
\midrule
  &  & <50 &  50--100 & >100 \\
  \pp 5~TeV & $\delta c_2\{2,|\Delta\eta|>2\} \times 10^{4}$ & 0.27 &  0.33 & 0.24 \\
  & $\delta c_2\{4\}\times 10^{6}$ & 4.10 & 0.64 & 0.17 \\
  & $\delta c_3\{2,|\Delta\eta|>2\} \times 10^{4}$  & --0.02 & --0.03 & 0.01   \\
  & $\delta c_4\{2,|\Delta\eta|>2\} \times 10^{4}$  & 0.01 & 0.01 & -   \\ \\
  &  & <50 &  50--100 & >100 \\
   \pp 13~TeV & $\delta c_2\{2,|\Delta\eta|>2\} \times 10^{4}$ & 0.31 &  0.20 & 0.19 \\
  & $\delta c_2\{4\}\times 10^{6}$ & 3.72 & 0.39 & 0.13 \\
  & $\delta c_3\{2,|\Delta\eta|>2\} \times 10^{4}$  & --0.04 & --0.01 & 0.01   \\
  & $\delta c_4\{2,|\Delta\eta|>2\} \times 10^{4}$  & 0.01 & 0.02 & -   \\ \\
   &  & <100 &  100--200 & >200 \\ 
    \pPb  & $\delta c_2\{2,|\Delta\eta|>2\} \times 10^{4}$ & --0.53 &  --0.51 & --0.56 \\
  & $\delta c_2\{4\}\times 10^{6}$ & --0.86 & 0.13 & 0.14 \\
  & $\delta c_2\{6\}\times 10^{7}$ & --0.29 & --0.01 & --0.02 \\
  & $\delta c_2\{8\}\times 10^{8}$ & --3.20 & 0.01 & 0.01 \\
  & $\delta c_3\{2,|\Delta\eta|>2\} \times 10^{4}$  & 0.09 & --0.05 & --0.10   \\
  & $\delta c_4\{2,|\Delta\eta|>2\} \times 10^{4}$  & --0.01 & --0.02 & --0.02   \\  \\
  &  & <100 &  100--200 & >200 \\ 
     \PbPb  & $\delta c_2\{2,|\Delta\eta|>2\} \times 10^{4}$ & --0.65 &  --0.97 & --1.22 \\
  & $\delta c_2\{4\}\times 10^{6}$ & --0.64 & 0.66 & 1.09 \\
  & $\delta c_2\{6\}\times 10^{7}$ & --0.30 & --0.19 & --0.41 \\
  & $\delta c_2\{8\}\times 10^{8}$ & --1.15 & 0.10 & 0.29 \\
  & $\delta c_3\{2,|\Delta\eta|>2\} \times 10^{4}$  & 0.04 & --0.07 & --0.12   \\
  & $\delta c_4\{2,|\Delta\eta|>2\} \times 10^{4}$  & --0.02 & --0.03 & --0.04   \\ 
\bottomrule
\end{tabular}
\end{center}
\end{table}
\begin{table}[h!]
\begin{center}
\caption{Systematic uncertainties related to the tracking efficiency uncertainty for multi-particle cumulants measured in different collision systems for $M_{\mathrm{ref}}$ with  $0.5 < \pT < 5$~GeV.  The maximum deviations in a given range of $N_{\mathrm{ch}}$ are listed.}
\label{tab:SystTE1}
\begin{tabular}{llrrr}  
\toprule
 \multicolumn{5}{ c }{Systematic uncertainties due to the tracking efficiency uncertainty} \\
\cline{1-5}
System	 & Systematic uncertainty & $\qquad$ $N_{\mathrm{ch}}$   & $\qquad$ $N_{\mathrm{ch}}$  & $\qquad$ $N_{\mathrm{ch}}$  \\
\midrule
  &  & <50 &  50--100 & >100 \\
  \pp 5~TeV & $\delta c_2\{2,|\Delta\eta|>2\} \times 10^{4}$ & 0.35 &  0.27 & 0.27 \\
  & $\delta c_2\{4\}\times 10^{6}$ & 5.11 & 1.38 & 0.70 \\
  & $\delta c_3\{2,|\Delta\eta|>2\} \times 10^{4}$  & --0.05 & --0.04 & --0.02   \\
  & $\delta c_4\{2,|\Delta\eta|>2\} \times 10^{4}$  & 0.02 & 0.02 & -   \\ \\
  &  & <50 &  50--100 & >100 \\
   \pp 13~TeV & $\delta c_2\{2,|\Delta\eta|>2\} \times 10^{4}$ & 0.36 &  0.24 & 0.21 \\
  & $\delta c_2\{4\}\times 10^{6}$ & 4.97 & 1.37 & 0.43 \\
  & $\delta c_3\{2,|\Delta\eta|>2\} \times 10^{4}$  & --0.06 & --0.03 & --0.01   \\
  & $\delta c_4\{2,|\Delta\eta|>2\} \times 10^{4}$  & 0.02 & 0.02 & 0.01   \\ \\
   &  & <100 &  100--200 & >200 \\ 
    \pPb  & $\delta c_2\{2,|\Delta\eta|>2\} \times 10^{4}$ & --0.15 &  --0.13 & --0.14 \\
  & $\delta c_2\{4\}\times 10^{6}$ & --0.60 & 0.04 & 0.06 \\
  & $\delta c_2\{6\}\times 10^{7}$ & --0.30 & --0.01 & --0.01 \\
  & $\delta c_2\{8\}\times 10^{8}$ & --3.81 & --0.01 & 0.01 \\
  & $\delta c_3\{2,|\Delta\eta|>2\} \times 10^{4}$  & 0.02 & --0.01 & --0.03   \\
  & $\delta c_4\{2,|\Delta\eta|>2\} \times 10^{4}$  & --0.01 & --0.01 & --0.01   \\ \\
  &  & <100 &  100--200 & >200 \\ 
     \PbPb  & $\delta c_2\{2,|\Delta\eta|>2\} \times 10^{4}$ & --0.18 &  --0.24 & --0.32 \\
  & $\delta c_2\{4\}\times 10^{6}$ & --0.66 & 0.19 & 0.38 \\
  & $\delta c_2\{6\}\times 10^{7}$ & --0.39 & --0.09 & --0.22 \\
  & $\delta c_2\{8\}\times 10^{8}$ & --0.77 & 0.09 & 0.25 \\
  & $\delta c_3\{2,|\Delta\eta|>2\} \times 10^{4}$  & 0.02 & --0.02 & --0.04   \\
  & $\delta c_4\{2,|\Delta\eta|>2\} \times 10^{4}$  & --0.01 & --0.01 & --0.02   \\ 
\bottomrule
\end{tabular}
\end{center}
\end{table}
\clearpage

Table~\ref{tab:SystTP} lists contributions to the systematic uncertainty from the pile-up effects, taken as the maximum difference between the base measurement and the results obtained with the higher or lower pile-up in \pp collisions and with the higher pile-up in the case of \pPb collisions  in a given range of $N_{\mathrm{ch}}$, for reference particles with  $0.3 < \pT < 3$~GeV. Table~\ref{tab:SystTP1} includes systematic uncertainties for reference particles with $0.5 < \pT < 5$~GeV.
\begin{table}[h!]
\begin{center}
\caption{Systematic uncertainties related to the pile-up for multi-particle cumulants measured in different collision systems for $M_{\mathrm{ref}}$ with  $0.3 < \pT < 3$~GeV.  }
\label{tab:SystTP}
\begin{tabular}{llrrr}  
\toprule
 \multicolumn{5}{ c }{Systematic uncertainties due to the pile-up} \\
\cline{1-5}
System	 & Systematic uncertainty & $\qquad$ $N_{\mathrm{ch}}$   & $\qquad$ $N_{\mathrm{ch}}$  & $\qquad$ $N_{\mathrm{ch}}$  \\
\midrule
  &  & <50 &  50--100 & >100 \\
  \pp 5~TeV & $\delta c_2\{2,|\Delta\eta|>2\} \times 10^{4}$ & 0.29 &  0.03 & --0.12 \\
  & $\delta c_2\{4\}\times 10^{6}$ & --0.66 & 0.50 & --0.62 \\
  & $\delta c_3\{2,|\Delta\eta|>2\} \times 10^{4}$  & --0.03 & --0.07 & --0.10   \\
  & $\delta c_4\{2,|\Delta\eta|>2\} \times 10^{4}$  & 0.12 & --0.05 & -   \\ \\
  &  & <50 &  50--100 & >100 \\
   \pp 13~TeV & $\delta c_2\{2,|\Delta\eta|>2\} \times 10^{4}$ & --0.02 &  --0.02 & 0.06 \\
  & $\delta c_2\{4\}\times 10^{6}$ & --0.05 & --0.05 & --0.14 \\
  & $\delta c_3\{2,|\Delta\eta|>2\} \times 10^{4}$  & --0.01 & 0.02 & 0.02   \\
  & $\delta c_4\{2,|\Delta\eta|>2\} \times 10^{4}$  & --0.01 & 0.02 & -   \\ \\
   &  & <100 &  100--200 & >200 \\ 
    \pPb  & $\delta c_2\{2,|\Delta\eta|>2\} \times 10^{4}$ & --0.01 &  0.01 & 0.03 \\
  & $\delta c_2\{4\}\times 10^{6}$ & --0.01 & --0.02 & --0.01 \\
  & $\delta c_2\{6\}\times 10^{7}$ & 0.01 & --0.01 & 0.01 \\
  & $\delta c_2\{8\}\times 10^{8}$ &--0.01 & 0.01 & 0.01 \\
  & $\delta c_3\{2,|\Delta\eta|>2\} \times 10^{4}$  & --0.01 & --0.01 & --0.01   \\
  & $\delta c_4\{2,|\Delta\eta|>2\} \times 10^{4}$  & --0.01 & --0.01 & 0.01   \\ 
\bottomrule
\end{tabular}
\end{center}
\end{table}
\begin{table}[h!]
\begin{center}
\caption{Systematic uncertainties related to the pile-up for multi-particle cumulants measured in different collision systems for $M_{\mathrm{ref}}$ with  $0.5 < \pT < 5$~GeV.  }
\label{tab:SystTP1}
\begin{tabular}{llrrr}  
\toprule
 \multicolumn{5}{ c }{Systematic uncertainties due to the pile-up} \\
\cline{1-5}
System	 & Systematic uncertainty & $\qquad$ $N_{\mathrm{ch}}$   & $\qquad$ $N_{\mathrm{ch}}$  & $\qquad$ $N_{\mathrm{ch}}$  \\
\midrule
  &  & <50 &  50--100 & >100 \\
  \pp 5~TeV & $\delta c_2\{2,|\Delta\eta|>2\} \times 10^{4}$ & 0.43 &  0.12 & 0.25 \\
  & $\delta c_2\{4\}\times 10^{6}$ & --3.43 & --0.58 & --2.17 \\
  & $\delta c_3\{2,|\Delta\eta|>2\} \times 10^{4}$  & 0.35 & 0.28 & --0.09   \\
  & $\delta c_4\{2,|\Delta\eta|>2\} \times 10^{4}$  & 0.27 & --0.42 & -   \\ \\
  &  & <50 &  50--100 & >100 \\
   \pp 13~TeV & $\delta c_2\{2,|\Delta\eta|>2\} \times 10^{4}$ & --0.03 &  0.01 & --0.13 \\
  & $\delta c_2\{4\}\times 10^{6}$ & --0.35 & --0.16 & --0.23 \\
  & $\delta c_3\{2,|\Delta\eta|>2\} \times 10^{4}$  & --0.01 & --0.02 & 0.03   \\
  & $\delta c_4\{2,|\Delta\eta|>2\} \times 10^{4}$  & 0.02 & 0.01 & --0.04   \\ \\
   &  & <100 &  100--200 & >200 \\ 
    \pPb  & $\delta c_2\{2,|\Delta\eta|>2\} \times 10^{4}$ & --0.03 &  0.02 & 0.01 \\
  & $\delta c_2\{4\}\times 10^{6}$ & --0.08 & --0.04 & --0.03 \\
  & $\delta c_2\{6\}\times 10^{7}$ & 0.03 & --0.02 & 0.01 \\
  & $\delta c_2\{8\}\times 10^{8}$ & --0.11 & 0.05 & 0.01 \\
  & $\delta c_3\{2,|\Delta\eta|>2\} \times 10^{4}$  & --0.01 & --0.01 & --0.01   \\
  & $\delta c_4\{2,|\Delta\eta|>2\} \times 10^{4}$  & 0.01 & --0.01 & --0.01   \\ 
\bottomrule
\end{tabular}
\end{center}
\end{table}
\clearpage

Tables~\ref{tab:SystTA} and \ref{tab:SystTA1} list contributions to the systematic uncertainty related to the two running periods for \pPb collisions at 5.02~TeV, taken as the maximum difference between the base measurement and the results obtained for the two running periods in a given range of $N_{\mathrm{ch}}$, for reference particles with  $0.3 < \pT < 3$~GeV and $0.5 < \pT < 5$~GeV, respectively. 
\begin{table}[h!]
\begin{center}
\caption{Systematic uncertainties related to the two running periods  (\pPb vs. Pb+$p$) for \pPb collisions for $M_{\mathrm{ref}}$ with  $0.3 < \pT < 3$~GeV.  }
\label{tab:SystTA}
\begin{tabular}{llrrr}  
\toprule
 \multicolumn{5}{ c }{Systematic uncertainties in \pPb results due to the two running periods} \\
\cline{1-5}
System $\qquad$	 & Systematic uncertainty & $\qquad$ $N_{\mathrm{ch}}$   & $\qquad$ $N_{\mathrm{ch}}$  & $\qquad$ $N_{\mathrm{ch}}$  \\
\midrule
   &  & <100 &  100--200 & >200 \\ 
    \pPb  & $\delta c_2\{2,|\Delta\eta|>2\} \times 10^{4}$ & --0.02 &  --0.15 & --0.08 \\
  & $\delta c_2\{4\}\times 10^{6}$ & 0.02 & 0.10 & 0.45 \\
  & $\delta c_2\{6\}\times 10^{7}$ & --0.31 & --0.22 & --0.08 \\
  & $\delta c_2\{8\}\times 10^{8}$ & 0.05 & 0.11 & 0.02 \\
  & $\delta c_3\{2,|\Delta\eta|>2\} \times 10^{4}$  & --0.12 & 0.05 & --0.03   \\
  & $\delta c_4\{2,|\Delta\eta|>2\} \times 10^{4}$  & 0.03 & --0.13 & 0.02   \\ 
\bottomrule
\end{tabular}
\end{center}
\end{table}
\begin{table}[h!]
\begin{center}
\caption{Systematic uncertainties related to the two running periods (\pPb vs. Pb+$p$)  for \pPb collisions  for $M_{\mathrm{ref}}$ with  $0.5 < \pT < 5$~GeV.  }
\label{tab:SystTA1}
\begin{tabular}{llrrr}  
\toprule
 \multicolumn{5}{ c }{Systematic uncertainties in \pPb results due to two running periods } \\
\cline{1-5}
 System	 $\qquad$ & Systematic uncertainty & $\qquad$ $N_{\mathrm{ch}}$   & $\qquad$ $N_{\mathrm{ch}}$  & $\qquad$ $N_{\mathrm{ch}}$   \\
\midrule
   &  & <100 &  100--200 & >200 \\ 
    \pPb  & $\delta c_2\{2,|\Delta\eta|>2\} \times 10^{4}$ & --0.06 &  --0.03 & --0.03 \\
  & $\delta c_2\{4\}\times 10^{6}$ & 0.21 & 0.91 & 1.17 \\
  & $\delta c_2\{6\}\times 10^{7}$ & --1.39 & --0.62 & --0.24 \\
  & $\delta c_2\{8\}\times 10^{8}$ & 0.07 & 0.39 & 0.06 \\
  & $\delta c_3\{2,|\Delta\eta|>2\} \times 10^{4}$  & --0.16 & 0.20 & 0.11   \\
  & $\delta c_4\{2,|\Delta\eta|>2\} \times 10^{4}$  & --0.09 & --0.05 & 0.12   \\ 
\bottomrule
\end{tabular}
\end{center}
\end{table}

\printbibliography


\clearpage
\appendix

\newpage 
\begin{flushleft}
{\Large The ATLAS Collaboration}

\bigskip

M.~Aaboud$^\textrm{\scriptsize 137d}$,
G.~Aad$^\textrm{\scriptsize 88}$,
B.~Abbott$^\textrm{\scriptsize 115}$,
J.~Abdallah$^\textrm{\scriptsize 8}$,
O.~Abdinov$^\textrm{\scriptsize 12}$$^{,*}$,
B.~Abeloos$^\textrm{\scriptsize 119}$,
S.H.~Abidi$^\textrm{\scriptsize 161}$,
O.S.~AbouZeid$^\textrm{\scriptsize 139}$,
N.L.~Abraham$^\textrm{\scriptsize 151}$,
H.~Abramowicz$^\textrm{\scriptsize 155}$
% ...
\bigskip
\\
% ...
$^{8}$ Department of Physics, The University of Texas at Arlington, Arlington TX, United States of America\\
% ...
$^{12}$ Institute of Physics, Azerbaijan Academy of Sciences, Baku, Azerbaijan\\
% ...
$^{88}$ CPPM, Aix-Marseille Universit{\'e} and CNRS/IN2P3, Marseille, France\\
% ...
$^{115}$ Homer L. Dodge Department of Physics and Astronomy, University of Oklahoma, Norman OK, United States of America\\
% ...
$^{119}$ LAL, Univ. Paris-Sud, CNRS/IN2P3, Universit{\'e} Paris-Saclay, Orsay, France\\
% ...
$^{137}$ $^{(a)}$ Facult{\'e} des Sciences Ain Chock, R{\'e}seau Universitaire de Physique des Hautes Energies - Universit{\'e} Hassan II, Casablanca; $^{(b)}$ Centre National de l'Energie des Sciences Techniques Nucleaires, Rabat; $^{(c)}$ Facult{\'e} des Sciences Semlalia, Universit{\'e} Cadi Ayyad, LPHEA-Marrakech; $^{(d)}$ Facult{\'e} des Sciences, Universit{\'e} Mohamed Premier and LPTPM, Oujda; $^{(e)}$ Facult{\'e} des sciences, Universit{\'e} Mohammed V, Rabat, Morocco\\
% ...
$^{139}$ Santa Cruz Institute for Particle Physics, University of California Santa Cruz, Santa Cruz CA, United States of America\\
% ...
$^{151}$ Department of Physics and Astronomy, University of Sussex, Brighton, United Kingdom\\
% ...
$^{155}$ Raymond and Beverly Sackler School of Physics and Astronomy, Tel Aviv University, Tel Aviv, Israel\\
% ...
$^{161}$ Department of Physics, University of Toronto, Toronto ON, Canada\\
% ...
$^{a}$ Also at Department of Physics, King's College London, London, United Kingdom\\
$^{b}$ Also at Institute of Physics, Azerbaijan Academy of Sciences, Baku, Azerbaijan\\
$^{c}$ Also at Novosibirsk State University, Novosibirsk, Russia\\
$^{d}$ Also at TRIUMF, Vancouver BC, Canada\\
$^{e}$ Also at Department of Physics {\&} Astronomy, University of Louisville, Louisville, KY, United States of America\\
% ...
$^{*}$ Deceased
\end{flushleft}


\end{document}